\documentclass[12pt,nofootinbib]{article}
\usepackage{graphicx}
\usepackage{amssymb}
\usepackage{amsmath}
\usepackage{color}
\usepackage[colorlinks=true,linkcolor=blue,citecolor=blue]{hyperref}

\newcommand{\bear}{\begin{array}}  
\newcommand {\eear}{\end{array}}
\newcommand{\bea}{\begin{eqnarray}}   
\newcommand{\eea}{\end{eqnarray}}
\newcommand{\beq}{\begin{equation}}   
\newcommand{\eeq}{\end{equation}}
\newcommand{\bef}{\begin{figure}}  \newcommand 
{\eef}{\end{figure}}
\newcommand{\bec}{\begin{center}}  \newcommand 
{\eec}{\end{center}}

\definecolor{orange}{cmyk}{0,0.61,0.87,0}
\definecolor{pinegreen}{cmyk}{0.92,0,0.59,0.25}
\definecolor{brown}{cmyk}{0,0.81,1,0.60}

\begin{document}

\begin{titlepage}

\begin{flushright}
%ICRR-Report-???\\
IPMU 11-0124 \\
UT-11-22\\
\end{flushright}

\vskip 1.35cm

%\preprint{ICRR-Report-???}
%\preprint{IPMU 09-????}

\begin{center}

{\Large \bf
Isocurvature perturbations in extra radiation
}

\vskip 1.2cm

Masahiro Kawasaki$^{a,b}$,
Koichi Miyamoto$^a$,
Kazunori Nakayama$^c$\\
and
Toyokazu Sekiguchi$^d$

\vskip 0.4cm

{\it $^a$Institute for Cosmic Ray Research,
University of Tokyo, Kashiwa 277-8582, Japan}\\
{\it $^b$Institute for the Physics and Mathematics of the Universe,
University of Tokyo, Kashiwa 277-8568, Japan}\\
{ \it $^c$Department of Physics, University of Tokyo, Bunkyo-ku, Tokyo 113-0033, Japan}\\
 {\it $^d$Department of Physics and Astrophysics, Nagoya University, Nagoya 464-8602, Japan}\\

 \date{\today}
\vspace{1.5cm}

\begin{abstract} 
Recent cosmological observations, including measurements of the CMB anisotropy and 
the primordial helium abundance, indicate the existence of an extra 
radiation component in the Universe beyond the standard three 
neutrino species.
In this paper we explore the possibility that the extra radiation has isocurvatrue fluctuations.
A general formalism to evaluate isocurvature perturbations in the extra radiation is provided
in the mixed inflaton-curvaton system, where the extra radiation is produced by the decay of both scalar fields.
We also derive constraints on the abundance of the
extra radiation and the amount of its isocurvature perturbation.
Current observational data favors the existence of an extra radiation 
component, but does not indicate its having isocurvature perturbation.
These constraints are applied to some particle physics motivated models.
If future observations detect isocurvature perturbations in the extra 
radiation, it will give us a hint to the origin of the extra radiation. 
\end{abstract}

%\pacs{95.35.+d}

%\maketitle

\end{center}
\end{titlepage}

\newpage
\tableofcontents

\newpage

%%%%%%%%%%%%%%%%%%%%%%%%%%%%%%%%%%%
\section{Introduction}
%%%%%%%%%%%%%%%%%%%%%%%%%%%%%%%%%%%

It is known that most of the energy content of the present Universe 
consist of ``dark'' components: dark matter and dark energy.
It is believed that the remaining components are well-known stuff: 
baryons, photons and neutrinos.
These are ingredients of the standard cosmology supported by 
cosmological observations.

However, there is no reason to exclude the existence of additional 
component other than listed above as long as its abundance is not 
large so that the cosmological evolution scenario is not much affected.
Actually, the WMAP measurement of the cosmic microwave background (CMB) 
anisotropy combined with observations of the baryon acoustic 
oscillation (BAO) and the Hubble parameter (H0)
constrains the effective number of neutrino species 
$N_{\rm eff} $ as $N_{\rm eff}=4.34_{-0.88}^{+0.86}$ (68\%C.L.)
\cite{Komatsu:2010fb}.
Including the Atacama Cosmology Telescope (ACT) data, 
the constraint is slightly improved as
$N_{\rm eff}=4.56 \pm 0.75$ (68\%C.L.)~\cite{Dunkley:2010ge}.
The recent results from South Pole Telescope (SPT), 
combined with WMAP, BAO and H0 indicate $N_{\rm eff} = 3.86\pm 0.42$ 
(68\%C.L.)~\cite{Keisler:2011aw}.
Thus these observations suggest the presence of an
extra radiation component beyond the standard three neutrino species
at the $2\sigma$ level
\footnote{
%%%
There are several studies which discuss 
how significant the deviation ($N_\mathrm{eff}\ne3.04$) is.
For detailed discussion, we refer to Refs.~\cite{GonzalezMorales:2011ty,Hamann:2011hu}
%%%
}.
(See also Refs.~\cite{Mangano:2006ur,Cirelli:2006kt,Ichikawa:2008pz,Simha:2008zj} for former analyses.)
Hereafter we call non-interacting relativistic energy component other than neutrinos as ``extra radiation''.

The CMB measurement is sensitive to the expansion rate of the Universe, 
or the total energy density of the Universe, at around the recombination epoch.
On the other hand, the big-bang nucleosynthesis (BBN) also gives a
constraint on the expansion rate at the cosmic temperature of 
$T\sim 1$MeV, where the weak 
interaction is frozen and the neutron-proton ratio is fixed.
Thus the observation of the primordial helium abundance gives 
a constraint 
on the extra radiation component at $T\sim 1$MeV.
Recently, it was reported that the primordial helium abundance 
derived from the observations of the extragalactic HII regions indicates 
an excess at the $2\sigma$ level compared with the theoretical 
expectation using the baryon number obtained from the WMAP 
data~\cite{Izotov:2010ca}.
(See also Ref.~\cite{Aver:2010wq}.)
Ref.~\cite{Hamann:2010bk} studied constraints on the effective number 
of neutrinos $N_{\rm eff}$ and the mass of extra radiation component using CMB (including WMAP 7yr, 
ACBAR, BICEP and QUaD), SDSS and H0 data,
and similarly claimed the existence of an extra radiation and 
that its mass should be smaller than $\sim 0.4$~eV.
If these observations are true,  we need a theory beyond the standard 
model in which a new light particle species exists.\footnote{%%
Discrepancy between the prediction and observation of the primordial helium abundance 
may be solved in the large lepton asymmetry 
scenario~\cite{Hansen:2001hi,Kawasaki:2002hq,Popa:2008tb,Mangano:2010ei}.
We do not pursue this issue in this paper.
}
Some cosmological scenarios and particle physics models are proposed
in order to explain $\Delta N_{\rm eff} \simeq 1$~\cite{Ichikawa:2007jv,Krauss:2010xg,Nakayama:2010vs,Fischler:2010xz,deHolanda:2010am}.

The most literature treating the extra radiation so far (implicitly) 
assumes that the extra radiation has the adiabatic perturbation.
However, the property of the fluctuation of the extra radiation 
depends on its production mechanism.
For example, there is a scenario that the decay of a scalar field 
nonthermally produces an extra radiation 
component~\cite{Ichikawa:2007jv,Fischler:2010xz}.
In this case the extra radiation has the fluctuation of the source 
scalar field, and if the scalar field is light during inflation,
it obtains large-scale quantum fluctuations.
Therefore, the extra radiation can have an 
independent perturbation from the adiabatic one: %perturbation :
it is an isocurvature perturbation. %mode
%Therefore, 
Investigating the isocurvature perturbation 
in the extra radiation may have a potential to distinguish the model 
of the extra radiation, if future observations further confirm 
$\Delta N_{\rm eff} \simeq 1$. 

Therefore, in this paper we study the effect of the isocurvature 
perturbation in the extra radiation
and derive constraints on them using the currently available datasets.
The effect on the CMB anisotropy is similar to that induced by 
the neutrino isocurvaure perturbation.
But we are mainly interested in the case where the energy density 
of the extra radiation itself causes significant change in the Hubble 
expansion rate.
These effects of the extra radiation with isocurvature perturbation 
are phenomenologically taken into account by scanning both the effective 
number of neutrino species $N_{\rm eff}$ and the neutrino isocurvature 
perturbation $S_\nu$.

First we develop formalism for calculating the primordial
isocurvature perturbation in the extra radiation
%the extra radiation isocurvature perturbation 
in Sec.~\ref{sec:formalism}.
For concreteness we restrict ourselves to the two-scalar field case, 
which generalizes the curvaton model~\cite{Lyth:2001nq}.
The extra radiation is assumed to be produced from decay of either or 
both of the scalar fields.
We also point out that a large non-Gaussianity may exist in the 
isocurvature perturbation of the extra radiation.
A similar study for the case of CDM/baryon isocurvature perturbations, 
including their non-Gaussianities, was done in 
Refs.~\cite{Kawasaki:2008sn,Kawasaki:2008jy,Langlois:2008vk,
Kawasaki:2008pa,Hikage:2008sk,Kawakami:2009iu,Hikage:2009rt,
Nakayama:2009cr,Langlois:2010dz,Langlois:2010fe,Langlois:2011hn}.
In Sec.~\ref{sec:signature}, we discuss the observational signatures 
of our model. 
In particular, we focus on the CMB angular power spectrum, 
which would be the best probe of our model for the time being.
%As we will see, our model can be best probed by the CMB angular power spectrum.
Then we derive observational constraints on such models with 
extra radiations and isocurvature perturbations using currently 
available datasets including WMAP, ACT, BAO and $H_0$ measurement
in Sec.~\ref{sec:constraint}.
They are applied to some particle physics-motivated models 
in Sec.~\ref{sec:model}.
We conclude in Sec.~\ref{sec:conc}.

We note that author in Ref.~\cite{Hu:1998gy} also
considered a specific model for isocurvature perturbations in extra 
radiation in the different context.

%%%%%%%%%%%%%%%%%%%%%%%%%%%%%%%%%%%
\section{Formalism}    \label{sec:formalism}
%%%%%%%%%%%%%%%%%%%%%%%%%%%%%%%%%%%

We consider a cosmological scenario of two scalar fields, both of which 
obtain large-scale quantum fluctuations and may generate isocurvature 
perturbations as well as adiabatic ones.
One is the inflaton field $\phi$ which initially dominates the energy 
density of the Universe during and soon after inflation.
We also introduce another scalar field $\sigma$.
We do not specify which of them is the dominant source of the adiabatic 
perturbation, since it depends much on models.
For convenience we call $\sigma$ the curvaton even if it may not 
dominantly generate the adiabatic perturbation. 

All the components of the Universe, photon, %radiation, 
neutrino, CDM, baryon and 
extra radiation arise from both $\phi$ and $\sigma$.
In general there exist isocurvature perturbations among these components.
But we particularly focus on the isocurvature perturbations between the 
extra radiation and the standard radiation.
We denote the extra radiation component by $X$, photon by $\gamma$, 
and neutrino by $\nu$.
We also denote the Standard Model radiation which 
exists prior to the neutrino decoupling by $r$, 
and the $X$ together with $\nu$ by ``DR'' (dark radiation).
We assume $X$ is produced by the decay of $\phi$ and/or $\sigma$ and
has no interaction with the other components throughout the whole history 
of the Universe after production.

%%%%%%%%%%%%%%%%%%%%%%%%%%%%%%%%%%%
\subsection{Nonlinear isocurvature perturbation}
%%%%%%%%%%%%%%%%%%%%%%%%%%%%%%%%%%%

We define $\zeta$ as the curvature perturbation on the uniform-density 
slice.
After the curvaton decay, it is the curvature perturbation 
on the slice where the 
total radiation energy density is spatially uniform.\footnote{%%
For the reason discussed later, we distinguish the curvature perturbation at different 
epochs: $\zeta$, $\tilde \zeta$ and $\hat\zeta$. They correspond to the
curvature perturbation after the curvaton decay, neutrino freezeout and $e^{\pm}$ annihilation, respectively. 
It is $\hat\zeta$ that is directly related to cosmological observations.
}
According to the $\delta N$ formalism~\cite{Sasaki:1995aw,Lyth:2004gb},
%it is the difference of the e-folding number between this slice and spatially flat slice,
it is related to the local e-folding number $N(t_f,t_i; \vec x)$ as
\begin{equation}
%	\hat\zeta(\vec x) = N(t_f,t_i; \vec x) - \bar N(t_f,t_i).
	\hat\zeta(\vec x) = N(t_f,t_i; \vec x) 
	- \ln\frac{a(t_{f})}{a(t_{i})},
\end{equation}
where $a(t)$ is the background scale factor. 
Here the hypersurface at $t=t_f$, in the radiation dominated era 
after the curvaton decays, 
is chosen to be the uniform density slice 
and the initial one at $t=t_i$ to be spatially flat slice.
The curvature perturbation $\zeta$ on large scales is conserved unless
there are isocurvature perturbations, or non-adiabatic pressure.
It is expanded by the scalar field fluctuations $\delta \phi_i$ as
\begin{equation}
	\hat\zeta = N_{\phi_i}\delta \phi_i + \frac{1}{2}N_{\phi_i \phi_j}  \delta \phi_i\delta \phi_j + \dots.
\end{equation}
In the two-field case, up to the second order, it is given by
\begin{equation}
	\hat\zeta = N_{\phi}\delta \phi + N_{\sigma}\delta \sigma 
	+ \frac{1}{2}N_{\phi \phi} \delta \phi \delta \phi 
	+ N_{\phi \sigma} \delta \phi \delta \sigma
	+  \frac{1}{2}N_{\sigma \sigma}  \delta \sigma \delta \sigma.
\end{equation}

Similarly we define the curvature perturbation of the fluid $i$, $\zeta_i$, as the curvature perturbation 
on the hypersurface where the energy density of $i$-th fluid is spatially uniform.
$\zeta_{i}$'s are conserved on sufficiently large spatial scales as long as there no interactions or
energy exchange between fluids~\cite{Lyth:2004gb}.
If the $i$-th fluid has the equation of state $p_i = w_i \rho_i$, $\zeta_i$ and $\zeta$ are related by 
\begin{equation}
	\rho_i(\vec x) = \bar \rho_i e^{3(1+w_i)(\zeta_i-\zeta)},
\end{equation}
where $\rho_i(\vec x)$ is evaluated on the uniform density slice.
Then we define the isocurvature perturbation between $i$-th and $j$-th fluids as~\cite{Wands:2000dp}
\begin{equation}
	S_{ij} \equiv 3(\zeta_i - \zeta_j).
\end{equation}

In this paper we are interested in the isocurvature perturbation in the extra radiation $X$.
For convenience, we define the isocurvature perturbation in the dark radiation, which includes both $X$ and neutrino, as
\begin{equation}
	\hat S_{\rm DR} \equiv 3(\zeta_{\rm DR} - \hat\zeta).
\end{equation}
It is formally expanded by the scalar field fluctuations $\delta \phi_i$ as
\begin{equation}
	\hat S_{\rm DR} = S_{\phi_i}\delta \phi_i + \frac{1}{2}S_{\phi_i \phi_j}  \delta \phi_i\delta \phi_j + \dots.
\end{equation}
In the two-field case, up to the second order, it is given by
\begin{equation}
	\hat S_{\rm DR} = S_{\phi}\delta \phi + S_{\sigma}\delta \sigma 
	+ \frac{1}{2}S_{\phi \phi} \delta \phi \delta \phi 
	+ S_{\phi \sigma} \delta \phi \delta \sigma
	+  \frac{1}{2}S_{\sigma \sigma}  \delta \sigma \delta \sigma.
\end{equation}

The power spectra of the curvature/isocurvature perturbations, 
and their correlation are given by
\begin{equation}
\begin{split}
	\langle \hat\zeta(\vec k_1)\hat\zeta(\vec k_2) \rangle 
		& \equiv (2\pi)^3\delta(\vec k_1+\vec k_2) P_{\zeta\zeta}(k_1), \\
	\langle \hat\zeta(\vec k_1)\hat S_{\rm DR}(\vec k_2) \rangle 
		& \equiv (2\pi)^3\delta(\vec k_1+\vec k_2) P_{\zeta S_{\rm DR}}(k_1), \\
	\langle \hat S_{\rm DR}(\vec k_1)\hat S_{\rm DR}(\vec k_2) \rangle 
		& \equiv (2\pi)^3\delta(\vec k_1+\vec k_2) P_{S_{\rm DR}S_{\rm DR}}(k_1) ,
\end{split}
\end{equation}
where
\begin{equation}
\begin{split}
	P_{\zeta\zeta} (k) &= [ N_\phi^2 + N_\sigma^2 ]P_{\delta \phi}(k), \\
	P_{\zeta S_{\rm DR}} (k) &= [ N_\phi S_\phi + N_\sigma S_\sigma ]P_{\delta \phi}(k), \\
	P_{S_{\rm DR}S_{\rm DR}} (k) &= [ S_\phi^2 + S_\sigma^2 ]P_{\delta \phi}(k),     \label{power}
\end{split}
\end{equation}
where we have neglected higher order terms, and the power spectrum of $\delta \phi_i$ is defined as
\begin{gather}
	\langle  \delta \phi_i(\vec k_1) \delta \phi_j(\vec k_2) \rangle 
	\equiv (2\pi)^3\delta(\vec k_1+\vec k_2) P_{\delta \phi}(k_1)\delta_{ij}, \\
	P_{\delta \phi}(k) = \frac{H_{\rm inf}^2}{2k^3}\left( \frac{k}{k_0} \right)^{n_s-1}.
\end{gather}
Here $H_{\rm inf}$ is the Hubble parameter during inflation,
$n_s$ is the scalar spectral index\footnote{
	The scalar spectral indices for $\phi$ and $\sigma$ do not coincide in general.
	In the following we assume they are the same just for simplicity.
} 
and $k_0$ is the pivot scale chosen as $k_0=0.002{\rm Mpc}^{-1}$.

The correlation parameter between the curvature and isocurvature 
perturbations is defined by
\begin{equation}
	\gamma \equiv \frac{P_{\zeta S_\mathrm{DR}}(k_0)}
	{\sqrt{P_{\zeta\zeta} (k_0)P_{S_\mathrm{DR}S_\mathrm{DR}}(k_0)}}
	= \frac{ N_\phi S_\phi + N_\sigma S_\sigma }{\sqrt{( N_\phi^2 + N_\sigma^2)(S_\phi^2 + S_\sigma^2)}}.
	\label{gamma}
\end{equation}
The uncorrelated isocurvature perturbation corresponds to $\gamma = 0$,
and totally (anti-) correlated one to $\gamma = (-)1$.
The effect of isocurvature perturbations on the CMB anisotropy depends on its magnitude as well as
the correlation parameter.
Thus what we need is to express $N_\phi, S_\phi, $ and so on, in terms of model parameters.

For later use, we also define the dimensionless power spectrum as
$\mathcal P_{AB}(k) \equiv (k^3/2\pi^2) P_{AB}(k)$ where 
the subscripts $A$ and $B$ are either $\zeta$ or $S_{\rm DR}$.
Using the quantities given above, they are expressed as
\begin{equation}
\begin{split}
	\mathcal P_{\zeta\zeta} (k) &= [ N_\phi^2 + N_\sigma^2 ] 
		\left( \frac{H_{\rm inf}}{2\pi} \right)^2 \left( \frac{k}{k_0} \right)^{n_s-1}, \\
	\mathcal P_{\zeta S_{\rm DR}} (k) &= [ N_\phi S_\phi + N_\sigma S_\sigma ]
		\left( \frac{H_{\rm inf}}{2\pi} \right)^2 \left( \frac{k}{k_0} \right)^{n_s-1}, \\
	\mathcal P_{S_{\rm DR}S_{\rm DR}} (k) &= [ S_\phi^2 + S_\sigma^2 ]
		\left( \frac{H_{\rm inf}}{2\pi} \right)^2 \left( \frac{k}{k_0} \right)^{n_s-1}.
	\label{dimless-power}
\end{split}
\end{equation}
In the following we formulate a method for calculating the extra radiation isocurvature perturbation
in a typical cosmological setup.

%%%%%%%%%%%%%%%%%%%%%%%%%%%%%%%%%%%%
\subsection{Extra radiation and isocurvature perturbation}  
\label{sec:ext}
%%%%%%%%%%%%%%%%%%%%%%%%%%%%%%%%%%%%

Now let us move to our concrete setup.
Here we consider ${\phi}$ as the inflaton field.
We want to know the final dark radiation (``DR'') isocurvature perturbation,
and hence we must connect it to the primordial perturbations of $\phi$ and $\sigma$
through some transition points where constituents of the cosmological fluids change.
Thus we evaluate the curvature perturbations step-by-step in the following.
The evolution of the fluids in our model is summarized in Table.~\ref{component}.
We assume that the $\sigma$ decays well before BBN begins.
Considering that the decay of $\sigma$ after BBN might produce significant energy in the form of radiations and hadrons
and upset the standard BBN, it is a reasonable assumption.
In the most part of this paper we take this assumption.\footnote{
The BBN constraint can be avoided if the $\sigma$ decays dominantly to $X$ particles even if its decay occurs after BBN.
For completeness, we will also discuss such a case in Appendix.
}

In general, the CDM and baryon can have isocurvature perturbations depending on their origins.
However, including them makes the analysis too complicated.
Thus we focus on the case where the extra radiation has an isocurvature perturbation and
CDM/baryon do not.
For example, if CDM/baryon are created from thermal bath after the curvaton decays, 
they only have adiabatic perturbations and our assumption is justified.

%%%%%%%%%%%%%%%% table %%%%%%%%%%%%%%%%%%%%%%
\begin{table}[t]
  \begin{center}
    \begin{tabular}{ | c | c | c |}
      \hline 
         epoch  & component &  energy transfer  \\ \hline \hline
         $\Gamma_\phi < H  $                                  & $\phi$, $\sigma$  & $\phi \to X^{(\phi)}+r^{(\phi)} $ \\ \hline 
         $\Gamma_\sigma < H < \Gamma_\phi$  & $X^{(\phi)}$, $r^{(\phi)}$, $\sigma$ & $\sigma \to X^{(\sigma)}+r^{(\sigma)} $ \\ \hline 
         $\Gamma_{\nu} < H < \Gamma_\sigma$ & $X$, $r$ & $r\to \nu + r_e$ \\ \hline
         $\Gamma_{e^\pm} < H <\Gamma_\nu$   & $X$, $\nu$, $r_e$  (${\rm DR}=X+\nu$) &  $e^\pm \to \gamma$   \\ \hline
                  $H<\Gamma_{e^\pm}$         & $X$, $\nu$, $\gamma$  (${\rm DR}=X+\nu$) &     \\ \hline

    \end{tabular}
    \caption{ 
    Energy components of the Universe at each epoch except for CDM and 
    baryon.
	Here $\Gamma_{\phi} (\Gamma_{\sigma})$ is the decay rate of the 
	inflaton (curvaton), and $\Gamma_{\nu}$ denotes the neutrino
	interaction 
	rate at the neutrino freezeout.
	$\Gamma_{e^\pm}$ denotes the Hubble parameter at the $e^\pm$ 
	annihilation. 
	Here $r_e$ denotes the plasma consisting of $\gamma$ and $e^\pm$.
    }
    \label{component}
  \end{center}
\end{table}
%%%%%%%%%%%%%%%%%%%%%%%%%%%%%%%%%%%%%%%%%%%%%% 

%%%%%%%%%%%%%%%%%%%%%%%%%%%%%%%%%%%
\subsubsection{At the curvaton decay}
%%%%%%%%%%%%%%%%%%%%%%%%%%%%%%%%%%%

Let us take the uniform density slice at the curvaton decay : $H(\vec x)= \Gamma_\sigma$,
where $\Gamma_\sigma$ is the decay rate of $\sigma$.
We make use of the sudden decay approximation in the following analysis~\cite{Sasaki:2006kq}.
On this slice, we have following relations
\begin{gather}
	\rho_X^{(\phi)}(\vec x) + (1-r_\sigma) \rho_\sigma (\vec x) = \rho_X(\vec x),\\
	\rho_r^{(\phi)}(\vec x) + r_\sigma \rho_\sigma (\vec x) = \rho_r(\vec x), \\
	\rho_r^{(\phi)}(\vec x)+\rho_X^{(\phi)}(\vec x)+ \rho_\sigma (\vec x) = \rho_{\rm tot}(\vec x) (=\bar \rho_{\rm tot}),
\end{gather}
where $r_\sigma$ is the branching fraction of $\sigma$ into particles other than the extra radiation $X$.
Superscript $(\phi)$ means that the corresponding component comes from the inflaton decay.
The curvature perturbation of each component is related to its background value as
\begin{gather}
	\rho_X^{(\phi)}(\vec x) =\bar \rho_X^{(\phi)} e^{4(\zeta_\phi-\zeta)},\\
	\rho_r^{(\phi)}(\vec x) =\bar \rho_r^{(\phi)} e^{4(\zeta_\phi-\zeta)},\\
	\rho_X(\vec x) =\bar \rho_X e^{4(\zeta_X-\zeta)},\\
	\rho_r(\vec x) =\bar \rho_r  e^{4(\zeta_r-\zeta)},\\	
	\rho_\sigma(\vec x) =\bar \rho_\sigma e^{3(\zeta_\sigma-\zeta)}.
\end{gather}
From these equations we obtain
%%
%\begin{gather}
%	A e^{4(\zeta_\phi-\zeta)} + B e^{3(\zeta_\sigma-\zeta)} = (A+B)e^{4(\zeta_X-\zeta)},\\
%	C e^{4(\zeta_\phi-\zeta)} + D e^{3(\zeta_\sigma-\zeta)} = (C+D)e^{4(\zeta_r-\zeta)},\\
%	(1-R_\sigma)e^{4(\zeta_\phi-\zeta)}+R_\sigma e^{3(\zeta_\sigma-\zeta)} = 1,
%\end{gather}
%%
%%
\begin{gather}
	R_X^{(\phi)} e^{4(\zeta_\phi-\zeta)} + R_X^{(\sigma)} e^{3(\zeta_\sigma-\zeta)} = R_X e^{4(\zeta_X-\zeta)},\\
	R_r^{(\phi)} e^{4(\zeta_\phi-\zeta)} + R_r^{(\sigma)} e^{3(\zeta_\sigma-\zeta)} = R_r e^{4(\zeta_r-\zeta)},\\
	(1-R_\sigma)e^{4(\zeta_\phi-\zeta)}+R_\sigma e^{3(\zeta_\sigma-\zeta)} = 1,
\end{gather}
where we have defined
\begin{equation}
	R_\sigma \equiv \bar \rho_\sigma / \bar \rho_{\rm tot} ~~~{\rm at~\sigma~decay}, \\
\end{equation}
and
\begin{equation}
\begin{split}
	R_X^{(\phi)}  &\equiv \bar \rho_X^{(\phi)} / \bar \rho_{\rm tot} = (1-r_\phi)(1-R_\sigma),\\  % =A
	R_X^{(\sigma)} & \equiv \bar \rho_X^{(\sigma)} / \bar \rho_{\rm tot} = (1-r_\sigma) R_\sigma,\\ % =B
	R_r^{(\phi)} & \equiv \bar \rho_r^{(\phi)} / \bar \rho_{\rm tot} = r_\phi (1-R_\sigma),\\ % =C
	R_r^{(\sigma)}  &\equiv \bar \rho_r^{(\sigma)} / \bar \rho_{\rm tot} = r_\sigma R_\sigma, % =D
	\label{RX-Rr}
\end{split}
\end{equation}
where $r_\phi$ is the branching fraction of $\phi$ into particles other than the 
extra radiation $X$,\footnote{%%
Precisely speaking, $1-r_\phi$ is not the branching ratio of 
$\phi$ into $X$ ($B_{\phi\to X}$).
It is given by 
$1-r_\phi = B_{\phi\to X}\epsilon / (1-B_{\phi\to X}(1-\epsilon))$
where $\epsilon \equiv 
\left[ g_*(H=\Gamma_\sigma)/g_*(H=\Gamma_\phi)\right]^{1/3}$
with $\Gamma_\phi$ denoting the inflaton decay width,
taking into account the change of relativistic degrees of freedom
between the inflaton decay and curvaton decay. 
}
and $R_X = R_X^{(\phi)}+R_X^{(\sigma)}=\bar\rho_X/\bar\rho_{\rm tot}$, 
$R_r =R_r^{(\phi)}+R_r^{(\sigma)} =\bar\rho_r/\bar\rho_{\rm tot}$.
Notice that $R_X + R_r = 1$.
All of these quantities are evaluated at the curvaton decay, $H=\Gamma_\sigma$.
As is already stated, these quantities are not constants in time because of the changes in the 
relativistic degrees of freedom $g_*$.
Solving these equations up to the second order in $\zeta_\phi$ and $\zeta_\sigma$, we obtain
%%
%\begin{equation}
%\begin{split}
%	\zeta &= (1-R)\zeta_\phi + R \zeta_\sigma + \frac{1}{2}R(1-R)(3+R)(\zeta_\phi-\zeta_\sigma)^2, \\
%	\zeta_X&=\frac{1}{4(A+B)}\left[ (4A+(1-R)B)\zeta_\phi +(3+R)B\zeta_\sigma  \right] \\
%		&~~ +\frac{B(3+R)}{8(A+B)^2}\left[ (1+R)(3-R)A + (1-R)R B \right](\zeta_\phi-\zeta_\sigma)^2,\\
%	\zeta_r&=\frac{1}{4(C+D)}\left[ (4C+(1-R)D)\zeta_\phi +(3+R)D\zeta_\sigma  \right] \\
%		&~~ +\frac{D(3+R)}{8(C+D)^2}\left[ (1+R)(3-R)C + (1-R)R D \right](\zeta_\phi-\zeta_\sigma)^2,
%\end{split}
%\end{equation}
%%
%%
\begin{equation}
\begin{split}
	\zeta &= \zeta_\phi + R (\zeta_\sigma -\zeta_\phi) + \frac{1}{2}R(1-R)(3+R)(\zeta_\phi-\zeta_\sigma)^2, \\
	\zeta_X&=\frac{1}{4R_X}\left[ (4R_X^{(\phi)}+(1-R)R_X^{(\sigma)})\zeta_\phi +(3+R)R_X^{(\sigma)}\zeta_\sigma  \right] \\
		&~~ +\frac{(3+R)R_X^{(\sigma)}}{8R_X^2}
		\left[ (1+R)(3-R)R_X^{(\phi)} + (1-R)R R_X^{(\sigma)} \right](\zeta_\phi-\zeta_\sigma)^2,\\
	\zeta_r&=\frac{1}{4R_r}\left[ (4R_r^{(\phi)}+(1-R)R_r^{(\sigma)})\zeta_\phi +(3+R)R_r^{(\sigma)}\zeta_\sigma  \right] \\
		&~~ +\frac{(3+R)R_r^{(\sigma)}}{8R_r^2}
		\left[ (1+R)(3-R)R_r^{(\phi)} + (1-R)R R_r^{(\sigma)} \right](\zeta_\phi-\zeta_\sigma)^2,
	\label{zeta_from_zetaphi}
\end{split}
\end{equation}
where we have defined 
\begin{equation}
	R \equiv \frac{3R_\sigma}{4-R_\sigma}  ~~~\leftrightarrow ~~~ R_\sigma= \frac{4R}{3+R}.
	\label{R-Rsigma}
\end{equation}
Note that $\zeta_X$ and $\zeta_r$ are conserved quantities at superhorizon scales~\cite{Wands:2000dp}.
In general, the curvature perturbation $\zeta$ is not conserved in the presence of non-adiabatic pressure.
%In the present case, however, both $X$ and $r$ have same equation of state and hence
%$\zeta$ is also conserved after the curvaton decays.
In the present case, both $X$ and $r$ behave as radiation,
and hence one may consider that $\zeta$ is conserved on sufficiently large scales.
More precisely, however, the radiation energy density $\rho_r$ does not scale as $a(t)^{-4}$
because of the changes in the relativistic degrees of freedom $g_*$.
Therefore, $\zeta$ is not conserved even after the curvaton decay.
The changes in $\zeta$ at the changes in $g_*$ depends on the fraction in which the $X$ dominates the Universe.

%%%%%%%%%%%%%%%%%%%%%%%%%%%%%%%%%%%
\subsubsection{At the neutrino freezeout}
%%%%%%%%%%%%%%%%%%%%%%%%%%%%%%%%%%%

Next let us take the uniform density slice at the freezeout of neutrinos $H=\Gamma_\nu$
($T\sim 1$MeV). %\footnote{
%	Precisely speaking, the hypersurface of the neutrino freezeout does not coincide with the uniform density slice.
%	This is because the freezeout occurs when $H = \Gamma_\nu \sim G_F^2 T^5$ 
%	where $G_F$ is the Fermi constant, and
%	the extra radiation has non-negligible energy density and affects the Hubble expansion rate 
%	on the left hand side,
%	while the temperature appearing on the right hand side is determined by the visible radiation.
%	This may produce ``induced'' isocurvature perturbation on the neutrino.
%	We do not consider it here because it is a higher order effect.
%}
On this slice we have the following relations
\begin{gather}
	(1-c_\nu)\rho_r(\vec x) = \rho_\gamma(\vec x),
	\label{eq:gamma_at_neu_freeze}\\
	c_\nu \rho_r(\vec x)  = \rho_\nu (\vec x), 
	\label{eq:neu_at_neu_freeze}\\
	\rho_r(\vec x)+\rho_X(\vec x) = \rho_{\rm tot}(\vec x) (=\bar \rho_{\rm tot}),
\end{gather}
where $c_\nu$ is the energy fraction of neutrinos. 
For the standard three neutrino species, it is given by
\begin{equation}
	c_\nu = \frac{\bar \rho_\nu}{\bar \rho_r} = \frac{21}{43}~~~~~{\rm at~neutrino~freezeout}.
\end{equation}
Here the symbol $\gamma$ should be interpreted as a thermal plasma consisting of photons electrons and positrons.
The curvature perturbation of each component is related to its background value as
\begin{gather}
	\rho_\nu(\vec x) =\bar \rho_\nu e^{4(\zeta_\nu-\tilde\zeta)},\\
	\rho_\gamma(\vec x) =\bar \rho_\gamma e^{4(\zeta_\gamma-\tilde\zeta)},
\end{gather}
where $\tilde\zeta$ denotes the curvature perturbation at this epoch 
and it is different from $\zeta$ in Eq.~(\ref{zeta_from_zetaphi}).
From Eqs. (\ref{eq:gamma_at_neu_freeze}) and (\ref{eq:neu_at_neu_freeze}) we find 
\begin{equation}
	\zeta_\gamma = \zeta_r, ~~~\zeta_\nu = \zeta_r ,
\end{equation}
which holds nonlinearly, as is expected.
We also define dark radiation (DR) as the sum of extra radiation $X$ and neutrino. 
It satisfies 
\begin{equation}
	\rho_X (\vec x)+c_\nu\rho_r (\vec x) = 
	\rho_X (\vec x)+\rho_\nu (\vec x) \equiv \rho_{\rm DR}(\vec x),   \label{rhoDR}
\end{equation}
on the uniform density slice at the neutrino freezeout.
The curvature perturbation of DR, $\zeta_{\rm DR}$, is given by
\begin{equation}
	\rho_{\rm DR}(\vec x) =\bar \rho_{\rm DR} e^{4(\zeta_{\rm DR}-\tilde\zeta)}.
\end{equation}
From Eq.~(\ref{rhoDR}), we find at first order
%%
%\begin{equation}
%	\zeta_{\rm DR} = \frac{\rho_X}{\rho_{\rm DR}}\zeta_X + \frac{\rho_\nu}{\rho_{\rm DR}}\zeta_\nu
%	= \frac{A+B}{A'+B'}\zeta_X + c_\nu\frac{C+D}{A'+B'}\zeta_r,
%\end{equation}
%%
%%
\begin{equation}
	\zeta_{\rm DR} = 
	\frac{\bar\rho_X}{\bar\rho_{\rm DR}}\zeta_X + \frac{\bar\rho_\nu}{\bar\rho_{\rm DR}}\zeta_\nu
	= \frac{\tilde R_X}{\tilde R_{\rm DR}}\zeta_X + \frac{c_\nu \tilde R_r}{\tilde R_{\rm DR}}\zeta_r,
\end{equation}
where we have defined $\tilde R_{\rm DR} \equiv \tilde R_{\rm DR}^{(\phi)}+\tilde R_{\rm DR}^{(\sigma)} = \bar\rho_{\rm DR}/\bar\rho_{\rm tot} |_{H=\Gamma_\nu}$,
with 
%%
%\begin{equation}
%	A' = A +c_\nu C, ~~~B' = B +c_\nu D,
%\end{equation}
%%
%%
\begin{equation}
\begin{split}
	\tilde R_{\rm DR}^{(\phi)} &= \tilde R_X^{(\phi)} +c_\nu \tilde R_r^{(\phi)},\\ %= (1-r_\phi(1-c_\nu))(1-R_\sigma),  \\
	\tilde R_{\rm DR}^{(\sigma)} &= \tilde R_X^{(\sigma)} +c_\nu \tilde R_r^{(\sigma)}.% = (1-r_\sigma(1-c_\nu))R_\sigma.
\end{split}
\end{equation}
Here quantities with a tilde means that they are evaluated at the neutrino freezeout.
Relations between quantities with and without a tilde are given by
\begin{equation}
	\tilde R_X = \frac{R_X}{R_X+R_r\left( \frac{g_*(H=\Gamma_\sigma)}{g_*(H=\Gamma_\nu)} \right)^{1/3}},
	~~~\tilde R_r = \frac{R_r \left( \frac{g_*(H=\Gamma_\sigma)}{g_*(H=\Gamma_\nu)} \right)^{1/3} }
	{ R_X+R_r\left( \frac{g_*(H=\Gamma_\sigma)}{g_*(H=\Gamma_\nu)} \right)^{1/3}  }.
\end{equation}
This is approximated as
\begin{equation}
	\tilde R_X \simeq R_X \left( \frac{g_*(H=\Gamma_\nu)}{g_*(H=\Gamma_\sigma)} \right)^{1/3},
	~~~\tilde R_r \simeq R_r,
\end{equation}
if $\rho_X (H=\Gamma_\nu) \ll \rho_r(H=\Gamma_\nu)$.
Here $g_*(H=\Gamma_\nu) = 10.75$ is the relativistic degrees of freedom at the neutrino freezeout.
Using these quantities, we find that $\tilde\zeta$ is given by
\begin{equation}
\begin{split}
	\tilde\zeta &= \zeta_\phi + \frac{3+R}{4}\left( 
		\frac{\tilde R_r R_r^{(\sigma)}}{R_r}+\frac{\tilde R_X R_X^{(\sigma)}}{R_X} \right) \\
		&\times \left[ (\zeta_\sigma-\zeta_\phi)
		+ \left\{\frac{(1+R)(3-R)}{2}-\frac{3+R}{2}\left(
		\frac{\tilde R_r R_r^{(\sigma)}}{R_r}+\frac{\tilde R_X R_X^{(\sigma)}}{R_X} \right) \right\}
		(\zeta_\sigma-\zeta_\phi)^2 \right].
	\label{zeta_tilde}
\end{split}
\end{equation}
It can be checked that it coincides with $\zeta$ in Eq.~(\ref{zeta_from_zetaphi})
for $\tilde R_r = R_r$ and $\tilde R_X = R_X$.
Therefore, as is expected, the relative difference between the evolutions of 
$\rho_X$ and $\rho_r$ results in the non-conservation of the curvature perturbation.  

Up to the second order in $\zeta_\phi$ and $\zeta_\sigma$, the DR isocurvature perturbation is given by
\begin{equation}
\begin{split}
	\tilde S_{\rm DR} &\equiv 3(\zeta_{\rm DR}-\tilde\zeta)  \\
	& = 3\frac{3+R}{4}\frac{\tilde R_r \tilde R_X}{\tilde R_{\rm DR}}
	\left( \frac{R_r^{(\sigma)}}{R_r}-\frac{R_X^{(\sigma)}}{R_X} \right)(1-c_\nu) 
	\Biggl [ -(\zeta_\sigma-\zeta_\phi) \Biggr. \\
	  &\left. +\left\{ \frac{(R-3)(R+1)}{2}+\frac{3+R}{2 \tilde R_{\rm DR}}
		\left(    \frac{\tilde R_r R_r^{(\sigma)} (c_\nu + \tilde R_{\rm DR}) }{R_r}
			+\frac{\tilde R_X R_X^{(\sigma)} (1+ \tilde R_{\rm DR})  }{R_X} 
		\right)
	\right\} (\zeta_\sigma-\zeta_\phi)^2
	\right].
	\label{SDR_tilde}
\end{split}
\end{equation}
%%

%%%%%%%%%%%%%%%%%%%%%%%%%%%%%%%%%%%
\subsubsection{After the $e^\pm$ annihilation}
%%%%%%%%%%%%%%%%%%%%%%%%%%%%%%%%%%%

Moreover, shortly after the neutrino freezeout, $e^{\pm}$ annihilation takes place.
This also slightly affects the curvature perturbation, since the radiation energy increases 
while other components ($X$ and $\nu$) are unaffected.
By taking the uniform density slice well after the $e^\pm$ annihilation, we find
\begin{equation}
	\rho_\gamma(\vec x)+\rho_\nu(\vec x)+\rho_X(\vec x) = \rho_{\rm tot}(\vec x) (=\bar \rho_{\rm tot}).
\end{equation}
In a similar way, we find the curvature perturbation $\hat \zeta$ at this epoch as
\begin{equation}
\begin{split}
	\hat\zeta &= \zeta_\phi + \frac{3+R}{4}\left( 
		\frac{\hat R_r R_r^{(\sigma)}}{R_r}+\frac{\hat R_X R_X^{(\sigma)}}{R_X} \right) \\
		&\times \left[ (\zeta_\sigma-\zeta_\phi)
		+ \left\{\frac{(1+R)(3-R)}{2}-\frac{3+R}{2}\left(
		\frac{\hat R_r R_r^{(\sigma)}}{R_r}+\frac{\hat R_X R_X^{(\sigma)}}{R_X} \right) \right\}
		(\zeta_\sigma-\zeta_\phi)^2 \right],
	\label{zeta}
\end{split}
\end{equation}
where quantities with a hat are defied at this epoch.
They are given by
\begin{equation}
	\hat R_X = \frac{\tilde R_X}{(1+\delta)\tilde R_r + \tilde R_X},~~~
	\hat R_r = \frac{(1+\delta)\tilde R_r}{(1+\delta)\tilde R_r + \tilde R_X},
\end{equation}
where
\begin{equation}
	1+\delta = \left( \frac{11}{4} \right)^{1/3}(1-c_\nu)+c_\nu \simeq 1.205. 
\end{equation}
Here the ``$r$'' should be interpreted as the sum of $\gamma$ and $\nu$.
The result (\ref{zeta}) has the same form as (\ref{zeta_tilde}) except for quantities with a tilde are
replaced with those with a hat.
Since $\zeta_{\rm DR}$ is conserved through the $e^\pm$ annihilation process, we can also evaluate the
DR isocurvature perturbation as
\begin{equation}
\begin{split}
	\hat S_{\rm DR} &\equiv 3(\zeta_{\rm DR}-\hat\zeta)  \\
	& = 3\frac{3+R}{4}\frac{\hat R_r \hat R_X}{\hat R_{\rm DR}}
	\left( \frac{R_r^{(\sigma)}}{R_r}-\frac{R_X^{(\sigma)}}{R_X} \right)(1-\hat c_\nu) 
	\Biggl [ -(\zeta_\sigma-\zeta_\phi) \Biggr. \\
	  &\left. +\left\{ \frac{(R-3)(R+1)}{2}+\frac{3+R}{2 \hat R_{\rm DR}}
		\left(    \frac{\hat R_r R_r^{(\sigma)} (\hat c_\nu + \hat R_{\rm DR}) }{R_r}
			+\frac{\hat R_X R_X^{(\sigma)} (1+ \hat R_{\rm DR})  }{R_X} 
		\right)
	\right\} (\zeta_\sigma-\zeta_\phi)^2
	\right],
	\label{SDR}
\end{split}
\end{equation}
where we have defined $\hat R_{\rm DR}=\bar \rho_{\rm DR}/\bar \rho_{\rm tot}$
evaluated after the $e^\pm$ annihilation, which is given by
\begin{equation}
\begin{split}
	\hat R_{\rm DR}= \hat R_X +\hat c_\nu \hat R_r,\\ %= (1-r_\phi(1-c_\nu))(1-R_\sigma),  \\
\end{split}
\end{equation}
and 
\begin{equation}
	\hat c_\nu = \frac{\bar \rho_\nu}{\bar \rho_\nu + \bar \rho_\gamma}  \simeq 0.405
	~~~{\rm after~e^\pm~annihilation.}
	\label{hatcnu}
\end{equation}
Notice that $\hat c_\nu (1+\delta) = \tilde c_\nu$.

%%%%%%%%%%%%%%%%%%%%%%%%%%%%%%%%%%%
\subsubsection{Connecting to primordial perturbations}
%%%%%%%%%%%%%%%%%%%%%%%%%%%%%%%%%%%

Finally we relate the curvature perturbation of $\phi$ and $\sigma$ to their quantum fluctuations,
$\delta \phi$ and $\delta \sigma$.
Let us take the uniform density slice when the curvaton $\sigma$ begins to oscillate.
Assuming that the curvaton energy density at this epoch is negligible
and the total energy density is dominated by the inflaton or the radiation coming from the inflaton decay,
the curvaton oscillation begins uniformly in space on this slice.
Just after that, the curvaton behaves as matter.\footnote{
	We assume the curvaton has a simple quadratic potential for simplicity.
	Otherwise, analytic estimation is difficult particularly for the analysis of non-Gaussian fluctuation
	\cite{Enqvist:2005pg,Kawasaki:2008mc}.
}
Thus we have
\begin{equation}
	\rho_\sigma (\vec x) = \bar \rho_\sigma e^{3(\zeta_\sigma - \zeta_\phi)}.
\end{equation}
From this we obtain
\begin{equation}
	\zeta_\sigma - \zeta_\phi = \frac{1}{3}\left( \frac{\delta \rho_\sigma}{\bar \rho_\sigma}
	 - \frac{1}{2}\frac{\delta \rho_\sigma^2}{\bar \rho_\sigma^2}  \right)
	 = \frac{1}{3}\left( \frac{2\delta \sigma}{\sigma_i}
	 - \frac{\delta \sigma^2}{\sigma_i^2}  \right),   \label{zetasigma}
\end{equation}
where $\sigma_i$ denotes the initial amplitude of the curvaton during inflation.
On the other hand, $\zeta_\phi$ is given by the standard formula
\begin{equation}
	\zeta_\phi = \frac{1}{M_{\rm P}^2}\frac{V}{V_\phi} \delta \phi 
	+ \frac{1}{2M_{\rm P}^2}\left( 1-\frac{VV_{\phi\phi}}{V_\phi^2}\right)(\delta\phi)^2,    \label{zetaphi}
\end{equation}
where $V$ is the inflaton potential and $V_{\phi} (V_{\phi\phi})$ are its first (second) derivative 
with respect to $\phi$.

Now we have obtained all ingredients for expressing curvature/isocurvature perturbations
in terms of model parameters.
From Eqs.~(\ref{zeta}), (\ref{zetaphi}) and (\ref{zetasigma}), we obtain
\begin{equation}
\begin{split}
	&N_\phi = \frac{1}{M_{\rm P}^2}\frac{V}{V_\phi}, \\
	&N_\sigma = \frac{3+R}{6\sigma_i}\left( 
		\frac{\hat R_r R_r^{(\sigma)}}{R_r}+\frac{\hat R_X R_X^{(\sigma)}}{R_X} \right),\\
	&N_{\phi\phi} = \frac{1}{M_{\rm P}^2}\left( 1-\frac{VV_{\phi\phi}}{V_\phi^2}\right),\\
	&N_{\sigma\sigma}=\frac{2}{9\sigma_i^2}\frac{3+R}{4}
		\left(\frac{\hat R_r R_r^{(\sigma)}}{R_r}+\frac{\hat R_X R_X^{(\sigma)}}{R_X} \right) \\
		&~~~~~\times \left[ 
		3+4R-2R^2 -2(3+R)\left(\frac{\hat R_r R_r^{(\sigma)}}{R_r}+\frac{\hat R_X R_X^{(\sigma)}}{R_X} \right)
		\right].
\end{split}
\end{equation}
From Eqs.~(\ref{SDR}) and (\ref{zetasigma}), we obtain
\begin{equation}
\begin{split}
	&S_\sigma = -\frac{3+R}{2\sigma_i}\frac{\hat R_r \hat R_X}{\hat R_{\rm DR}}
	\left( \frac{R_r^{(\sigma)}}{R_r}-\frac{R_X^{(\sigma)}}{R_X} \right)(1-\hat c_\nu)
	,\\
	&S_{\sigma\sigma} = 
	\frac{3+R}{2\sigma_i^2}\frac{\hat R_r \hat R_X}{\hat R_{\rm DR}}
	\left( \frac{R_r^{(\sigma)}}{R_r}-\frac{R_X^{(\sigma)}}{R_X} \right)\frac{(1-\hat c_\nu)}{3} \\
	&~~~\times \left[ 2R^2-4R-3 + \frac{2(3+R)}{\hat R_{\rm DR}}
	     \left(    \frac{\hat R_r R_r^{(\sigma)} (\hat c_\nu + \hat R_{\rm DR}) }{R_r}
			+\frac{\hat R_X R_X^{(\sigma)} (1+ \hat R_{\rm DR})  }{R_X}  \right)
	\right]
	,\\
	&S_\phi = S_{\phi\phi} = 0.
\end{split}
\end{equation}

For later use, we summarize quantities relevant for calculating the CMB anisotropy.
We define the effective number of neutrino species, $N_{\rm eff}$, by
\begin{equation}
	\bar\rho_{\rm DR}=\bar\rho_\nu + \bar\rho_X = 
	N_{\rm eff}\frac{7}{8}\left( \frac{4}{11} \right)^{4/3}\bar\rho_\gamma,
	\label{rhoDR0th}
\end{equation}
where quantities are evaluated after the $e^\pm$ annihilation.
The extra effective number of neutrino species, $\Delta N_{\rm eff}$, is then given by
\begin{equation}
	\Delta N_{\rm eff} =3\frac{\bar\rho_X}{\bar\rho_\nu} = \frac{3\hat R_X}{\hat c_\nu \hat R_r}
	= \frac{3\tilde R_X}{c_\nu \tilde R_r}
	= \frac{43}{7}\frac{\tilde R_X}{\tilde R_r}.
	\label{DeltaNeff}
\end{equation}
While our primary interest in this paper resides in the Gaussian perturbations
and hence the primordial power spectrum, 
the primordial bispectrum generated from non-Gaussian perturbations 
will be briefly discussed in the next subsection.
The isocurvature perturbation in the dark radiation is then given by
\begin{equation}
	\hat S_{\rm DR} = S_\sigma \delta\sigma= 
	-\frac{\delta \sigma}{\sigma_i}\frac{3+R}{2}\frac{\hat R_r \hat R_X}{\hat R_{\rm DR}}
	\left( \frac{R_r^{(\sigma)}}{R_r}-\frac{R_X^{(\sigma)}}{R_X} \right)(1-\hat c_\nu).
	\label{SDRfinal}
\end{equation}
The curvature perturbation is given by
\begin{equation}
	\hat\zeta = N_\phi \delta\phi+N_\sigma\delta\sigma = \zeta_\phi +
	\frac{3+R}{6}\left( 
		\frac{\hat R_r R_r^{(\sigma)}}{R_r}+\frac{\hat R_X R_X^{(\sigma)}}{R_X} \right)
	 \frac{\delta\sigma}{\sigma_i}.
	 \label{zetafinal}
\end{equation}
The correlation parameter (\ref{gamma}) in this case is given by
\begin{equation}
	\gamma = {\rm sign}(S_\sigma) \frac{N_\sigma}{\sqrt{N_\phi^2+N_\sigma^2}}.   \label{cor}
\end{equation}
%%

%%%%%%%%%%%%%%%%%%%%%%%%%%%%%%%%%%%%%%%%
\subsection{Non-Gaussianity in isocurvature perturbations of the extra radiation }
%%%%%%%%%%%%%%%%%%%%%%%%%%%%%%%%%%%%%%%%

In this paper we do not go into details of analysis on non-Gaussianities,
but it may be worth mentioning them briefly here
since, to the best of our knowledge,
no literature has discussed non-Gaussianities in the extra radiation isocurvature perturbation
or the neutrino isocurvture perturbation.
For this purpose, we have expressed all quantities up to the second order in $\delta \phi$ and $\delta \sigma$.

Non-Gaussianities of the cosmological perturbations are characterized by their bispectra.
In the present case we have isocurvature perturbations in the extra radiation as well as the 
adiabatic perturbation, and hence non-Gaussianities may appear in various combinations.
We define the bispectrum of the curvature/isocurvature perturbations through their
three point correlation functions as
\begin{equation}
\begin{split}
	\langle \hat\zeta(\vec k_1)\hat\zeta(\vec k_2)\hat\zeta(\vec k_3) \rangle 
		& \equiv (2\pi)^3\delta(\vec k_1+\vec k_2 + \vec k_3) B_{\zeta\zeta\zeta}(k_1,k_2,k_3), \\
	\langle \hat\zeta(\vec k_1)\hat\zeta(\vec k_2)\hat S_{\rm DR}(\vec k_3) \rangle + ({\rm cyclic})
		& \equiv (2\pi)^3\delta(\vec k_1+\vec k_2 + \vec k_3) B_{\zeta\zeta S}(k_1,k_2,k_3), \\
	\langle \hat\zeta(\vec k_1)\hat S_{\rm DR}(\vec k_2)\hat S_{\rm DR}(\vec k_3) \rangle  + ({\rm cyclic})
		& \equiv (2\pi)^3\delta(\vec k_1+\vec k_2 + \vec k_3) B_{\zeta SS}(k_1,k_2,k_3), \\
	\langle \hat S_{\rm DR}(\vec k_1)\hat S_{\rm DR}(\vec k_2)\hat S_{\rm DR}(\vec k_3) \rangle 
		& \equiv (2\pi)^3\delta(\vec k_1+\vec k_2 + \vec k_3) B_{SSS}(k_1,k_2,k_3), 
\end{split}
\end{equation}
where
\begin{equation}
\begin{split}
	B_{\zeta\zeta\zeta}(k_1,k_2,k_3) &= [ N_\phi^2N_{\phi\phi} + N_\sigma^2 N_{\sigma\sigma}]
		\left[ P_{\delta \phi}(k_1)P_{\delta \phi}(k_2)+({\rm 2~perms.}) \right], \\
	B_{\zeta\zeta S}(k_1,k_2,k_3) &= [ N_\phi^2 S_{\phi\phi} + N_\sigma^2 S_{\sigma\sigma} 
		+2N_\phi S_\phi N_{\phi\phi} + 2N_\sigma S_\sigma N_{\sigma\sigma}  ]
		\left[ P_{\delta \phi}(k_1)P_{\delta \phi}(k_2)+({\rm 2~perms.}) \right], \\
	B_{\zeta SS}(k_1,k_2,k_3) &= [ S_\phi^2 N_{\phi\phi} + S_\sigma^2 N_{\sigma\sigma} 
		+2N_\phi S_\phi S_{\phi\phi} + 2N_\sigma S_\sigma S_{\sigma\sigma}  ]
		\left[ P_{\delta \phi}(k_1)P_{\delta \phi}(k_2)+({\rm 2~perms.}) \right], \\
	B_{SSS}(k_1,k_2,k_3) &= [ S_\phi^2S_{\phi\phi} + S_\sigma^2 S_{\sigma\sigma}]
		\left[ P_{\delta \phi}(k_1)P_{\delta \phi}(k_2)+({\rm 2~perms.}) \right].
\end{split}
\end{equation}
Here we have neglected contributions from purely non-Gaussian parts,
e.g., terms which are proportional to $N_{\phi\phi}^3, N_{\phi\phi}^2S_{\phi\phi}$, etc.
The first one, $B_{\zeta\zeta\zeta}$, is same as that usually studied in the mixed inflaton-curvaton
system~\cite{Ichikawa:2008iq}.
Others are bispectra that arise only when the extra radiation has isocurvature perturbations.
Also one should notice that the first two terms and the remaining two terms in 
$B_{\zeta\zeta S}$ and $B_{\zeta SS}$ have different effects on the CMB anisotropy.
Non-Gaussian imprints of the extra radiation components on the CMB anisotropy
may also be important for constraining or confirming models of extra radiation.
We leave these issues for future work.

%%%%%%%%%%%%%%%%%%%%%%%%%%%%%%%%%%%
\section{Observational signature of isocurvature perturbations in
the extra radiation}
\label{sec:signature}
%%%%%%%%%%%%%%%%%%%%%%%%%%%%%%%%%%%

Isocurvature perturbations in the extra radiation
generated in the early Universe affect the late-time structure formation
and hence the CMB anisotropy. 
Since extra radiations and active massless neutrinos are
cosmologically equivalent, the initial condition for the structure formation
with $S_\mathrm{DR}\ne 0$ can be identified as 
the neutrino isocurvature perturbation mode~\cite{Bucher:1999re}.
In this section, the neutrino isocurvature density (NID) perturbation collectively means the
isocurvature perturbation of the dark radiation (DR), which is sum of the active neutrinos and 
the extra radiation particle $X$.
Therefore, all the following analyses are applied to the case of 
the standard neutrino isocurvature perturbation
if there are no extra radiation particles and the dark radiation 
consists only of the active neutrinos:
$S_{\rm DR}=S_{\nu} (= 3(\zeta_\nu-\zeta))$ and 
$\hat R_{\rm DR}=R_\nu (=\rho_\nu / \rho_{\rm tot})$.
Our formulation includes more general case where the extra radiation 
$X$ significantly 
contributes to the relativistic energy density and the isocurvature 
perturbation.
%In this section, we do not distinguish extra radiations and
%the active neutrinos, and they are simply called ``neutrinos''
%with no distinction, instead of ``dark radiations'' adopted in the previous section.
In addition, throughout this section, we assume a vanilla flat $\Lambda$CDM model
as the unperturbed Universe, and all the associated cosmological parameters
are fixed to the WMAP 7-year mean value~\cite{Komatsu:2010fb}, except the 
effective number of neutrino species, which can deviate from the standard
value due to the existence of the extra radiation, i.e. $N_\mathrm{eff}\neq3.04$. 

So far, several authors have investigated the initial perturbation evolution
(i.e. at superhorizon scales deep in the radiation dominated (RD) epoch)
for the NID mode~\cite{Bucher:1999re,Trotta:2004qj}.
Perturbation evolution at subhorizon scales or in later epoch 
can be calculated numerically by adopting Boltzmann codes 
(e.g. CAMB~\cite{Lewis:1999bs}). 
This allows us to constrain the amplitude of 
the NID mode from current cosmological observations. 
For the time being, the best probe of the NID mode 
is the CMB temperature anisotropy. Thus we concentrate
on the CMB signals of the NID mode in this section.

Although the adiabatic (AD) and the CDM 
isocurvature (CI)\footnote{%%
%%%
We omit the baryon isocurvature mode 
since its CMB power spectrum completely degenerates with one for CI 
(See e.g. Refs.~\cite{Gordon:2002gv,Kawasaki:2011ze}).
}
%%%
modes have been investigated in many literatures (e.g. Ref.~\cite{Hu:1994uz}) 
and well understood, there are few papers which offer physical 
understanding of the perturbation evolution beyond the initial behavior
for the NID mode 
(See also Ref. \cite{Zunckel:2010mm}). 
In the following, we investigate the semi-analytical description 
of the perturbation evolution for the NID mode.

%%%%%%%%%%%%%%%%%%%%%%%%%%%%%%%%%%%
\subsection{Perturbation equations}
\label{sec:pert_eq}
%%%%%%%%%%%%%%%%%%%%%%%%%%%%%%%%%%%

We firstly summarize the evolution equations of 
the cosmological perturbation at a linear level.
Scalar perturbations in the flat Universe are to be discussed, 
which can be expanded in terms of the eigen function
$Q(\vec k, \vec x)=\exp[i\vec k\cdot \vec x]$. We consider a single 
Fourier mode with a wave number $\vec k$.
In addition, we adopt the conformal Newtonian gauge, 
where physics can be intuitively understood in various cases.
In this gauge, the perturbed metric is given by
\begin{equation}
ds^2=
a(\eta)^2\left[
-(1+2\Psi Q)d\eta^2
+\delta_{ij}(1+2\Phi Q)dx^idx^j
\right], 
\end{equation}
where $\eta$ is the conformal time, $a(\eta)$ 
is the scale factor, and $\Psi$ and $\Phi$ are 
the gravitational potentials.
The perturbed energy-momentum 
tensor of a fluid is given by
\begin{eqnarray}
T^0_0&=&-(1+\delta Q)\rho, \\
T^0_i&=&VQ_i(\rho+p), \\
T^j_0&=&-VQ^j(\rho+p), \\
T^i_j&=&-(\delta^i_j(1+\pi_LQ)+\Pi Q^i_j)p,
\end{eqnarray}
where $\delta=\delta \rho/\rho$, $V$, $\pi_L=\delta p/p$ and $\Pi$ 
are the density, bulk velocity, pressure and 
anisotropic stress perturbations, respectively.
Note that $Q_i=-i\hat k_i Q$ and $Q_{ij}=-(\hat k_i\hat k_j-\delta_{ij}/3)Q$
represent (traceless parts of) the spatial derivatives of $Q$.
Our notation is basically same as that adopted in Ref.~\cite{Hu:1994uz}. 
%In the following, dots represent derivatives respective to $\eta$.

The Einstein equation gives a set of perturbation equations for the metric, 
\begin{eqnarray}
\label{eq:metric}
k^2\Phi+3\mathcal H(\dot\Phi-\mathcal H\Psi)&=&4\pi G a^2\bar\rho_{\rm tot}\delta_{\rm tot},
\\
k(\dot\Phi-\mathcal H\Psi)&=&-4\pi G a^2(\bar\rho_{\rm tot}+\bar p_{\rm tot})V_{\rm tot},\\
\ddot\Phi+\mathcal H(2\dot\Phi-\dot\Psi)-(2\dot{\mathcal H}+\mathcal H^2)\Psi
+\frac{k^2}{3}(\Phi+\Psi)&=&4\pi G a^2 \bar p_{\rm tot} \pi_{L\,{\rm tot}},\\
k^2(\Phi+\Psi)&=&-8\pi G a^2 \bar p_{\rm tot}\Pi_{\rm tot}
\end{eqnarray}
where $\mathcal H=\dot a/a$ is the conformal hubble expansion rate
and the subscript ``tot" represents the total fluid which consists of 
CDM, baryons, photons and neutrinos.
Here and hereafter the dot represents a derivative with respective to $\eta$.

On the other hand, the conservation law of energy-momentum gives
the perturbation equations of the fluids. For CDM they
are given by
\begin{eqnarray}
\dot\delta_c=-kV_c-3\dot\Phi,\quad
\dot V_c=-\mathcal H V_c+k\Psi.
\label{eq:CDM}
\end{eqnarray}
As for baryons, we obtain
\begin{eqnarray}
\dot\delta_b=-kV_b-3\dot\Phi,\quad
\dot V_b=-\mathcal H V_b+k\Psi+\frac{a\sigma_Tn_e}{f}(V_\gamma-V_b),
\label{eq:baryons}
\end{eqnarray}
where
$f\equiv3\bar\rho_b/4\bar\rho_\gamma$ roughly 
denotes the ratio of the baryon and photon energy density
and the cross section of the Thomson scattering and
the number density of electrons are denoted by
$\sigma_T$ and $n_e$, respectively.
We also obtain perturbation equations for photons
\begin{eqnarray}
\dot\delta_\gamma=-\frac{4}{3}kV_\gamma-4\dot\Phi,\quad
\dot V_\gamma=k(\delta_\gamma+\Psi-\frac{1}{6}\Pi_\gamma)-
a\sigma_Tn_e(V_\gamma-V_b),
\label{eq:photons}
\end{eqnarray}
and DR 
\begin{eqnarray}
\dot\delta_{\rm DR}=-\frac{4}{3}kV_{\rm DR}-4\dot\Phi,\quad
\dot V_{\rm DR}=k(\delta_{\rm DR}+\Psi-\frac{1}{6}\Pi_{\rm DR}).
\label{eq:neutrinos}
\end{eqnarray}
Note that here we have omitted evolution equations for the anisotropic stress
and higher order multipole moments of photons and DR, which are
not very important for our discussion. (However, these are included
in numerical calculations.)

%%%%%%%%%%%%%%%%%%%%%%%%%%%%%%%%%%%
\subsection{Initial conditions}
\label{sec:initial}
%%%%%%%%%%%%%%%%%%%%%%%%%%%%%%%%%%%

By solving equations \eqref{eq:metric}-\eqref{eq:neutrinos} 
at superhorizon scales deep in RD
epoch, we obtain several independent initial
solutions for the structure formation, which can be characterized by 
the gauge-invariant curvature perturbations $\zeta$ 
and isocurvature perturbations $S_{ij}$
\footnote{
%%%
In general, there can be a neutrino velocity (NIV) mode, which 
cannot be characterized by $\zeta$ nor $S_{ij}$ \cite{Bucher:1999re}.
We do not consider the NIV mode in this paper.
%%%
}.
While our primary interest resides in the NID mode,
other initial modes (i.e. AD and CI) are also discussed 
in a parallel manner for reference. 

Now let us summarize the initial conditions for
the AD, CI and NID modes. In Ref.~\cite{Bucher:1999re}, 
they are derived in the synchronous gauge,
and we can obtain the initial conditions in the Newtonian gauge
by performing a gauge transformation.
Since we are mostly interested in the CMB signals
of these modes, here we restrict ourselves to the
initial solutions for only the perturbations in the photon fluid
and the metric. For later convenience, we define
$k_\mathrm{eq}=\Omega_m H_0/\sqrt{\Omega_r}$, which roughly
corresponds to the horizon scale at the matter-radiation equality.

The AD mode is characterized with nonzero initial $\zeta$. 
Up to $\mathcal O (\eta)$, the initial condition for the AD mode is 
given as
\begin{eqnarray}
\delta_\gamma/\hat \zeta(0)&=&
\frac{20}{15 + 4 \hat R_{\rm DR}} + \frac{5(15 + 16 \hat R_{\rm DR})}
{2(15 +4 \hat R_{\rm DR})(15 + 2 \hat R_{\rm DR})}k_\mathrm{eq}\eta,\\
V_\gamma/\hat\zeta(0)&=& 
-\frac{5}{15 + 4 \hat R_{\rm DR}}k\eta,\\
\Psi/\hat\zeta(0)&=&
-\frac{10}{15 + 4 \hat R_{\rm DR}}
+\frac{25 (3 - 8 \hat R_{\rm DR})}{8(15 + 4 \hat R_{\rm DR})(15 + 2 \hat R_{\rm DR})}k_\mathrm{eq}\eta,\\
\Phi/\hat\zeta(0)&=&
\frac{10+4\hat R_{\rm DR}}{15 + 4 \hat R_{\rm DR}}
-\frac{5(15 + 16 \hat R_{\rm DR})}{8(15 + 4 \hat R_{\rm DR})(15 + 2 \hat R_{\rm DR})}k_\mathrm{eq}\eta,
\end{eqnarray}
where $\hat R_{\rm DR}=\frac{\bar\rho_{\rm DR}}{\bar\rho_\gamma+\bar\rho_{\rm DR}}$ 
is the fraction of DR (= neutrinos + $X$) in the radiation energy 
density, 
which can be rewritten in terms of $N_\mathrm{eff}$ defined in Eq.~(\ref{rhoDR0th}) as
\begin{equation}
	\hat R_{\rm DR}=\frac{N_\mathrm{eff}\frac{7}{8}\left(\frac{4}{11}\right)^{4/3}}
	{1+N_\mathrm{eff}\frac{7}{8}\left(\frac{4}{11}\right)^{4/3}}.
\end{equation}
We note that the initial density perturbation of photons $\delta_\gamma$ and
the metric perturbations $\Psi,~\Phi$ are all nonzero and comparable
in the amplitude. As is well known, this fact results in 
the cosine-like acoustic oscillation of the photon-baryon fluid,
which will be discussed in more detail in the next subsection.

The CI mode is characterized by a nonzero 
initial CDM isocurvature perturbation
$\hat S_c\equiv3(\zeta_c-\hat\zeta)=3(\delta_c-\frac{4}{3}\delta_\gamma)$, 
where $\zeta_c$ is the curvature perturbation on the uniform density slice of CDM. 
In the same way as the AD mode, the initial condition for the CI mode can be obtained as 
\begin{eqnarray}
\delta_\gamma/\hat{S}_c(0)&=&
-\frac{\hat R_c (15 + 4 \hat R_{\rm DR})}{2(15 + 2 \hat R_{\rm DR})}k_\mathrm{eq}\eta,\\
V_\gamma/\hat{S}_c(0)&=& \mathcal O(\eta^2),\\
\Psi/\hat{S}_c(0)&=&
\frac{\hat R_c (-15 + 4 \hat R_{\rm DR})}{8 (15 + 2 \hat R_{\rm DR})}k_\mathrm{eq}\eta,\\
\Phi/\hat{S}_c(0)&=&
\frac{\hat R_c (15 + 4 \hat R_{\rm DR})}{8 (15 + 2 \hat R_{\rm DR})}k_\mathrm{eq}\eta,
\end{eqnarray}
where $\hat R_c$ is the fraction of CDM in the energy density
of matters, i.e. $\hat R_c=\bar\rho_c/(\bar\rho_c+\bar\rho_b)$.
In the CI mode, $\delta_\gamma,~\Psi$ and $\Phi$ all
vanish initially, which results in the sine-like acoustic 
oscillation of the photon-baryon fluid.

The NID mode is characterized by a non-vanishing
initial DR isocurvature perturbation $\hat S_{\rm DR}$.
%\footnote{
%%%
%As a reminder, this is what we wrote as $S_\mathrm{DR}$ in the previous section.
%%%
%}
The initial condition for the NID mode is given as
\begin{eqnarray}
\delta_\gamma/\hat S_{\rm DR}(0)&=&
-\frac{\hat R_{\rm DR} (8 \hat R_{\rm DR} + 11)}{\hat R_\gamma(15 + 4 \hat R_{\rm DR})} 
+\frac{\hat R_{\rm DR} (2 \hat R_{\rm DR} - 15)}{(2 \hat R_{\rm DR} + 15) 
(4 \hat R_{\rm DR} + 15)}k_\mathrm{eq}\eta, 
\label{eq:dg_nid}\\
V_\gamma/\hat S_{\rm DR}(0)&=& 
-\frac{19 \hat R_{\rm DR} }{4 \hat R_\gamma (4 \hat R_{\rm DR} + 15)}k\eta, \\
\Psi/\hat S_{\rm DR}(0)&=&-\frac{\hat R_{\rm DR}}{15 + 4 \hat R_{\rm DR}} -
\frac{\hat R_{\rm DR}(2 \hat R_{\rm DR}-15)}{4(4\hat R_{\rm DR}+15)(2\hat R_{\rm DR}+15)}k_\mathrm{eq}\eta, \\
\Phi/\hat S_{\rm DR}(0)&=&-\frac{2 \hat R_{\rm DR}}{4 \hat R_{\rm DR}+15} 
- \frac{\hat R_{\rm DR}(2 \hat R_{\rm DR} - 75)}{4(4\hat R_{\rm DR}+15)(2\hat R_{\rm DR}+15)}
k_\mathrm{eq}\eta,
\label{eq:phi_nid}
\end{eqnarray}
where $\hat R_{\gamma}=1-\hat{R}_{\rm DR}$.
This NID initial condition is distinct from both the AD and CI modes in several respects.
First of all, photons have non-vanishing initial 
density perturbations $\delta_\gamma(0)\ne0$ in the NID mode, which is 
in contrast to the CI mode, where
$\delta_\gamma$ vanishes initially. On the other hand, while $\delta_\gamma$
does not vanish initially for the AD mode also, the gravitational potentials for the NID mode
are initially much smaller than those for the AD mode. 
This is because the density perturbations in photons and
DR cancel out each other for the NID mode, and
the density perturbations in the total radiations vanish.
However, due to the presence of the anisotropic stress in DR, 
initial $\Psi$ and $\Phi$ do not vanish exactly. 
In a nutshell, the NID initial condition can be characterized by
non-vanishing $\delta_\gamma$ and small $\Phi$ and $\Psi$.

%%%%%%%%%%%%%%%%%%%%%%%%%%%%%%%%%%%
\subsection{Acoustic oscillation}
\label{sec:acoustic}
%%%%%%%%%%%%%%%%%%%%%%%%%%%%%%%%%%%

Now we discuss how evolution of the photon perturbation in the NID mode
differs from those in the AD and CI modes. Before recombination, 
as perturbations enter the horizon $k\simeq \mathcal H$, 
the photon fluid undergoes the acoustic oscillation. 
Tight-coupling approximation
is a good approximation above the diffusion scale of photons.
In the limit of tight-coupling, the photons and baryons can
be treated as a single fluid with sound velocity $c_s=
1/\sqrt{3(1+f)}$ 
%where $f\equiv 3\rho_b/(4\rho_\gamma)$ roughly 
%denotes the ratio of the baryon and photon energy density
and obey equations of motion given by
\begin{eqnarray}
\dot\delta_\gamma&=&-\frac{4}{3}kV_\gamma-4\dot\Phi, \\
\dot V_\gamma&=&-\frac{\dot f}{1+f}V_\gamma
+\frac{3}{4}kc_s^2\delta_\gamma+k\Psi, 
\end{eqnarray}
which are combined into a single second order differential equation
\begin{equation}
\ddot \delta_\gamma+\frac{\dot f}{1+f}\dot \delta_\gamma
+k^2c_s^2\delta_\gamma=4F, \label{eq:acoustic}
\end{equation}
where 
\begin{equation}
F=-\ddot\Phi-\frac{\dot f}{1+f}\dot\Phi-\frac{k^2}{3}\Psi.
\label{eq:force}
\end{equation}
The left hand side of Eq.~\eqref{eq:acoustic} describes the acoustic 
oscillation of the photon-baryon fluid and $4F$ in the right hand side 
can be regarded as an external force.

%%%%%%%%%%%%%%%%%%%%
\begin{figure}[th]
  \begin{center}
    \begin{tabular}{ccc}
	\includegraphics[scale=.45]{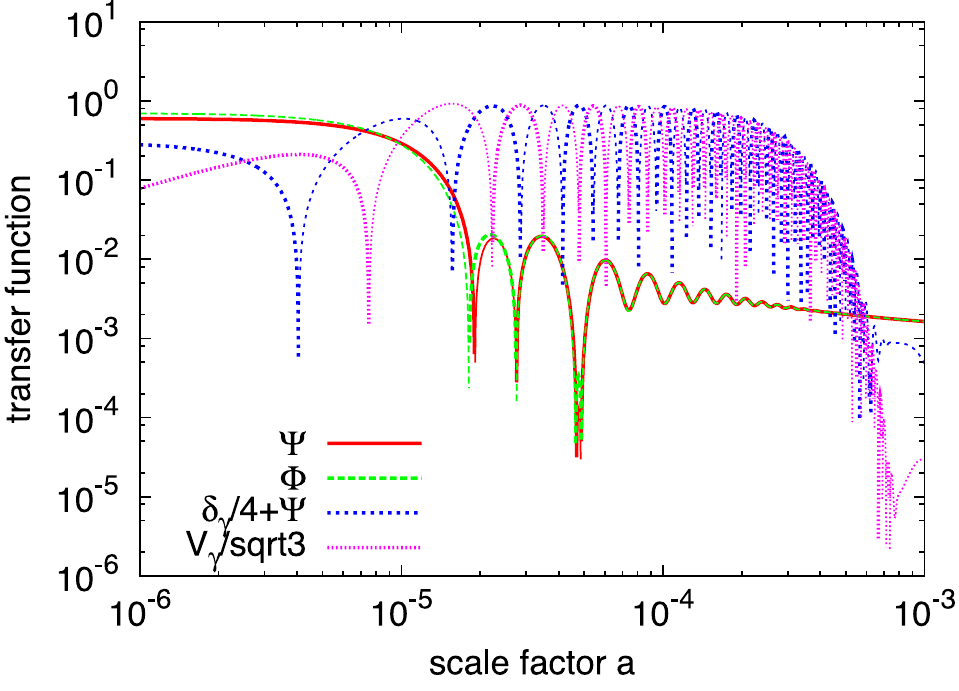} &
	\includegraphics[scale=.45]{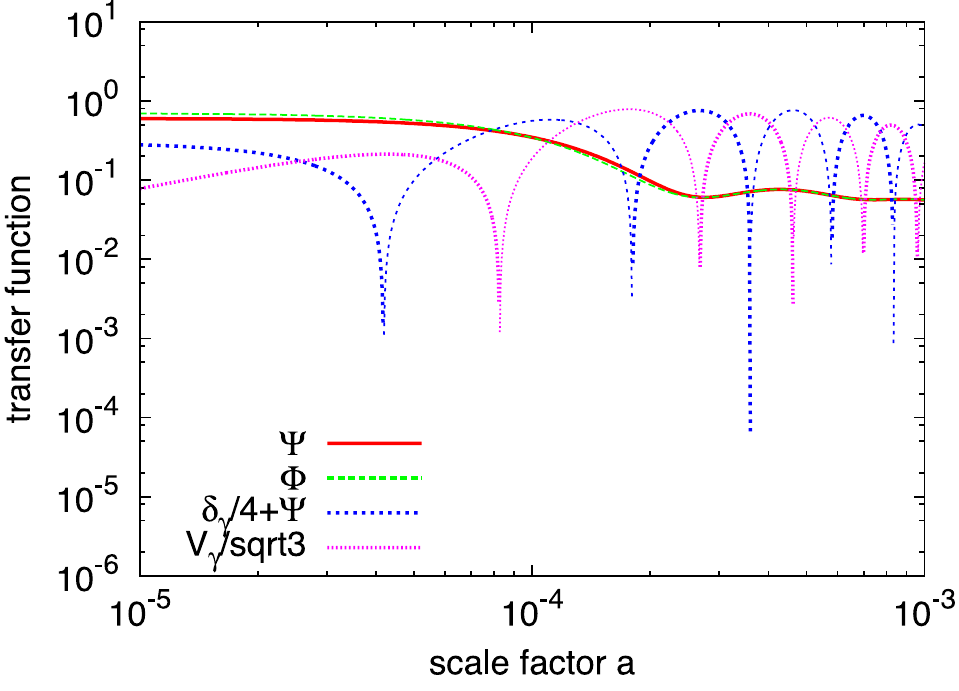} &
	\includegraphics[scale=.45]{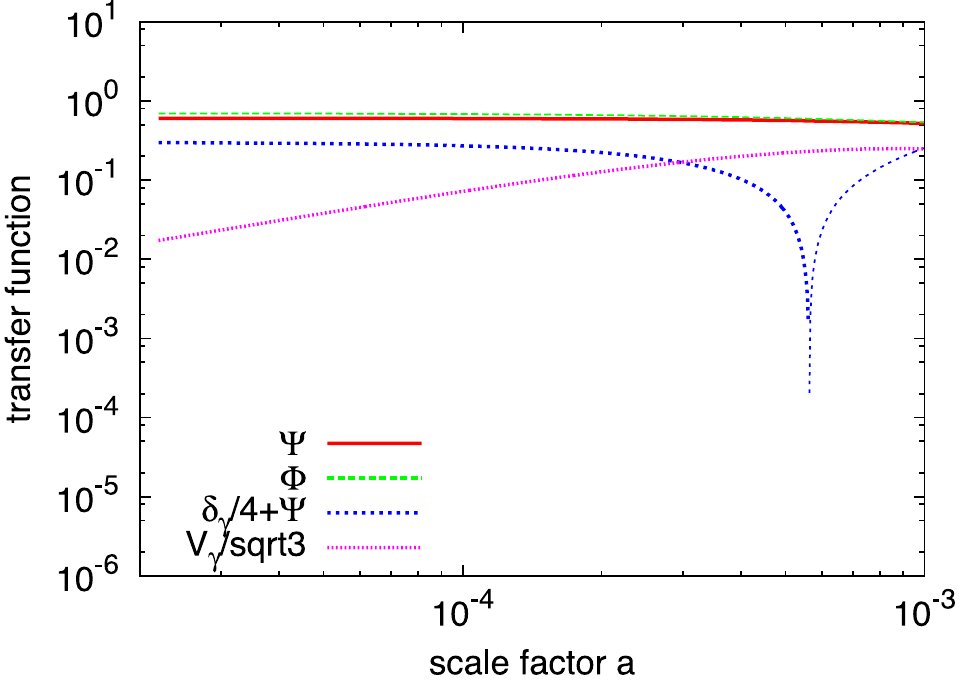} \\
	\includegraphics[scale=.45]{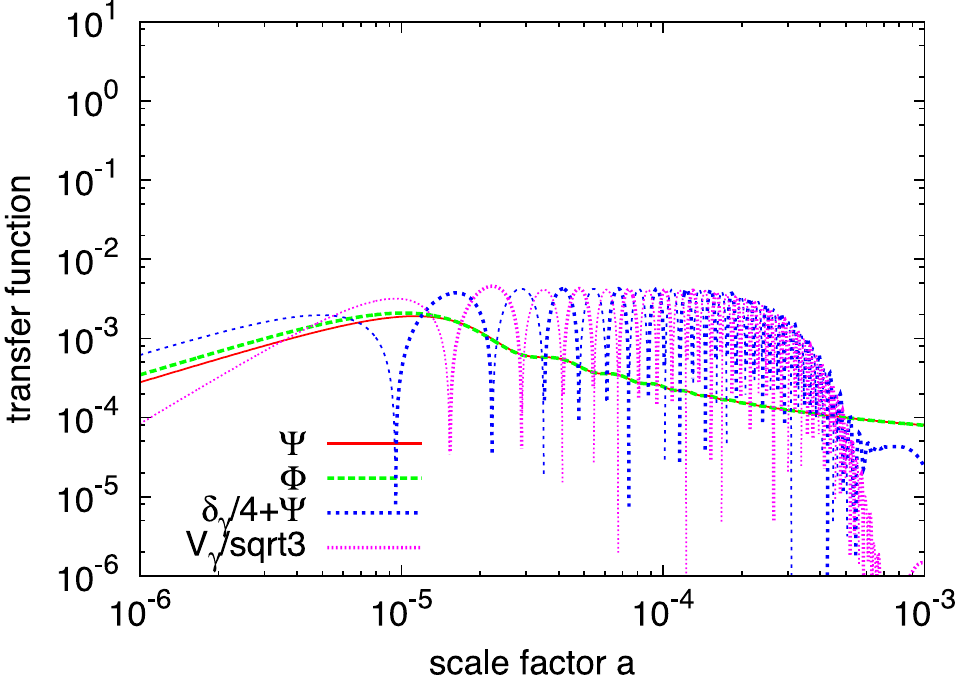} &
	\includegraphics[scale=.45]{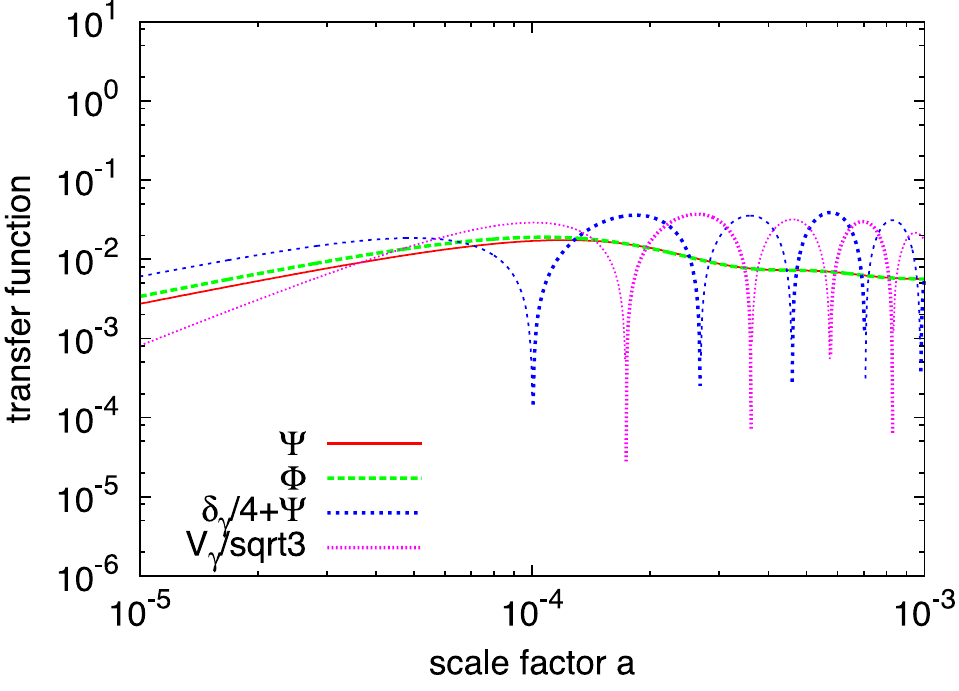} &
	\includegraphics[scale=.45]{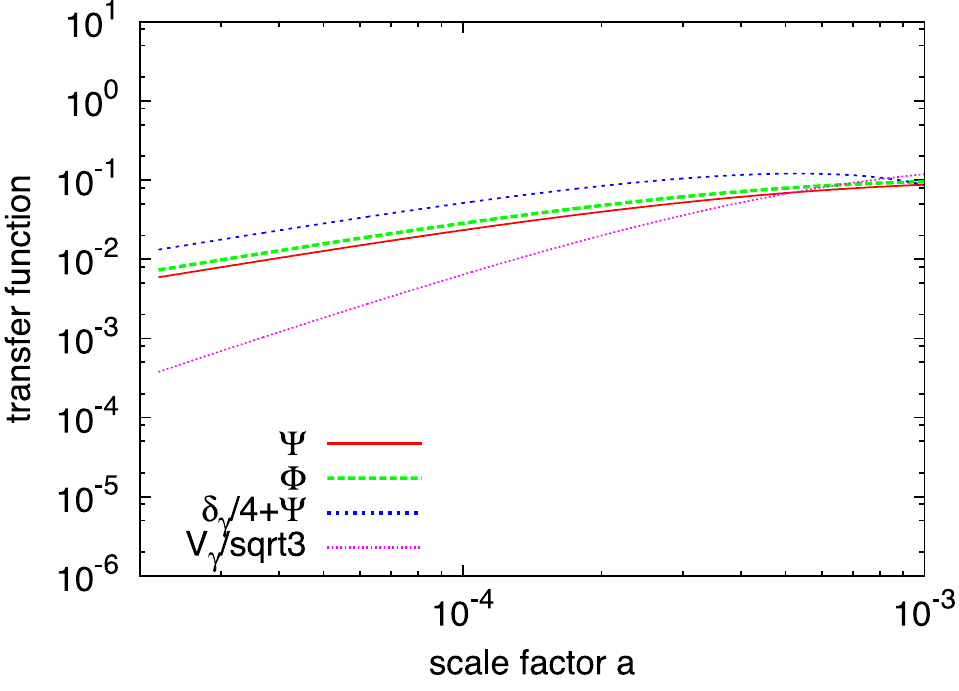} \\
	\includegraphics[scale=.45]{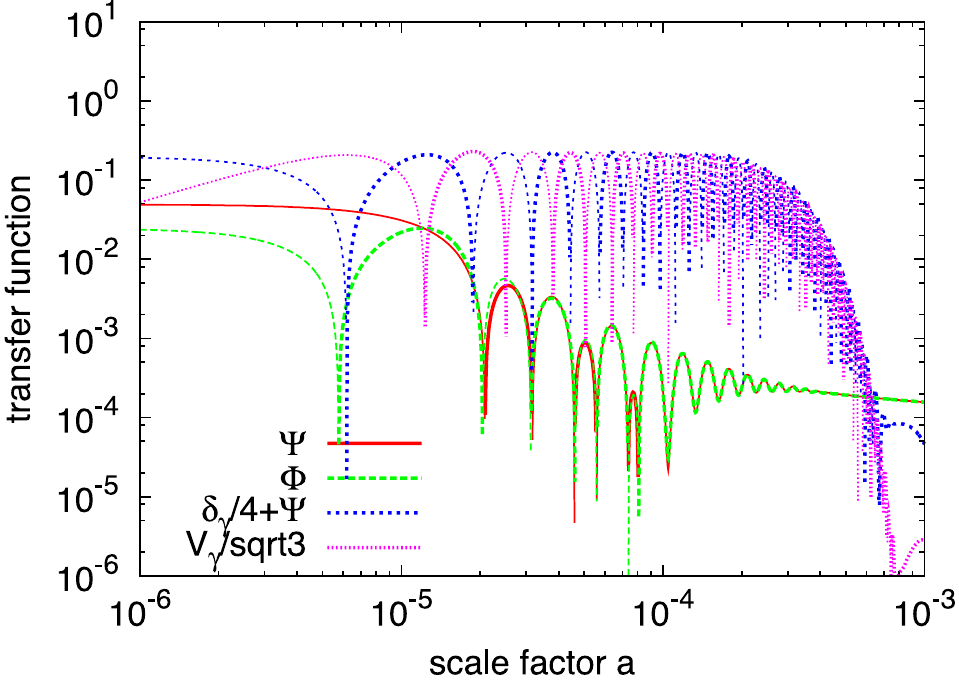} &
	\includegraphics[scale=.45]{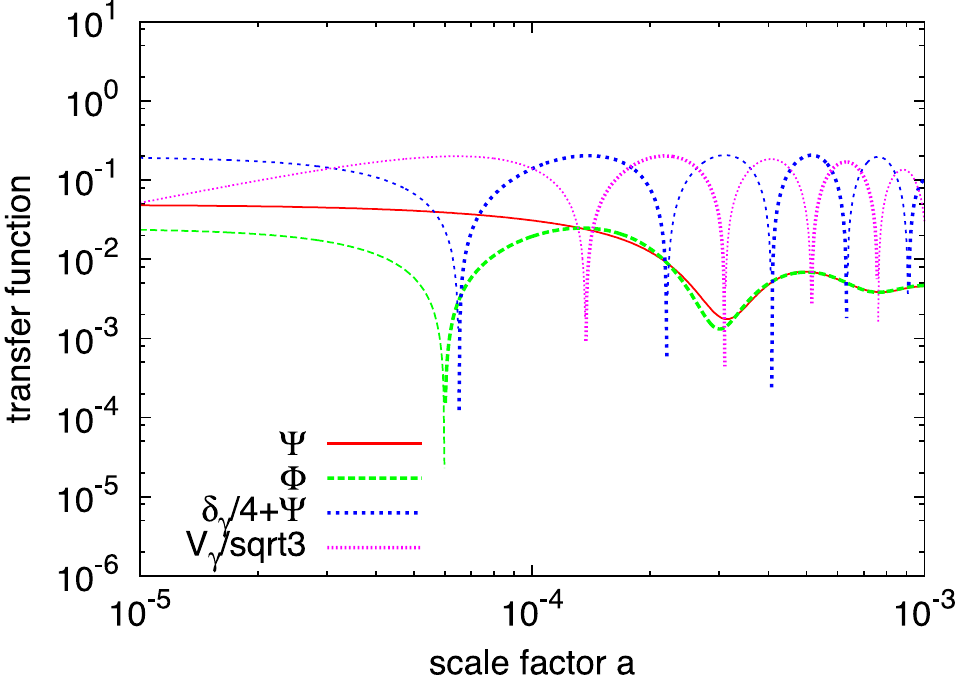} &
	\includegraphics[scale=.45]{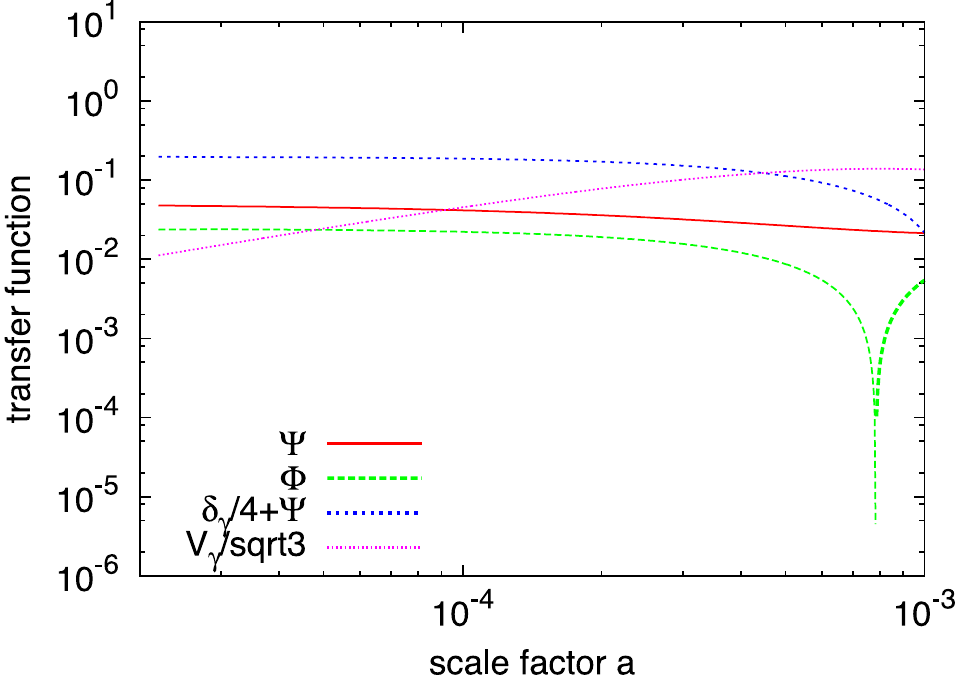}
    \end{tabular}
  \end{center}
\caption{Perturbation evolution of the photon fluid and the metric.
We showed the metric perturbations $\Psi$ (solid red) and $\Phi$ (dashed green), 
the photon effective temperature perturbation 
$\delta_\gamma/4+\Psi$ (dotted blue) and 
the photon velocity perturbation $V_\gamma/\sqrt3$(dot-dashed magenta).
Positive and negative values are represented by thick and thin lines, respectively.
From the top to bottom, each row corresponds to 
AD, CI and NID modes, in order.
From the left to right, each column corresponds to 
three different scales $k=1$, $0.1$, $0.01$ Mpc$^{-1}$ in order.
}
\label{fig:transfers}
\end{figure}
%%%%%%%%%%%%%%%%%%%%

Fig.~\ref{fig:transfers} shows perturbation evolution of the
photon fluid and the metric for the AD, CI and NID modes at 
three different scales ($k=1,~0.1$ and $0.01$ Mpc$^{-1}$), 
which are calculated by using a modified version of CAMB.
In the figure, we plotted the metric perturbations $\Psi$ and $\Phi$,
the photon effective temperature perturbation
$\delta_\gamma/4+\Psi$ and the photon velocity 
perturbation $V_\gamma/\sqrt3$. As can be see from the figure, 
when perturbations enter the horizon, the photon fluids
starts the acoustic oscillations, and at subhorizon scales
the effective temperature $\delta_\gamma/4+\Psi$ keeps
oscillating with an almost constant amplitude as long as 
it damps inside the diffusion scale. On the other hand, 
the photon velocity perturbation $V_\gamma/\sqrt3$ 
oscillates with the same amplitude but a phase shifted by $\pi/2$
compared with the effective temperature perturbation.
We also note that gravitational potentials decay during
the horizon crossing in RD. 

Now we discuss the differences in the photon acoustic
oscillation among the three different initial modes.
The most distinct one that is found in Fig.~\ref{fig:transfers}
would be the difference in the oscillation phase for CI mode against other 
two modes. This originates from the fact that in CI mode
oscillation starts  with vanishing $\delta_\gamma$
at $\eta=0$, while oscillations of the other two modes start with 
non-vanishing $\delta_\gamma$. On the other hand, 
the differences between AD and NID modes may be less
noticeable as their oscillation phases look similar.  
The most significant difference can be 
found in the change in the oscillation amplitudes between the horizon crossing. 
The amplitude of the acoustic oscillation
gets boosted by the decay of gravitational potentials in the AD mode,
which is less effective in the NID mode due
to the initial smallness of the gravitational potentials.

In order to verify the above discussion, we adopt the
semi-analytic solution of the acoustic oscillation derived in Ref.~\cite{Hu:1994uz}. 
Under the adiabatic approximation, 
the WKB solution of Eq.~\eqref{eq:acoustic} can be obtained as
\begin{eqnarray}
\left[1+f(\eta)\right]^{1/4}\delta_\gamma(\eta)&=&
\delta_\gamma(0)\cos kr_s(\eta)+\frac{\sqrt3}{k}
\left[\dot\delta_\gamma(0)+\frac{1}{4}\dot{f}(0)
\delta_\gamma(0)\right]\sin kr_s(\eta)\notag\\
&&+\frac{4\sqrt3}{k}\int^\eta_0d\eta^\prime
\left[1+f(\eta^\prime)\right]^{3/4}\sin\left[kr_s(\eta)-kr_s(\eta^\prime)\right]F(\eta^\prime),
\label{eq:WKB}
\end{eqnarray}
where
\begin{equation}
r_s(\eta)=\int_0^\eta d\eta^\prime c_s(\eta^\prime)
\end{equation}
is the sound horizon of the photon-baryon fluid.
In Eq.~\eqref{eq:WKB}, the first and second terms
correspond to the free damped oscillation which reflect the
initial condition of photon density and velocity perturbations.
On the other hand, the third term corresponds to the forced oscillation
driven by the gravitational potentials. In the following discussion,
the first, second, and third terms are called the
cosine, sine and forced terms, respectively.

\begin{figure}
  \begin{center}
	\includegraphics[scale=.6]{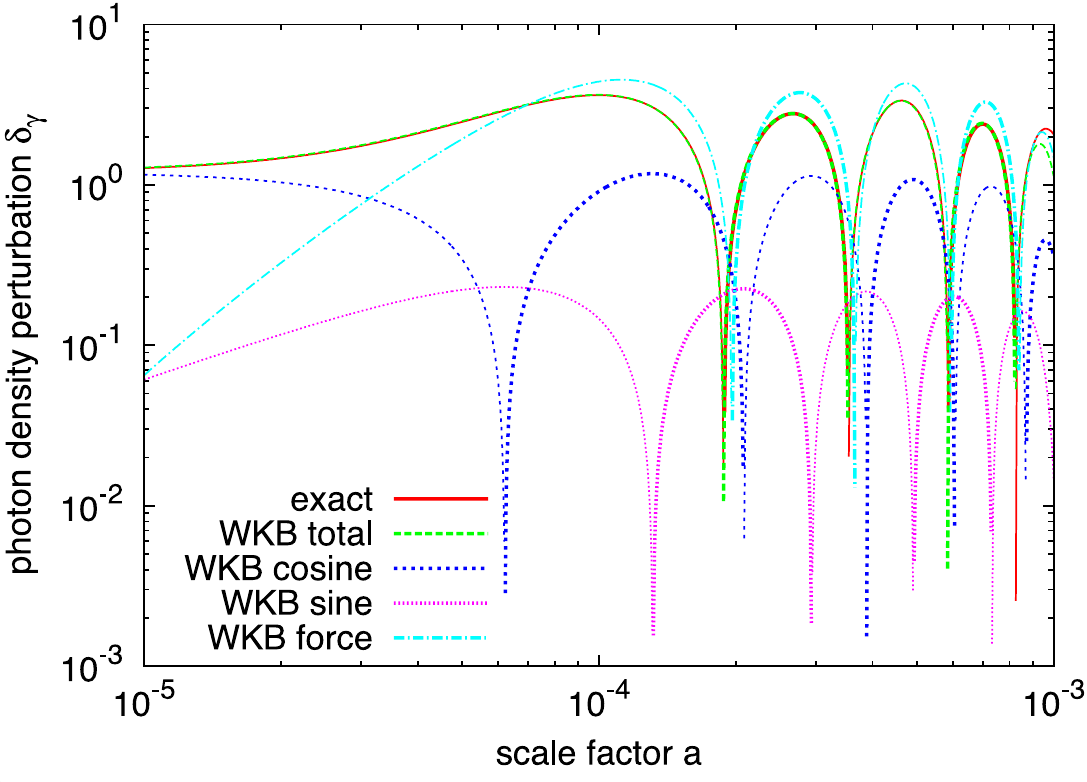} \\
	\includegraphics[scale=.6]{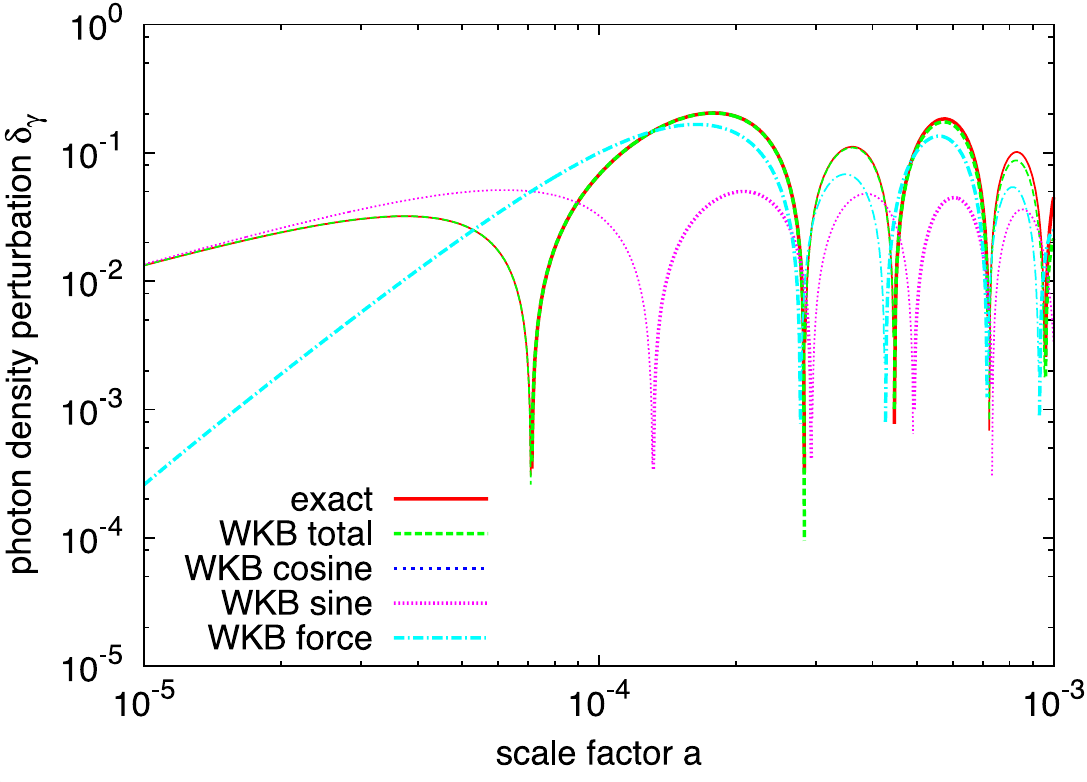} \\
	\includegraphics[scale=.6]{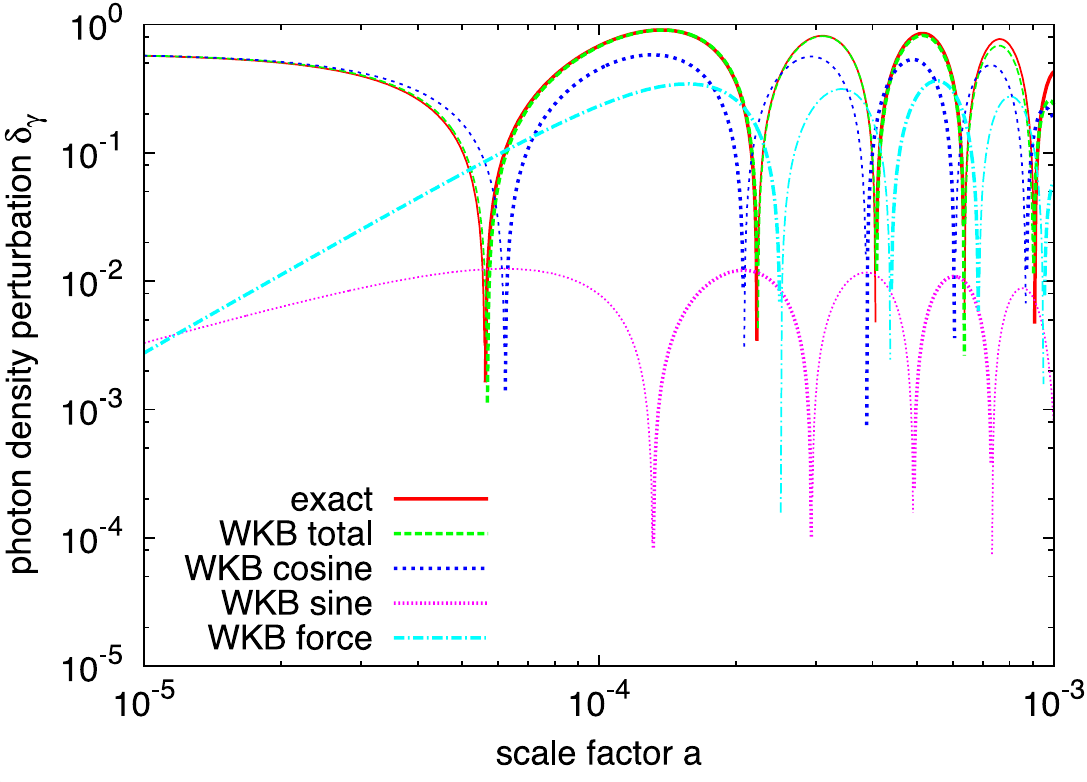}
  \end{center}
\caption{Time evolutions of the photon density perturbation $\delta_\gamma$
for AD (top), CI (middle) and NID (bottom) modes at $k=0.1~\mathrm{Mpc}^{-1}$.
Plotted are the exact solution (solid red) the total WKB solution (dashed green), 
and three terms consisting the total WKB solution including
the cosine term (dotted blue), sine term (dot-dahsed magenta) and term from the 
external force (dot-dot-dahsed light blue).
Positive and negative values are represented 
by thick and thin lines, respectively.
}
\label{fig:WKB}
\end{figure}

In Fig.~\ref{fig:WKB}, the time evolution
of the photon density perturbation are plotted for the AD, CI and NID modes at
$k=0.1\mathrm{Mpc}^{-1}$.
In the figure, we have shown exact numerical solutions for $\delta_\gamma$
as well as the WKB solutions Eq.~\eqref{eq:WKB} obtained by 
adopting the numerically evaluated source term $F$. 
We can see that the WKB solution is in excellent agreement
with the exact one, regardless of initial conditions.
This allows us to understand correctly
the physical origin that drives the evolution of photon perturbations
by comparing the different terms in Eq.~\eqref{eq:WKB}.
For this purpose, we have also plotted the 
WKB cosine, sine and forced terms separately in the figure.

%%%%%%%%%%%%%%%%%%%%
%%%%%%%%%%%%%%%%%%%%

For the AD mode, we see that the forced term dominates
the acoustic oscillation and imprint of the initial cosine-like 
oscillation induced by $\delta_\gamma(0)$ is almost invisible. 
However, the external force also induces an acoustic oscillation 
with cosine-like phase, as the gravitational potentials also oscillate 
in a similar manner around the horizon crossing.
Totally, the acoustic oscillation for AD mode can be regarded as
a kind of forced oscillation with a cosine-type phase.

In the case of CI mode, the source term is also dominant 
as can be seen in Fig.~\ref{fig:WKB}. Since the gravitational
potentials grow proportionally to $\eta$ at superhorizon scales 
as the density perturbations in CDM generates the curvature perturbation, 
they induce a sine-type acoustic oscillation
rather than a cosine-type one. The acoustic oscillation
in the CI mode can be regarded as a forced oscillation with a sine-type phase.

On the other hand, in the case of NID mode, 
the cosine term dominates the acoustic oscillation
and the forced term is subdominant. This is in contrast to the previous two modes.
This result comes from the fact that the
gravitational potentials are initially small
and grow little before the horizon crossing in NID mode.
Therefore, the acoustic oscillation for the NID mode
can be regarded as free oscillation with a cosine-type phase. 

Smallness of the gravitational potentials also
affect the perturbation evolution of photons
after the recombination, when
photons become collisionless and free-stream
in the potentials which are dominantly formed 
by the CDM density perturbations. While gravitational potentials
is constant deep in the matter dominated (MD) epoch, 
it is not fully MD just after the recombination, where
the potentials decay during the horizon crossing.
Effective temperature of free-streaming photons are 
boosted (or suppressed) due to the decay of the potentials $\Phi-\Psi$, which
is well known as the early integrated Sachs-Wolfe (ISW) effect.
In the AD and CI modes, the early ISW effect is significant, 
since their gravitational potentials  are relatively large 
at the time of horizon crossing. 
Contrastively, the gravitational potentials are small 
in the NID mode, and hence their decay
affects the photon perturbations little.

%%%%%%%%%%%%%%%%%%%%%%%%%%%%%%%%%%%
\subsection{CMB temperature power spectrum}
\label{sec:cls}
%%%%%%%%%%%%%%%%%%%%%%%%%%%%%%%%%%%

\begin{figure}[th]
  \begin{center}
	\includegraphics[scale=1.]{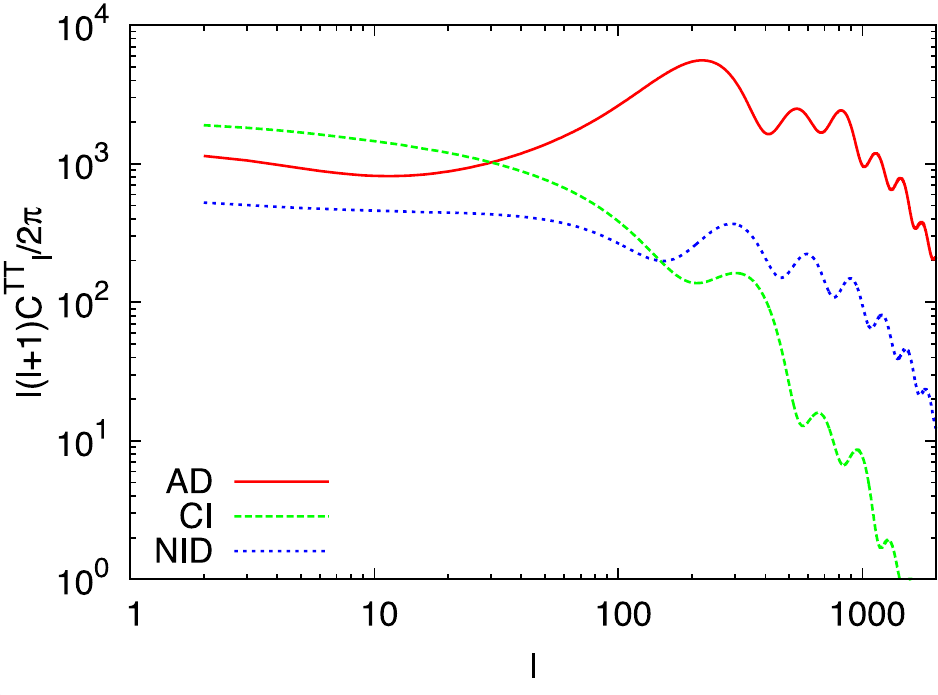}
  \end{center}
\caption{
	The CMB temperature power spectrum
	($C^\mathrm{TT}_\ell$) for the AD, CI and NID modes.
}
\label{fig:cls}
\end{figure}

In Fig.~\ref{fig:cls}, we plot the CMB temperature power spectrum
($C^\mathrm{TT}_\ell$) for the NID mode
as well as those for other AD and CI modes for reference.
The spectrum of the NID mode has distinct features coming from 
the fact that 
the evolution of photon perturbations in the NID mode 
is different from those in the AD and CI modes as we have seen above.

First of all, the positions of acoustic peaks in $C^\mathrm{TT}_\ell$
for the NID mode are more similar to those for the AD mode than
those for the CI mode, which reflects the phase of acoustic oscillation
in the each initial mode. In addition, the amplitude at the first 
acoustic peak
($\ell\simeq300$) in the NID mode is almost same as that at the 
Sachs-Wolfe (SW) plateau 
($\ell\lesssim50$). This reflects the smallness of the gravitational 
potentials in the NID mode, where decay of the potentials does not 
boost the photon temperature significantly before and after 
the recombination.
This should be contrasted to the AD mode, where the amplitude of 
$C^\mathrm{TT}_\ell$ at the acoustic regime is higher than
that at the SW plateau. 

Finally, we would like to give some brief comments on the effects 
of $\Delta N_\mathrm{eff}> 0$
on the CMB power spectrum. At the sensitivity of current and 
near-future observations such as Planck, two characteristic changes 
from the case of $\Delta N_\mathrm{eff}=0$ would be important.
One is that the RD epoch lasts longer and hence the expansion rate 
of the Universe becomes larger given at fixed redshifts.
This enhances the early ISW effect and causes shifts in 
the acoustic peaks towards smaller angular scales.
Another is that the free-streaming of neutrinos affects 
the perturbation evolution more significantly. This enhances the 
decay of the gravitational potentials before perturbations enter 
the sound horizon of the photon-baryon fluid and dampens 
the photon perturbations in the RD epoch. 
Thus, this effect suppresses the amplitude of $C^\mathrm{TT}_\ell$ 
at small angular scales. For more detailed discussion, 
we refer to e.g. Refs.~\cite{Bashinsky:2003tk,Ichikawa:2008pz}.

%%%%%%%%%%%%%%%%%%%%
%%%%%%%%%%%%%%%%%%%%

%%%%%%%%%%%%%%%%%%%%%%%%%%%%%%%%%%%
\section{Constraints from current observations}
\label{sec:constraint}
%%%%%%%%%%%%%%%%%%%%%%%%%%%%%%%%%%%

In this section we derive constraints on the isocurvature perturbations 
in the extra radiation using current cosmological observations.
We adopt CMB data of WMAP 7-year result~\cite{Gold:2010fm,Larson:2010gs,Jarosik:2010iu}
and ACT~\cite{Hajian:2010fj,Das:2010ga,Dunkley:2010ge} at small angular scales. 
While CMB power spectrum is the best probe for neutrino isocurvature
perturbations, CMB itself contains considerable amount of parameter
degeneracies among $S_\mathrm{DR}$, $N_\mathrm{eff}$ and 
other cosmological parameters. In order to solve the degeneracies, 
we also include data from the baryon acoustic oscillation (BAO)
in the power spectrum of SDSS galaxies~\cite{Percival:2009xn} and 
the direct measurement of the Hubble constant (H0)~\cite{Riess:2009pu}.
Hereafter, we will refer to sets of combined datasets of WMAP+ACT 
and WMAP+ACT+BAO+H0 as ``CMB" and ``ALL", respectively.

The initial condition for structure formation in our model is 
a mixture of the AD and NID modes and these two initial modes 
can be in general correlated. Correlation functions of the initial modes, 
or equivalently the primordial perturbation spectra, form a symmetric matrix.
We assume that the primordial perturbation spectra
can be represented by power-law with same spectral indices.
Then the primordial power spectra can be
parametrized as follows, which is widely seen in literatures:
\begin{eqnarray}
\begin{pmatrix}
\mathcal P_{\zeta\zeta}(k)&
\mathcal P_{\zeta S_\mathrm{DR}}(k)\\
\mathcal P_{S_\mathrm{DR}\zeta}(k)&
\mathcal P_{S_\mathrm{DR}S_\mathrm{DR}}(k)\\
\end{pmatrix}
=A_s \left(\frac{k}{k_0}\right)^{n_s-1}
\begin{pmatrix}
1-\alpha & \gamma\sqrt{\alpha(1-\alpha)}\\
\gamma\sqrt{\alpha(1-\alpha)} & \alpha 
\end{pmatrix},
\end{eqnarray}
where $\mathcal P_{AB}$ is the (cross-)power spectrum of
initial perturbations $A$ and $B$ defined in Eq.~(\ref{dimless-power}),
and $\gamma$ denotes the correlation parameter given in (\ref{gamma}).
%their correlation function by the relation
%\begin{equation}
%\langle A(\vec k)B(\vec k^\prime)\rangle
%=\frac{2\pi^2}{k^3}(2\pi)^3\delta^{(3)}(\vec k+\vec k^\prime)
%\mathcal P_{AB}(k).
%\end{equation}
%The dimensionless power spectrum $\mathcal P_{AB}(k)$ are 
%related to the power spectrum $P_{AB}(k)$
%defined in (\ref{power}) through $\mathcal P_{AB}(k)=(k^3/2\pi^2) P_{AB}(k)$.

%%%
\begin{table}%[ht]
  \begin{center}
  \begin{tabular}{llc}
    \hline
    \hline
    parameters & symbols & prior ranges \\
    \hline
    baryon density parameter & $\omega_b$ & $0.005\to0.1$\\
    CDM density parameter & $\omega_c$ & $0.01\to0.99$\\
    angular scale of sound horizon & $\theta_s$ & $0.5\to10$\\
    optical depth of reionization & $\tau$ & $0.01\to0.8$\\
    effective number of neutrino species & $N_\mathrm{eff}$ & $(3.046\to10)$\\
    spectral index of primordial power spectra & $n_s$ & $0.5\to1.5$\\
    amplitude of primordial power spectra & $\ln[10^{10}A_s]$ & $2.7\to4$\\
    fraction of NID mode in primordial perturbations & $\alpha$ & $0\to1$\\
    correlation between AD and NID modes & $\gamma$ & $(-1\to1)$\\
    \hline
    template amplitude of thermal SZ effect & $A_{SZ}$ & $0\to2$\\
    template amplitude of Poisson distributed sources & $A_p$ & $0\to30$\\
    template amplitude of clustered dust & $A_{cl}$ & $0\to15$\\
    \hline
    \hline
  \end{tabular}
  \caption{Model parameters we constrain in our analysis. 
  First nine from the top are the primary cosmological parameters, 
  and rest three are the nuisance ones. The last column shows
  the range of top-hat priors for the parameters we adopt in the analysis.
  Note that the prior ranges of $N_\mathrm{eff}$ and $\gamma$, which
  are shown with parentheses, are only adopted when they are varied
  in the analysis (see text for more details).
  }
  \label{tbl:priors}
  \end{center}
\end{table}
%%%

In a most general case, the parameter space we explore consists 
of nine primary cosmological parameters 
$(\omega_b,~\omega_c,~\theta_s,~\tau,
~N_\mathrm{eff},~n_s,~A_s,~\alpha,~\gamma)$. 
Definitions of these parameters and top-hat priors we adopt in
the parameter estimation are listed in Table~\ref{tbl:priors}.
In particular, we take $N_\mathrm{eff}\ge 3.046$
since our model assumes that there are always the Standard Model neutrinos 
which are fully thermalized in the early Universe.
In order to take account of foregrounds, 
we also include three nuisance parameters $A_{SZ}$, $A_{p}$, $A_{cl}$, 
which measure the amplitudes of power spectra from the Sunyaev-Zel'dovich (SZ) effect,
point source, and clustered dust, respectively.
We adopt the same template power spectra for these foregrounds
as in the cosmological parameter estimation of Ref. \cite{Dunkley:2010ge}.
In particular, templates for the SZ effect and clustered dust are based on 
Ref. \cite{Sehgal:2009xv}.

There are several works where constraints on the NID mode
are investigated \cite{Trotta:2002iz, Moodley:2004nz, 
Beltran:2004uv,Trotta:2006ww, Kawasaki:2007mb}. 
$N_{\rm eff}$ is however fixed to 
the standard value $3.04$ in these analyses, so that
our analysis explore a new parameter space which has 
not been investigated so far.

Parameter estimation is performed using a modified version of
the publicly available CosmoMC code~\cite{Lewis:2002ah}. 
Convergence of a Markov chain Monte Carlo (MCMC) analysis
is diagnosed by the Gelman-Rubin test $R-1<0.1$.

%%%%%%%%%%%%%%%%%%%%%%%%%%%%%%%%%%%
\subsection{Fixed $\gamma$}
\label{sec:fixed_corr}
%%%%%%%%%%%%%%%%%%%%%%%%%%%%%%%%%%%

%%%
\begin{table}%[ht]
  \begin{center}
  \begin{tabular}{lrr}
    \hline
    \hline
    parameters & CMB & ALL\\
    \hline
    $100\omega_b$ & $2.35^{+0.10}_{-0.10}$ & $2.25^{+0.05}_{-0.05}$\\
    $\omega_c$ & $0.135^{+0.016}_{-0.016}$ & $0.133^{+0.012}_{-0.012}$\\
    $\theta_s$ & $1.038^{+0.003}_{-0.003}$ & $1.037^{+0.003}_{-0.003}$\\
    $\tau$ & $0.089^{+0.007}_{-0.007}$ & $0.085^{+0.007}_{-0.006}$\\
    $N_\mathrm{eff}$ (95\%CL) & $\le6.5$ & $\le5.3$\\
    $n_s$ & $1.02^{+0.03}_{-0.03}$ & $0.987^{+0.015}_{-0.015}$\\
    $\ln[10^{10}A_s]$ & $3.19^{+0.05}_{-0.05}$ & $3.17^{+0.04}_{-0.04}$\\
    $\alpha$ (95\%CL) & $\le0.16$ & $\le0.11$\\
    \hline
%    $A_{SZ}$ & $^{+}_{-}$ & $^{+}_{-}$\\
%    $A_p$  & $^{+}_{-}$ & $^{+}_{-}$\\
%    $A_{cl}$  & $^{+}_{-}$ & $^{+}_{-}$\\
    \hline
  \end{tabular}
  \caption{Constraints for the uncorrelated case ($\gamma=0$). 
  We present mean values and 68 \%CL intervals for 
  cosmological parameters except for $N_\mathrm{eff}$ and $\alpha$,
  for which we present 95 \%CL intervals since they are not bounded
  from below.}
  \label{tbl:ucorr}
  \end{center}
\end{table}
%%%

%%%%%%%%%%%%%%%%%%%%
\begin{figure}
\begin{center}
\tabcolsep=10mm
\begin{tabular}{ccc}
\hspace{-10mm}\includegraphics[scale=0.9]{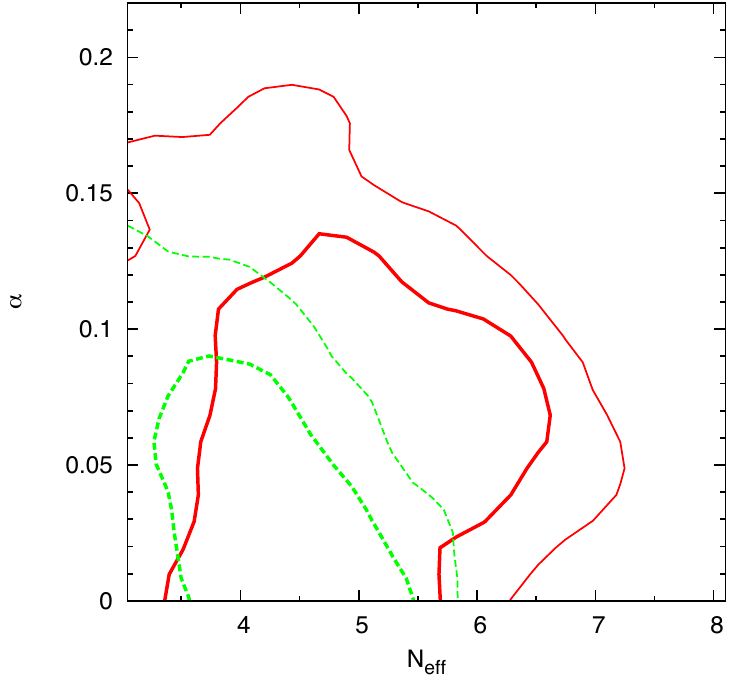}\\
\hspace{-10mm}\includegraphics[scale=0.9]{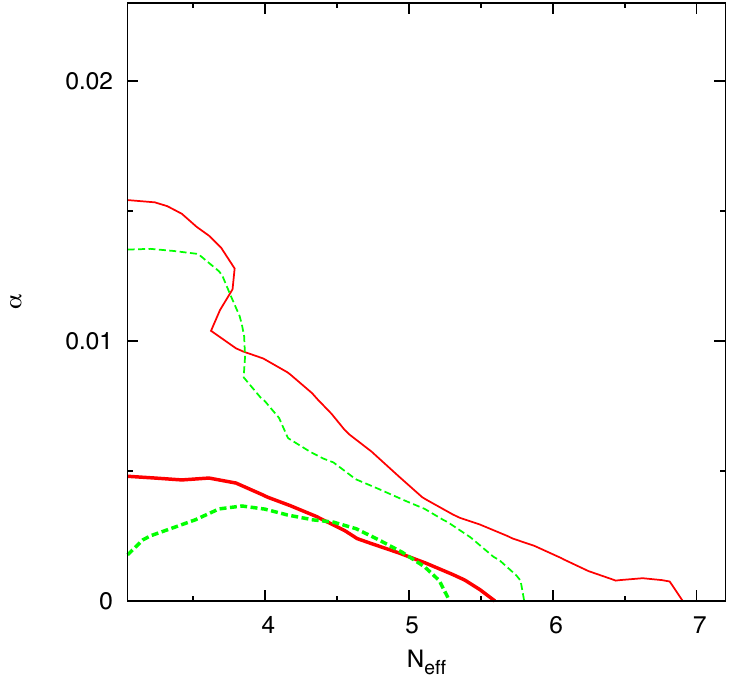}\\
\hspace{-10mm}\includegraphics[scale=0.9]{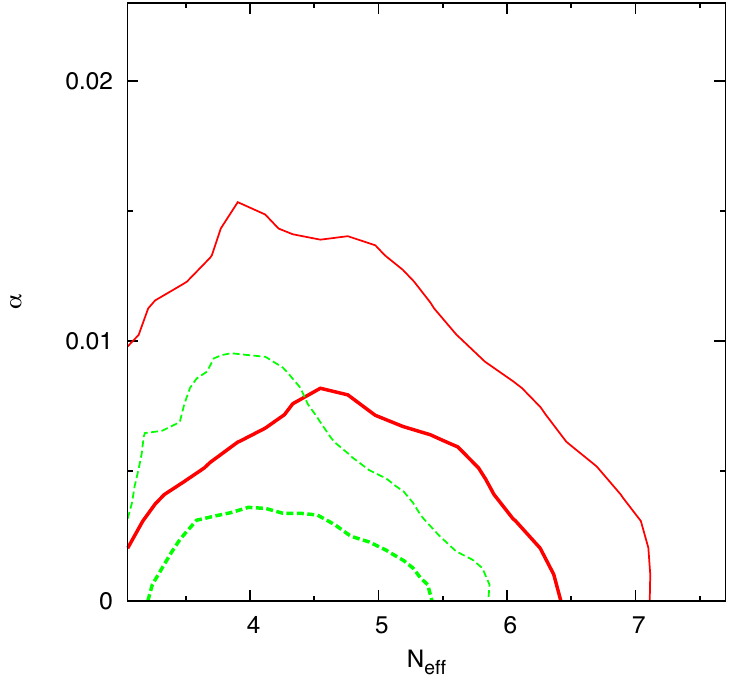}
\end{tabular}
\caption{68 \% and 95 \% CL constraints in $N_\mathrm{eff}$-$\alpha$ plane
from the CMB (red solid) and ALL (green dashed) datasets. From top to bottom, shown are the
constraints for the uncorrelated ($\gamma=0$), totally correlated ($\gamma=1$)
and totally anti-correlated ($\gamma=-1$) cases. Note that the scales are not same among 
three panels.}
\label{fig:N_alpha}
\end{center}
\end{figure}
%%%%%%%%%%%%%%%%%%%%

Before varying all the nine primary parameters, we
first fix $\gamma$ to $0$ or $\pm1$ and vary other parameters.
This allows us to investigate the constraints for the uncorrelated 
and totally (anti-)correlated cases separately. 

Let us see from the constraints for the uncorrelated case ($\gamma=0$).
Constraints on parameters are summarized in Table.~\ref{tbl:ucorr}.
In particular, we obtain constraints $N_\mathrm{eff}\le5.3$
and $\alpha\le 0.11$ at 95 \%CL from the ALL dataset. 
We also show 2d constraints in $N_\mathrm{eff}$-$\alpha$ plane 
in Fig.~\ref{fig:N_alpha}.
We see there is no strong degeneracy between $N_\mathrm{eff}$
and $\alpha$, so that constraints on these two parameters can be
discussed separately. As is often discussed, 
current CMB observations essentially constrain the epoch of matter-radiation equality.
Consequently, when $N_\mathrm{eff}$ is freely varied, $\omega_c$
and $N_\mathrm{eff}$ are strongly degenerated. This degeneracy
is partially solved by including BAO and H0, which determine
$\Omega_m$ and $H_0$ and hence lead to a better constraint
on $N_\mathrm{eff}$. 
On the other hand, BAO and H0 data
are also effective in the determination of $\omega_b$, which is strongly
degenerated with $n_s$ because these two affect the 
relative height of the first and second peaks in a similar way.
$\alpha$ is mostly degenerated 
with $\omega_b$ and $n_s$ because mixture of 
the NID mode $\alpha\ne 0$ affects the relative height 
of acoustic peaks with little effect on the peak positions.
Thus, inclusion of BAO and H0 
improves the constraint on $\alpha$ by solving 
its degeneracy with $\omega_b$ and $n_s$.\footnote{%
Please notice that if we are allowed to include an extra dark 
radiation, i.e., $N_{\rm eff} > 3$
the best fit value for the spectral index increases and 
the Harrison-Zeldovich spectrum with $n_s=1$ is consistent with 
the present observational data. This is true even if the isocurvature 
perturbation does not exist.}  
%%

%%%
\begin{table}%[ht]
  \begin{center}
  \begin{tabular}{lrr}
    \hline
    \hline
    parameters & CMB & ALL\\
    \hline
    $100\omega_b$ & $2.17^{+0.06}_{-0.06}$ & $2.19^{+0.05}_{-0.05}$\\
    $\omega_c$ & $0.133^{+0.014}_{-0.014}$ & $0.130^{+0.012}_{-0.012}$\\
    $\theta_s$ & $1.032^{+0.003}_{-0.003}$ & $1.033^{+0.003}_{-0.003}$\\
    $\tau$ & $0.087^{+0.007}_{-0.007}$ & $0.088^{+0.006}_{-0.007}$\\
    $N_\mathrm{eff}$ (95\%CL) & $\le5.4$ & $\le5.1$\\
    $n_s$ & $0.97^{+0.02}_{-0.02}$ & $0.972^{+0.015}_{-0.015}$\\
    $\ln[10^{10}A_s]$ & $3.16^{+0.04}_{-0.04}$ & $3.15^{+0.04}_{-0.04}$\\
    $\alpha$ (95\%CL) & $\le0.012$ & $\le0.011$\\
    \hline
%    $A_{SZ}$ & $^{+}_{-}$ & $^{+}_{-}$\\
%    $A_p$  & $^{+}_{-}$ & $^{+}_{-}$\\
%    $A_{cl}$  & $^{+}_{-}$ & $^{+}_{-}$\\
%    \hline
  \end{tabular}
  \caption{Same table as in Table.~\ref{tbl:ucorr} but for the
  totally correlated case ($\gamma=1$).}
  \label{tbl:tcorr}
  \end{center}
\end{table}
%%%

%%%
\begin{table}%[ht]
  \begin{center}
  \begin{tabular}{lrr}
    \hline
    \hline
    parameters & CMB & ALL\\
    \hline
    $100\omega_b$ & $2.31^{+0.07}_{-0.07}$ & $2.25^{+0.05}_{-0.04}$\\
    $\omega_c$ & $0.135^{+0.016}_{-0.016}$ & $0.135^{+0.011}_{-0.011}$\\
    $\theta_s$ & $1.041^{+0.003}_{-0.003}$ & $1.039^{+0.003}_{-0.003}$\\
    $\tau$ & $0.089^{+0.007}_{-0.007}$ & $0.083^{+0.006}_{-0.006}$\\
    $N_\mathrm{eff}$ (95\%CL) & $\le6.2$ & $\le5.2$\\
    $n_s$ & $1.01^{+0.02}_{-0.02}$ & $0.987^{+0.013}_{-0.014}$\\
    $\ln[10^{10}A_s]$ & $3.10^{+0.04}_{-0.04}$ & $3.11^{+0.04}_{-0.04}$\\
    $\alpha$ (95\%CL) & $\le0.013$ & $\le0.007$\\
    \hline
%    $A_{SZ}$ & $^{+}_{-}$ & $^{+}_{-}$\\
%    $A_p$  & $^{+}_{-}$ & $^{+}_{-}$\\
%    $A_{cl}$  & $^{+}_{-}$ & $^{+}_{-}$\\
    \hline
  \end{tabular}
  \caption{Same table as in Table.~\ref{tbl:ucorr} but for the
  anti-correlated case ($\gamma=-1$).}
  \label{tbl:acorr}
  \end{center}
\end{table}
%%%

Let us then turn our attention to the totally correlated ($\gamma=1$) 
and anti-correlated ($\gamma=-1$) cases. Constraints 
on parameters for these two cases are summarized in 
Tables~\ref{tbl:tcorr}-\ref{tbl:acorr} and 2d constraints in 
$N_\mathrm{eff}$-$\alpha$ plane are also shown
in Fig.~\ref{fig:N_alpha}. From the ALL dataset, 
we obtain $N_\mathrm{eff}\le5.4$ and $\alpha\le0.011$ for
the totally correlated case and 
$N_\mathrm{eff}\le5.2$ and $\alpha\le0.007$ for
the totally anti-correlated case, respectively.
We first note that the constraints on 
$\alpha$ for the correlated cases are an 
order of magnitude tighter than those for the uncorrelated case.
This is because given fixed $\alpha$, the NID mode affects the 
CMB power spectrum more significantly
for the correlated cases than the uncorrelated one due to 
the power $P_{\zeta S_\mathrm{DR}}(k)$
arising from the cross correlation of the AD and NID modes.
For small enough $\alpha$ of order e.g. $10^{-3}$, 
changes in the CMB power spectrum are 
somewhat similar in magnitude but opposite in the sign.
However, data adopted in the analysis are not very constraining
and relatively large $\alpha\simeq 0.01$ is allowed,
where degeneracies of $\alpha$ with other parameters
are different between the totally correlated and anti-correlated cases.
For the totally correlated case ($\gamma=1$) $\alpha$ 
is not degenerated with other parameters strongly.
This is the reason why the CMB constraint on $\alpha$ is little
improved by including BAO and H0 for this case.
On the other hand, 
$\alpha$ considerably is degenerated with $\omega_c$ and $n_s$
for the case of anti-correlated case ($\gamma=-1$).
Better determination of $H_0$ and $\Omega_m$ due to inclusion 
of BAO and H0 directly tightens the constraint on $\omega_c$.
Furthermore, $n_s$ itself is also strongly degenerated with
$\omega_c$ and hence becomes more tightly constrained.
Such the improvement in determination of $\omega_c$ and 
$n_s$ finally results in better constraints on $\alpha$.

%%%%%%%%%%%%%%%%%%%%%%%%%%%%%%%%%%%
\subsection{Fixed $N_\mathrm{eff}$}
%%%%%%%%%%%%%%%%%%%%%%%%%%%%%%%%%%%

As we have discussed in Introduction, there are several
cosmological observations which independently show preference for
a somewhat larger $N_\mathrm{eff}\simeq4$ over the
standard $N_\mathrm{eff}=3.04$ at around 2$\sigma$ level.
When we look over the constraints obtained so far, 
it can be seen that such preference still survives, 
though with somewhat reduced significance, even if
we include the possibility for the extra radiation to have isocurvature fluctuations
which can be uncorrelated or totally (anti-)correlated with $\zeta$.
This motivate us to investigate a generally correlated 
case with a fixed $N_\mathrm{eff}\ne3.04$, for which 
we devote the rest of this section.

\begin{figure}[t]
\begin{center}
\includegraphics[scale=0.6]{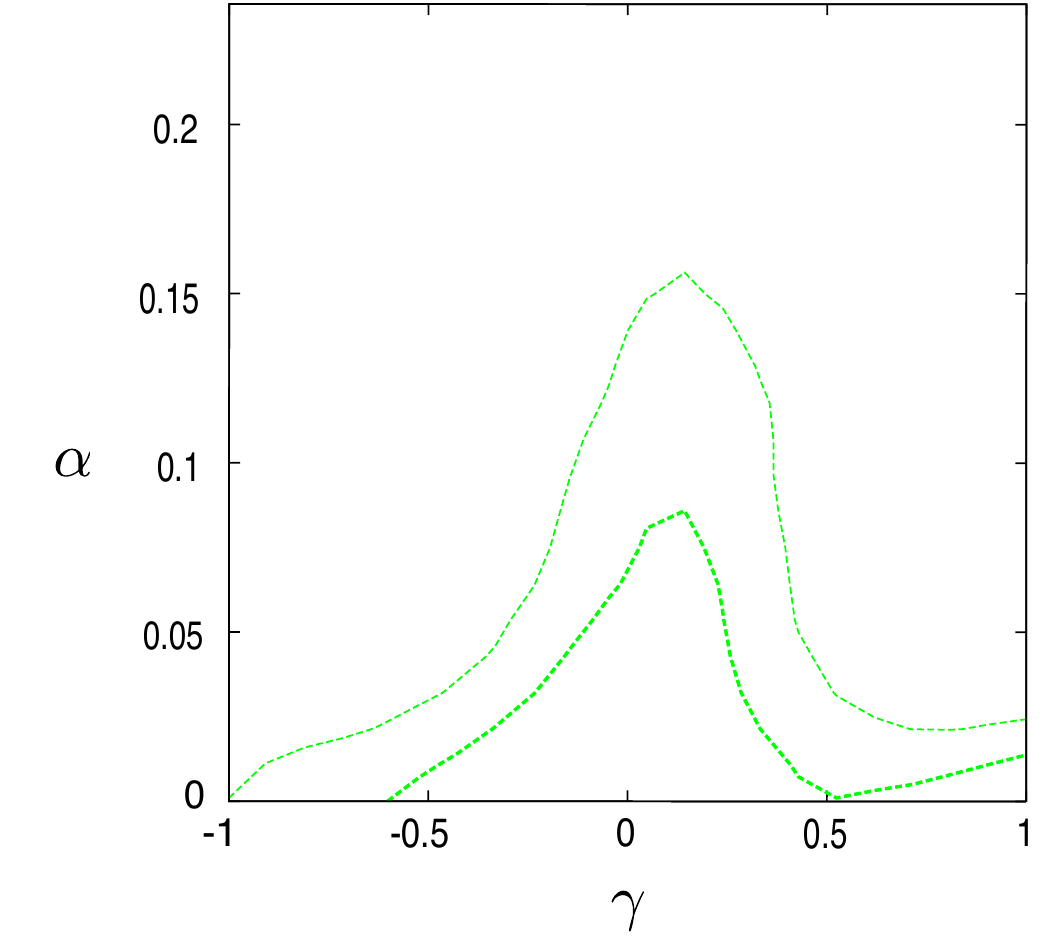}
\caption{68 \% and 95 \% CL constraints in $\gamma$-$\alpha$ plane
from ALL datasets.}
\label{fig:alpha_gamma}
\end{center}
\end{figure}

We perform  MCMC analysis with $\gamma$ being varied and 
$N_\mathrm{eff}=4$ being fixed.
We impose a top-hat prior on $\gamma$ with a range $[-1,1]$,
and priors for other parameters are as 
listed in Table~\ref{tbl:priors}. 
This time we use only the ALL dataset and the resultant
constraints are summarized in Table~\ref{tbl:corr}.
In addition, the 2d constraint in the $\gamma$-$\alpha$ plane
is shown in Fig.~\ref{fig:alpha_gamma}.
As can be seen from the figure, we do not find
any evidence for $\alpha>0$ and the
extra radiation which is assumed to exist is consistent 
with purely AD perturbations. When $\gamma$ is marginalized, 
we obtain an upper bound $\alpha\le0.12$ at 95\% CL, which is almost 
the same as one obtained for the uncorrelated case.
We note that the constraint on $\gamma$ is of little importance,
as long as $\alpha$ is consistent with zero.

%%%
\begin{table}%[ht]
  \begin{center}
  \begin{tabular}{lrr}
    \hline
    \hline
    parameters & ALL\\
    \hline
    $100\omega_b$ & $2.24^{+0.05}_{-0.05}$\\
    $\omega_c$ & $0.129^{+0.004}_{-0.004}$\\
    $\theta_s$ & $1.037^{+0.003}_{-0.004}$\\
    $\tau$ & $0.086^{+0.007}_{-0.007}$\\
    $n_s$ & $0.983^{+0.013}_{-0.012}$\\
    $\ln[10^{10}A_s]$ & $3.16^{+0.06}_{-0.05}$\\
    $\alpha$ (95\%CL) & $\le0.12$\\
    $\gamma$ & $0.025^{+0.123}_{-0.080}$\\
    \hline
%    $A_{SZ}$ & $^{+}_{-}$ & $^{+}_{-}$\\
%    $A_p$  & $^{+}_{-}$ & $^{+}_{-}$\\
%    $A_{cl}$  & $^{+}_{-}$ & $^{+}_{-}$\\
    \hline
  \end{tabular}
  \caption{Constraints in the case that $N_{\rm eff}$ is set to $4$ and $\gamma$ is varied. 
  We present mean values and 68 \%CL intervals for 
  cosmological parameters except for $\alpha$,
  for which we present 95 \%CL intervals since it is not bounded
  from below.}
  \label{tbl:corr}
  \end{center}
\end{table}
%%%

%%%%%%%%%%%%%%%%%%%%
%%%%%%%%%%%%%%%%%%%%

%new analysis%%%%%%%%

%%%%%%%%%%%%%%%%%%%%%%%%%%%%%%%%%%%
\subsection{Varying $N_\mathrm{eff}$ and $\gamma$}
%%%%%%%%%%%%%%%%%%%%%%%%%%%%%%%%%%%

We finally present constraints for the most general case, 
where all nine primary parameters including $N_\mathrm{eff}$, 
$\alpha$ and $\gamma$ are varied simultaneously.
Parameter constraints from ALL dataset are presented in 
Table~\ref{tbl:varying_all}.
Compared with the case with fixed $N_\mathrm{eff}=4$ in Table~\ref{tbl:corr},
we can find that there is no significant degradation of constraints on 
parameters except for $\omega_c$, which is significantly degenerated
with $N_\mathrm{eff}$. 
On the other hand, upper bound on $N_\mathrm{eff}$ is 
also as tight as those obtained for the case of fixed $\gamma$ in Section 
\ref{sec:fixed_corr}. 

\begin{figure}[t]
\begin{center}
  \begin{tabular}{lrr}
  \includegraphics[scale=0.6]{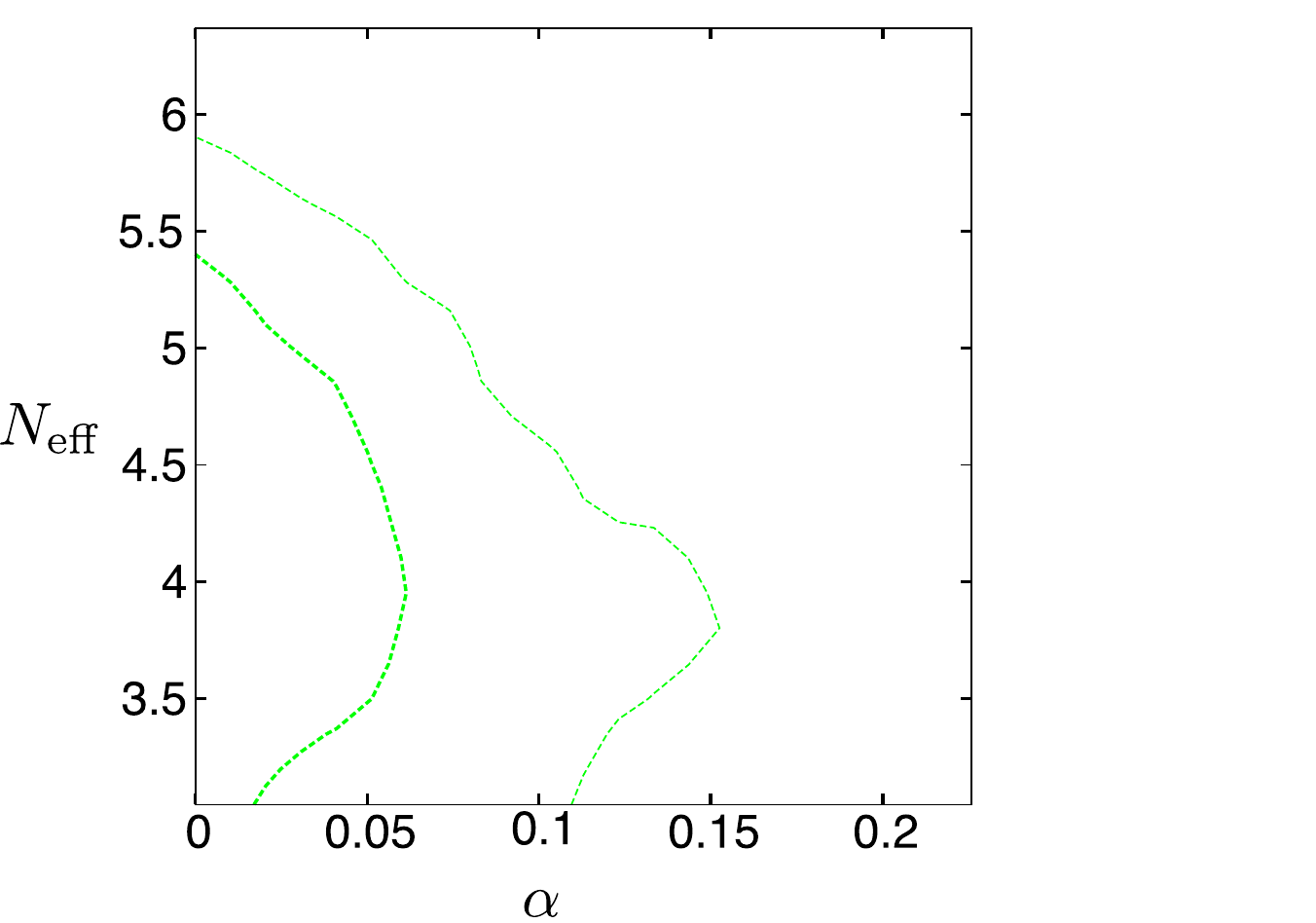} &
  \hspace{-15mm}\includegraphics[scale=0.6]{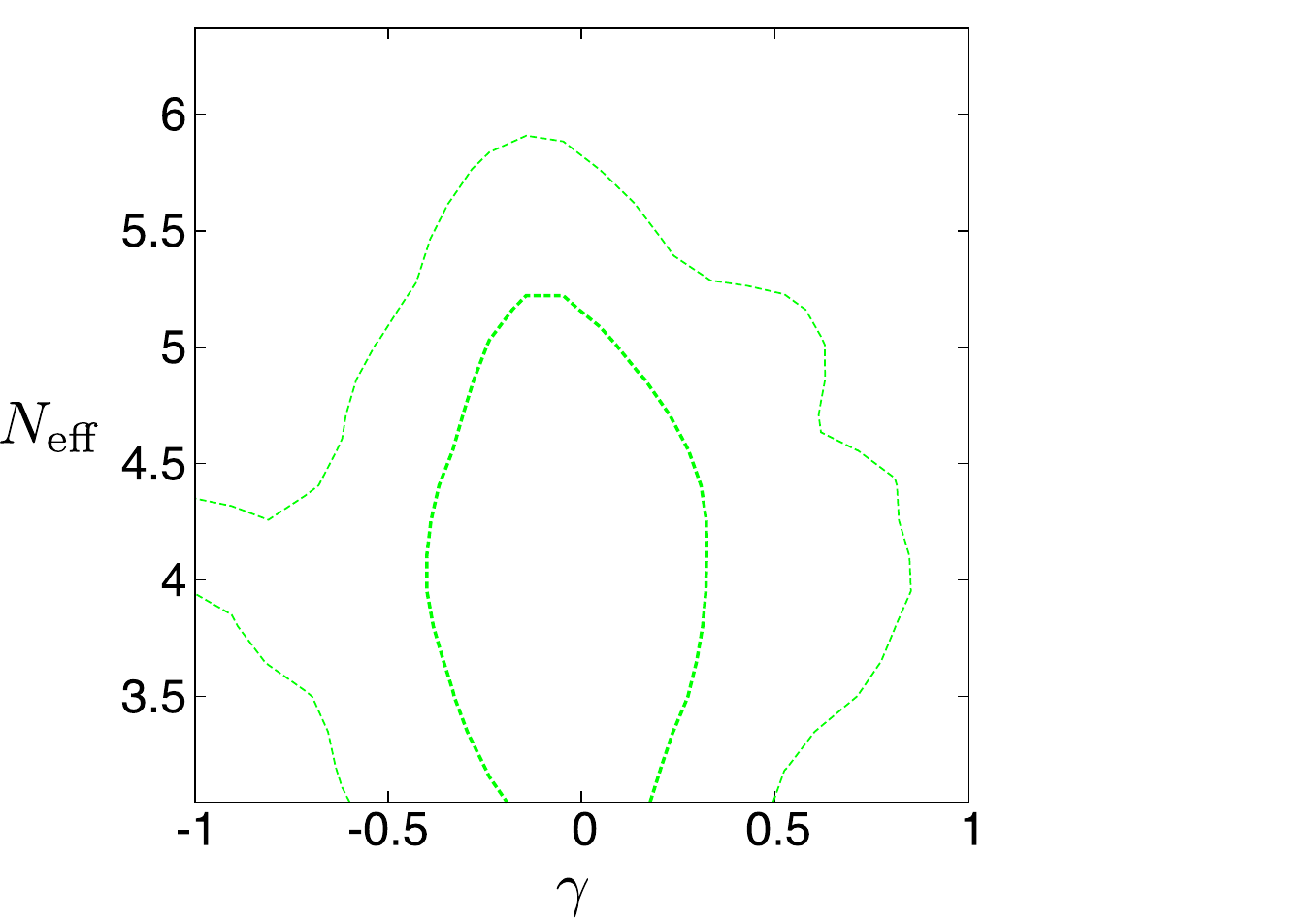} \\
  & \hspace{-15mm}\includegraphics[scale=0.6]{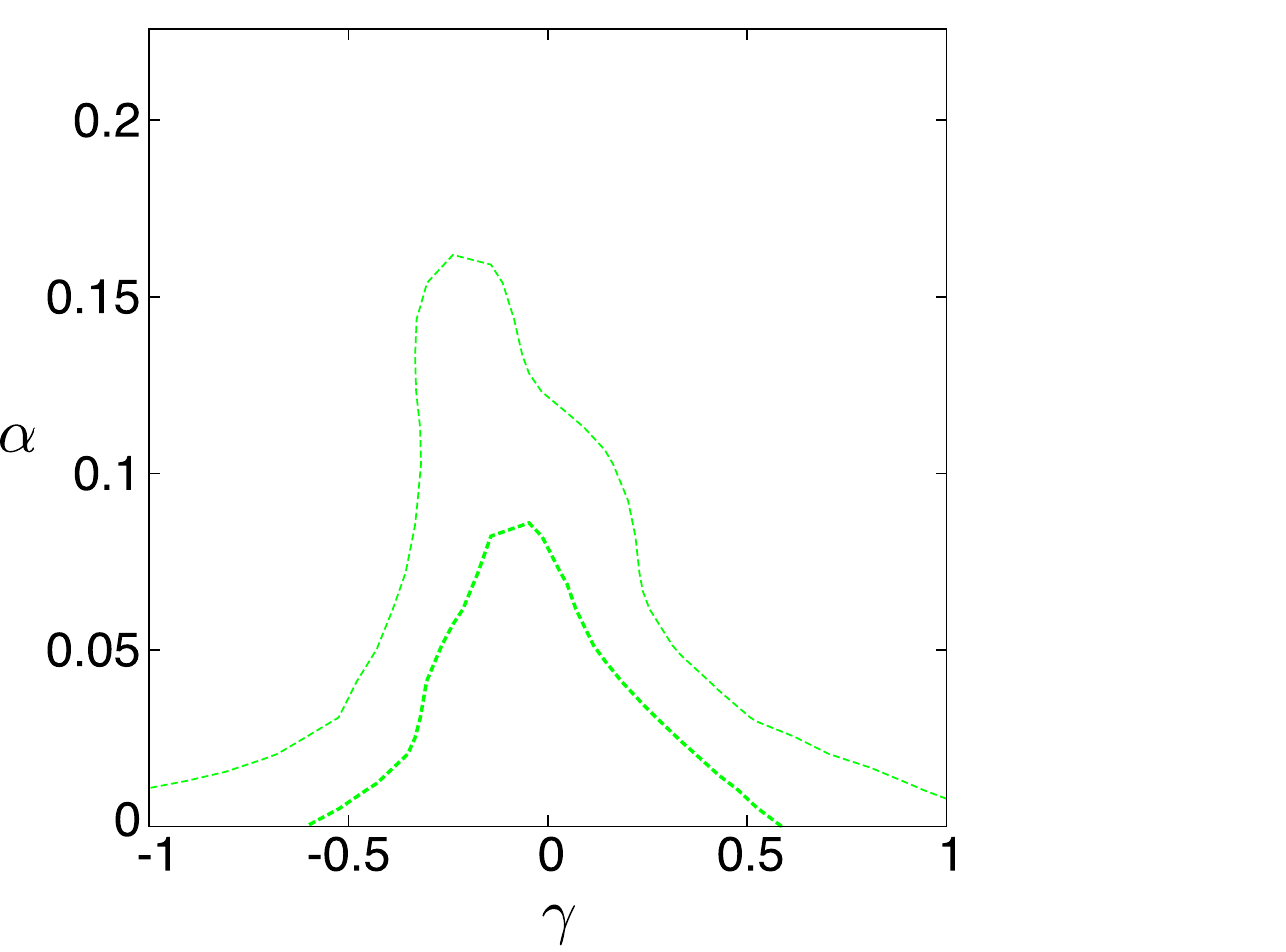}\\
  \end{tabular}
\caption{68 \% and 95 \% CL constraints on $(N_\mathrm{eff}, \alpha, \gamma)$ 
for generally correlated case from ALL dataset.}
\label{fig:varying_all}
\end{center}
\end{figure}

Marginalized constraints on
$N_\mathrm{eff}$ and $\alpha$ are given as
$N_\mathrm{eff}\le 5.1$ and $\alpha\le 0.11$
at 95 \%CL. 
Following the results so far, 
we again do not find any signature for existence of
non-vanishing isocurvature perturbations in extra radiations.
2d marginalized constraints for $N_\mathrm{eff}$, $\alpha$ and $\gamma$
are presented in Fig.~\ref{fig:varying_all}.

%%%
\begin{table}%[ht]
  \begin{center}
  \begin{tabular}{lrr}
    \hline
    \hline
    parameters & ALL\\
    \hline
    $100\omega_b$ & $2.24^{+0.05}_{-0.05}$\\
    $\omega_c$ & $0.131^{+0.011}_{-0.012}$\\
    $\theta_s$ & $1.037^{+0.003}_{-0.003}$\\
    $\tau$ & $0.083^{+0.006}_{-0.006}$\\
    $N_\mathrm{eff}$ (95\%CL) & $\le5.1$\\
    $n_s$ & $0.983^{+0.016}_{-0.015}$\\
    $\ln[10^{10}A_s]$ & $3.14^{+0.05}_{-0.05}$\\
    $\alpha$ (95\%CL) & $\le0.11$\\
    $\gamma$ & $0.07^{+0.11}_{-0.14}$\\
    \hline
    \hline
  \end{tabular}
  \caption{Constraints for the generally correlated case ($-1\le\gamma\le1$)
  from ALL dataset.
  We present mean values and 68 \%CL intervals for cosmological parameters 
  except for $N_\mathrm{eff}$ and $\alpha$, for which we present 95 \%CL intervals 
  since they are not bounded from below.}
  \label{tbl:varying_all}
  \end{center}
\end{table}
%%%

%%%%%%%%%%%%%%%%%%%%
%%%%%%%%%%%%%%%%%%%

%new analysis%%%%%%%%

%%%%%%%%%%%%%%%%%%%%%%%%%%%%%%%%%%%
\section{Implications}     \label{sec:model}
%%%%%%%%%%%%%%%%%%%%%%%%%%%%%%%%%%%

We have formulated a method for calculating the isocurvature perturbations in the extra radiation
and derived observational constraints on them.
Now let us apply the results to some concrete cases.
Before discussing a concrete model, we derive formulae for some 
limiting cases for later convenience.

%%%%%%%%%%%%%%%%%%%%%%%%%%%%%%%%%%%
\subsection{Limiting cases}    
%%%%%%%%%%%%%%%%%%%%%%%%%%%%%%%%%%%

%%%%%%%%%%%%%%%%%%%%%%%%%%%%%%%%%%%
\subsubsection{Uncorrelated case}    
%%%%%%%%%%%%%%%%%%%%%%%%%%%%%%%%%%%

First, we suppose that the inflaton $\phi$ does not decay into $X$ ($r_\phi = 1$)
and the adiabatic perturbation is dominantly produced by the inflaton $(N_\phi \gg N_\sigma)$.
The $\sigma$ decays into $X$ with branching fraction of $1-r_\sigma$.
Then one can easily imagine that the $X$ has an isocurvature perturbation 
that traces the primordial perturbation of $\sigma$.

In this case, from Eq.~(\ref{DeltaNeff}) we have
\begin{equation}
	\Delta N_{\rm eff} \simeq \frac{3(1-r_\sigma)R_\sigma}{c_\nu(1-(1-r_\sigma)R_\sigma) }
	\left( \frac{g_*(H=\Gamma_\nu)}{g_*(H=\Gamma_\sigma)} \right)^{1/3}.
	\label{DeltaN_uncorr}
\end{equation}
and
\begin{equation}
	\hat S_{\rm DR} \simeq \frac{(1-r_\sigma)(1-R_\sigma)(1-c_\nu)}{\tilde R_{\rm DR}}
	\left( \frac{g_*(H=\Gamma_\nu)}{g_*(H=\Gamma_\sigma)} \right)^{1/3}
	\frac{2R\delta\sigma}{\sigma_i}.
	\label{S_uncorr}
\end{equation}
with 
\begin{equation}
	\tilde R_{\rm DR} \simeq (1-r_\sigma)R_\sigma
	\left( \frac{g_*(H=\Gamma_\nu)}{g_*(H=\Gamma_\sigma)} \right)^{1/3}
	 + c_\nu (1-R_\sigma + r_\sigma R_\sigma).
\end{equation}
%%
%%
%\begin{equation}
%	S_{\rm DR} = \frac{(1-r_\sigma)(1-R_\sigma)(1-c_\nu)}{c_\nu(1-R_\sigma)+(1-(1-c_\nu)r_\sigma)R_\sigma} 
%	\frac{2R\delta\sigma}{\sigma_i}.
%	\label{S_uncorr}
%\end{equation}
%%
In deriving (\ref{S_uncorr}), we have approximated parameters as 
$\hat R_X \simeq \tilde R_X \ll 1$ and $\hat R_r \simeq \tilde R_r$.
Note that the condition $(N_\phi \gg N_\sigma)$ leads to the constraint $2R\delta\sigma/\sigma_i \ll 10^{-4}$.
In that limit the correlation parameter (\ref{cor}) is small : $|\gamma| \ll 1$.
Thus the isocurvature perturbation is nearly uncorrelated with the curvature perturbation.
It should be noticed that $S_{\rm DR}$ vanishes in the limit $R_\sigma \to 1$,
since the inflaton contribution to the curvature perturbation becomes zero, as is clear from (\ref{zeta_from_zetaphi}).

\begin{figure}[t]
\begin{center}
\includegraphics[scale=0.4]{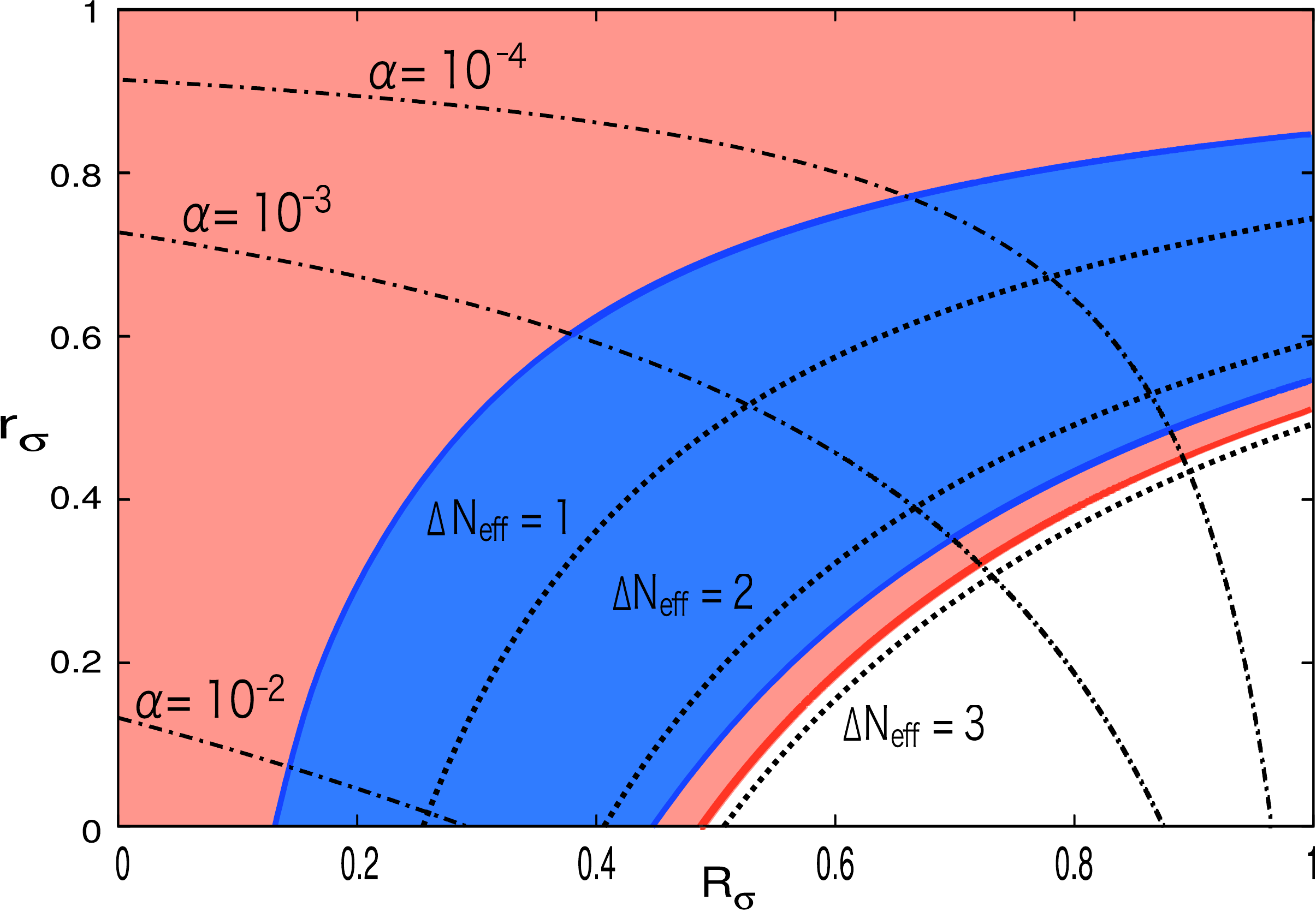}
\caption{ 
	Constraints on $R_\sigma$-$r_\sigma$ plane for uncorrelated case.
	1$\sigma$ and 2$\sigma$ allowed regions are shown by blue and orange, respectively. 
	Contours of $\alpha$ and $\Delta N_{{\rm eff}}$ are also shown.
	In this figure $2R\delta\sigma/\sigma_i \simeq 5\times 10^{-6}$ is fixed.
	}
\label{fig:r-omega_uncorr}
\end{center}
\end{figure}

Fig.~\ref{fig:r-omega_uncorr} shows
constraints on $R_\sigma$-$r_\sigma$ plane for uncorrelated case.
This is drawn by translating the results shown in the top panel of Fig.~\ref{fig:N_alpha} 
into the constraint on $R_\sigma$ and $r_\sigma$, using (\ref{DeltaNeff}) and (\ref{SDRfinal}).
1$\sigma$ and 2$\sigma$ allowed regions are shown by blue and orange, respectively.
Contours of $\alpha$ and $\Delta N_{{\rm eff}}$ are also shown.
In this figure $2R\delta\sigma/\sigma_i = 0.1\times \zeta_{\phi} \simeq 5\times 10^{-6} $ 
and $g_*(H=\Gamma_{\sigma})=100$ are fixed.

%%%%%%%%%%%%%%%%%%%%
%%%%%%%%%%%%%%%%%%%%

%%%%%%%%%%%%%%%%%%%%%%%%%%%%%%%%%%%
\subsubsection{Totally anti-correlated case}    
%%%%%%%%%%%%%%%%%%%%%%%%%%%%%%%%%%%

Next, suppose that the $\sigma$ does not decay into $X$, i.e., $r_\sigma =1$,
and $\sigma$ dominantly produces the curvature perturbation ($N_\phi \ll N_\sigma$) :
$\sigma$ truly takes a role of curvaton.
The inflaton $\phi$ is assumed to decay into $X$ with branching fraction of $1-r_\phi$.
Then the extra radiation $X$ produced by the inflaton decay is expected to have 
isocurvature perturbation correlated with the curvature one.

In this case we have
\begin{equation}
	\Delta N_{\rm eff} \simeq \frac{3(1-r_\phi)(1-R_\sigma)}{c_\nu(r_\phi(1-R_\sigma)+R_\sigma) }
	\left( \frac{g_*(H=\Gamma_\nu)}{g_*(H=\Gamma_\sigma)} \right)^{1/3},
	\label{DeltaN_anticorr}
\end{equation}
and
\begin{equation}
	\hat S_{\rm DR} = -\frac{(1-r_\phi)(1-R_\sigma)(1-c_\nu)}{\tilde R_{\rm DR}}
	\left( \frac{g_*(H=\Gamma_\nu)}{g_*(H=\Gamma_\sigma)} \right)^{1/3}
	\frac{2R\delta\sigma}{\sigma_i}.
	\label{S_anticorr}
\end{equation}
with 
\begin{equation}
	\tilde R_{\rm DR} \simeq (1-r_\phi)(1-R_\sigma) 
	\left( \frac{g_*(H=\Gamma_\nu)}{g_*(H=\Gamma_\sigma)} \right)^{1/3}
	+ c_\nu (r_\phi+R_\sigma - r_\phi R_\sigma).
\end{equation}
%%
%%
%\begin{equation}
%	S_{\rm DR} = -\frac{(1-r_\phi)(1-R_\sigma)(1-c_\nu)}{(1-R_\sigma)(1-(1-c_\nu)r_\phi)+c_\nu R_\sigma} 
%	\frac{2R\delta\sigma}{\sigma_i}.
%\end{equation}
%%
In deriving (\ref{S_anticorr}), we have approximated parameters as 
$\hat R_X \simeq \tilde R_X \ll 1$ and $\hat R_r \simeq \tilde R_r$.
Notice that we have the condition $2R\delta\sigma/\sigma_i \simeq1.5\times 10^{-4}$
in order for the curvaton to produce the curvature perturbation.
In this limit, the correlation parameter (\ref{cor}) becomes $\gamma \simeq -1$ :
the isocurvature perturbation is almost totally anti-correlated with the curvature perturbation.
Also we need $R \gtrsim 0.01$ in order not to have too large non-Gaussianity.

\begin{figure}[t]
\begin{center}
\includegraphics[scale=0.4]{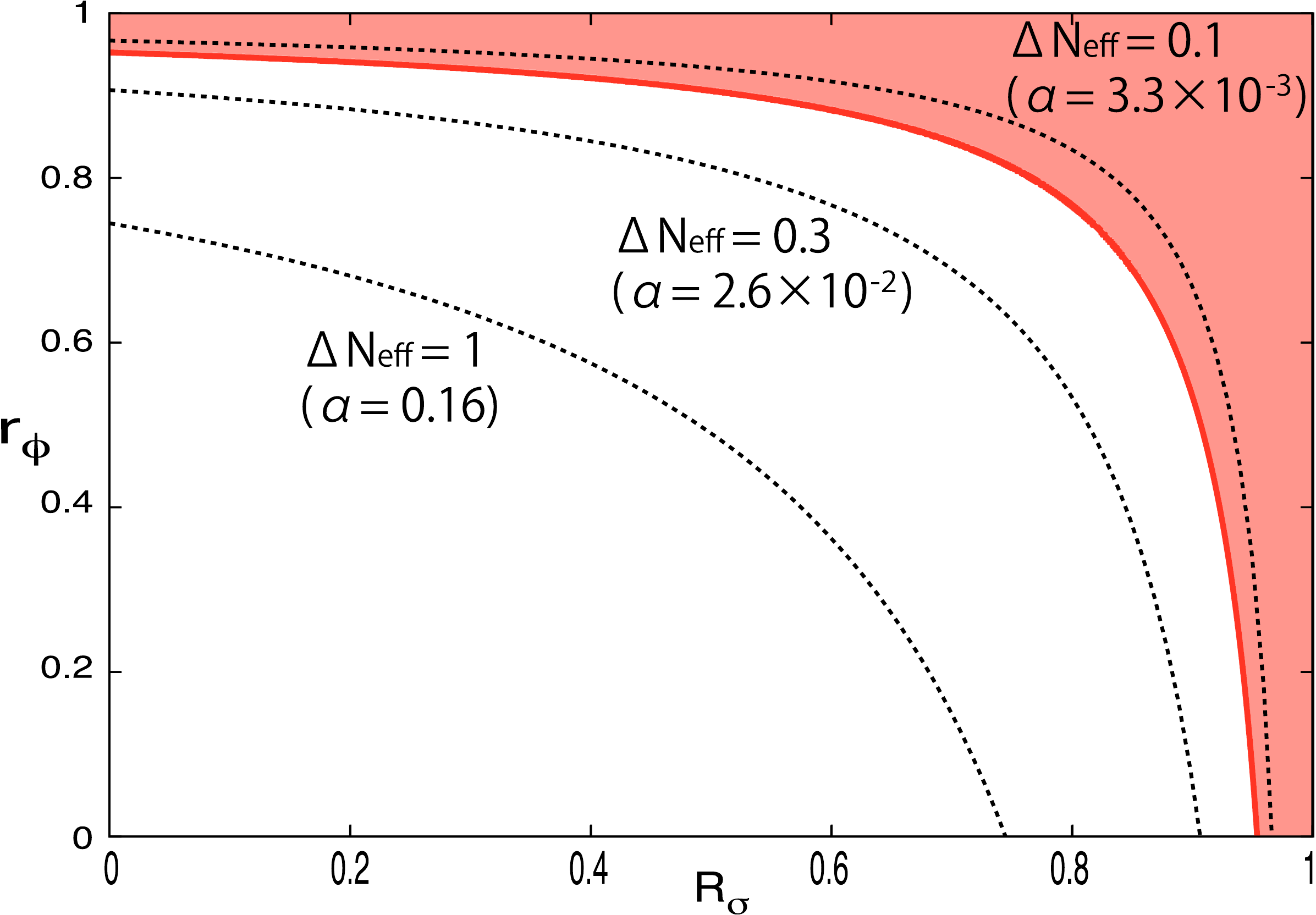}
\caption{ 
	Constraints on $R_\sigma$-$r_\phi$ plane for totally anti-correlated case. 
	2$\sigma$ allowed regions are shown by orange. There is no 1$\sigma$ allowed region. 
}
\label{fig:r-omega_anticorr}
\end{center}
\end{figure}

Fig.~\ref{fig:r-omega_anticorr} shows
constraints on $R_\sigma$-$r_\phi$ plane for totally anti-correlated case.
This figure is obtained by converting bottom panel of Fig.~\ref{fig:N_alpha}
using (\ref{DeltaNeff}) and (\ref{SDRfinal}).
2$\sigma$ allowed regions are shown by orange. 
Any values of $R_\sigma$ and $r_\phi$ are excluded at 1$\sigma$ level.
In this figure, contours of both $\alpha$ and $\Delta N_{{\rm eff}}$ are shown.
Note that in the case of $r_\sigma =1$, 
the following relation between $\alpha$ and $\Delta N_{\rm eff}$ is satisfied.
\begin{equation}
\sqrt{\frac{\alpha}{1-\alpha}}=-\frac{\hat{S}_{\rm DR}}{\hat{\zeta}}=3(1-\hat{c}_{\nu})\frac{\Delta N_{\rm eff}}{3+\Delta N_{\rm eff}},
\end{equation}
which is derived from (\ref{DeltaNeff}), (\ref{SDRfinal}) and (\ref{zetafinal}).
Here $g_*(H=\Gamma_{\sigma})$ is set to $100$ in this figure.

%%%%%%%%%%%%%%%%%%%%
%%%%%%%%%%%%%%%%%%%%

%%%%%%%%%%%%%%%%%%%%%%%%%%%%%%%%%%%
\subsection{Particle physics model}    
%%%%%%%%%%%%%%%%%%%%%%%%%%%%%%%%%%%

A model to explain $\Delta N_{\rm eff} \simeq 1$ was proposed 
in Ref.~\cite{Ichikawa:2007jv}.
It is based on the supersymmetric (SUSY) extension of the axion model.
The axion is a pseudo Nambu-Goldstone boson associated with the 
spontaneous breakdown of the
Peccei-Quinn (PQ) symmetry, which is introduced in order to solve 
the strong CP problem in quantum chromo dynamics~\cite{Kim:1986ax}.
In a SUSY axion model, there exists a scalar field $\sigma$, called 
saxion, which is a scalar partner of the PQ 
axion~\cite{Rajagopal:1990yx}.
The saxion can naturally have a mass of the order of the gravitino, 
which is the superpartner of the graviton,
and it can be $\mathcal O$(keV)-$\mathcal O$(TeV).
The saxion can have a large initial amplitude $\sigma_i$ during 
inflation, and may obtain quantum fluctuations of 
$\delta \sigma \sim H_{\rm inf}/(2\pi)$ since its mass is much smaller 
than the Hubble scale during inflation.\footnote{%%
The Hubble-induced mass term of the saxion is assumed to be suppressed.
Otherwise, its quantum fluctuation is significantly suppressed.
}
Therefore, the saxion is a candidate for the curvaton.
The saxion decay mode depends on the model (see e.g. 
Refs.~\cite{Chun:1995hc,Asaka:1998ns,Abe:2001cg,Kawasaki:2007mk,
Nakamura:2008ey,Kim:2008yu}).
We consider following two cases : the saxion dominantly decays into two 
axions $(\sigma \to aa)$ and into the Higgs boson pair $(\sigma \to hh)$.
The former situation is typically realized in the KSVZ axion model~\cite{Kim:1979if}
and leads to the uncorrelated isocurvature perturbation in the relativistic axion particles
if the curvature perturbation is dominantly sourced by the inflaton.
The latter possibility appears in the DFSZ axion model~\cite{Dine:1981rt}.
In this case, the saxion can take a role of the curvaton
and thermally produced axions after the inflaton decay have correlated isocurvature perturbations.
In the axion model, the axion also has an isocurvature perturbation 
which results in the CDM isocurvature mode.
We do not discuss it in detail since it can be reduced by tuning the initial misalignment angle.

%%%%%%%%%%%%%%%%%%%%%%%%%%%%%%%%%%%
\subsubsection{SUSY KSVZ axion model}    
%%%%%%%%%%%%%%%%%%%%%%%%%%%%%%%%%%%

First, let us consider the case where the saxion decays into two axions.
This is often the case in the KSVZ axion model~\cite{Kim:1979if}.
The decay rate of this process is given by
\begin{equation}
	\Gamma(\sigma \to 2a) = \frac{f^2}{64\pi}\frac{m_\sigma^3}{f_a^2},
\end{equation}
where $f_a$ denotes the PQ symmetry breaking scale, which is constrained as 
$10^9$GeV $\lesssim f_a \lesssim 10^{12}$GeV and $f$ is a typically $\mathcal O(1)$ constant.
Hereafter we take $f=1$.
The temperature at the saxion decay is calculated as
\begin{equation}
	T_\sigma \simeq 100~{\rm MeV}
	\left( \frac{100}{g_*(T_\sigma)} \right)^{1/4} 
	\left( \frac{m_\sigma}{100{\rm GeV}} \right)^{3/2}
	\left(\frac{10^{12}{\rm GeV}}{f_a} \right).
\end{equation}
The saxion also decays into gluons if kinematically allowed.
The branching ratio of the saxion into gluons, $r_\sigma$, is given by
\begin{equation}
	r_\sigma \simeq \frac{2\alpha_s^2}{\pi^2} \sim 3\times 10^{-3},
\end{equation}
and hence very small.
Nonthermal axions produced by the saxion decay is regarded as an extra radiation $X$,
since the PQ axion has only extremely weak interaction with ordinary matter and its mass is very small.
The abundance of the saxion coherent oscillation is given by
\begin{equation}
\begin{split}
	\frac{\rho_\sigma}{s} = \frac{1}{8}T_{\rm R}\left( \frac{\sigma_i}{M_{\rm P}} \right)^2
	 \simeq 2\times 10^{-8}{\rm GeV}\left( \frac{T_{\rm R}}{10^6{\rm GeV}} \right)
	 \left( \frac{\sigma_i}{10^{12}{\rm GeV}} \right)^2, 
\end{split}
\end{equation}
for $m_\sigma > \Gamma_\phi$ and
\begin{equation}
\begin{split}
	\frac{\rho_\sigma}{s} = \frac{1}{8}T_{\rm osc}\left( \frac{\sigma_i}{M_{\rm P}} \right)^2
	 \simeq 2\times 10^{-6}{\rm GeV}\left( \frac{m_{\sigma}}{100{\rm GeV}} \right)^{1/2}
	 \left( \frac{\sigma_i}{10^{12}{\rm GeV}} \right)^2, 
\end{split}
\end{equation}
for $m_\sigma < \Gamma_\phi$
where $s$ denotes the entropy density, $T_{\rm R}$ the reheating temperature after inflation
which is related to the inflaton decay rate $\Gamma_\phi$ as $T_{\rm R}=(10/\pi^2 g_*)^{1/4}\sqrt{\Gamma_\phi M_P}$
and $T_{\rm osc} \equiv (10/\pi^2 g_*)^{1/4}\sqrt{m_\sigma M_P}$.
Thus $R \simeq (3/4)R_\sigma = 3\rho_\sigma / (4\rho_{\rm tot})$ at the $\sigma$ decay is estimated as
\begin{equation}
\begin{split}
	R = \frac{1}{8}\frac{T_{\rm R}}{T_{\sigma}}\frac{\sigma_i^2}{M_{\rm P}^2} 
	\simeq 2\times 10^{-4}\left( \frac{T_{\rm R}}{10^6{\rm GeV}} \right)
	\left( \frac{1{\rm GeV}}{m_\sigma} \right)^{3/2}
	\left( \frac{f_a}{10^{12}{\rm GeV}} \right)^3
	\left( \frac{\sigma_i}{f_a} \right)^2,    \label{R_saxion}
\end{split}
\end{equation}
for $R\ll 1$ and $m_\sigma > \Gamma_\phi$ and
\begin{equation}
\begin{split}
	R = \frac{1}{8}\frac{T_{\rm osc}}{T_{\sigma}}\frac{\sigma_i^2}{M_{\rm P}^2} 
	\simeq 2\times 10^{-3}
	\left( \frac{1{\rm GeV}}{m_\sigma} \right)
	\left( \frac{f_a}{10^{12}{\rm GeV}} \right)^3
	\left( \frac{\sigma_i}{f_a} \right)^2,    \label{R_saxion2}
\end{split}
\end{equation}
for $R\ll 1$ and $m_\sigma < \Gamma_\phi$.
Assuming that the inflaton decays only to the visible sector $(r_\phi = 1)$,
$\Delta N_{\rm eff}$ and $S_{\rm DR}$ are obtained by substituting (\ref{R_saxion})
into Eqs.~(\ref{DeltaN_uncorr}) and (\ref{S_uncorr}).
The $\Delta N_{\rm eff}$ is estimated as
\begin{equation}
	\Delta N_{\rm eff} \simeq \frac{43}{7} \frac{R_\sigma}{1-R_\sigma} 
	\left( \frac{g_*(H=\Gamma_\nu)}{g_*(H=\Gamma_\sigma)} \right)^{1/3}.
\end{equation}
Therefore, both $\Delta N_{\rm eff}$ and $\hat S_{\rm DR}$ can have sizable values
depending on model parameters.

%%%%%%%%%%%%%%%%%%%%%%%%%%%%%%%%%%%
\subsubsection{SUSY DFSZ axion model}    
%%%%%%%%%%%%%%%%%%%%%%%%%%%%%%%%%%%

Next let us consider the case where the saxion dominantly decays into a Higgs boson pair,
as is realized in the DFSZ axion model~\cite{Dine:1981rt}.
In this case, the saxion can take a role of the curvaton.
The decay rate of this process is given by
\begin{equation}
	\Gamma(\sigma \to 2h) = \frac{1}{8\pi}\frac{m_\sigma^3}{f_a^2}\left( \frac{\mu}{m_\sigma}\right)^4,
\end{equation}
where $\mu$ denotes the so-called $\mu$-parameter in SUSY, which gives the higgsino mass.
If this is the dominant mode, the invisible branching ratio into an axion pair is estimated as
\begin{equation}
	1-r_\sigma = \frac{1}{8}\left( \frac{m_\sigma}{\mu}\right)^4,
\end{equation}
which can be very small if $\mu > m_\sigma$, and we have $r_\sigma \simeq 1$.
However, even if we neglect the nonthermal axions produced by the saxion decay,
there is another contribution from thermal production during reheating.
For simplicity we consider the case that axions are thermalized after the inflaton decays,
which occurs if the reheating temperature satisfies~\cite{Graf:2010tv}
\begin{equation}
	T_{\rm R} \gtrsim T_{\rm D}\equiv 9.6\times 10^6~{\rm GeV}\left( \frac{f_a}{10^{10}{\rm GeV}} \right)^{2.246}.
\end{equation}
If the reheating temperature is higher than this, axions are thermalized for $T>T_{\rm D}$
and decouple from thermal bath at $T\simeq T_{\rm D}$.
At this moment, the fraction of the axion energy density relative to the total radiation energy density is
given by $g_*(T=T_{\rm D})^{-1}$.
Thus we can effectively regard that the inflaton decayed into axions with branching fraction of 
$\simeq g_*(T=T_{\rm D})^{-1}$.
Therefore, we can replace $1-r_\phi$, which is defined at the curvaton (saxion) decay, with
\begin{equation}
	1-r_\phi = \frac{1}{g_*(T=T_{\rm D})} \left( \frac{g_*(H=\Gamma_\sigma)}{g_*(T=T_{\rm D})} \right)^{1/3}.
\end{equation}
From Eq.~(\ref{DeltaN_anticorr}), we obtain a well known expression 
for the $\Delta N_{\rm eff}$ for thermalized species (see e.g., Ref.~\cite{Nakayama:2010vs})
\begin{equation}
	\Delta N_{\rm eff} \simeq 
	 \frac{3(1-r_\phi)}{c_\nu}\left( \frac{g_*(H=\Gamma_\nu)}{g_*(H=\Gamma_\sigma)} \right)^{1/3}
	= \frac{4}{7} \left( \frac{g_*(H=\Gamma_\nu)}{g_*(T=T_{\rm D})} \right)^{4/3},
\end{equation}
for $R_\sigma \ll 1$. 
Otherwise, the saxion decay dilutes the thermally produced axions to a negligible level. 
Those thermal axions have an anti-correlated isocurvature perturbation to the curvature perturbation,
since the latter is assumed to be produced dominantly by the curvaton.
The magnitude of the isocurvature perturbation in this case is given by (\ref{S_anticorr}).

%%%%%%%%%%%%%%%%%%%%%%%%%%%%%%%%%%%
\section{Conclusions and discussion}      \label{sec:conc}
%%%%%%%%%%%%%%%%%%%%%%%%%%%%%%%%%%%

In this paper, we have investigated isocurvature perturbations
in an extra radiation component. We have firstly formulated how
primordial isocurvature perturbations between
the extra radiation and the Standard Model radiation 
can be generated from fluctuations of two scalar fields
in the inflationary Universe. We gave a detailed
description of what isocurvature perturbations
in the dark radiation are generated.
Our derivation of the isocuravature perturbations is based on 
the $\delta N$ formalism and fully non-linear. 
We have also pointed out that 
non-Gaussianities can  be generated 
in such isocurvature perturbations, which
are to be investigated in more detail in future works.
It would be straightforward to extend our formulation
for cases with three or more fields. We have also
discussed observational signatures of 
the isocurvature perturbations in the extra radiation.
We have pointed out that our model leads to distinct features
in the CMB power spectrum which are different from those predicted 
by the ordinary curvature and CDM/baryon 
isocurvature perturbations. We have shown that
that our model can be constrained from 
current cosmological observations.
Roughly speaking, CMB combined with BAO and the direct measurement
of $H_0$ gives constraints on the abundance of the
extra radiation and the amplitude of its isocurvature perturbations
as $3\le N_\mathrm{eff}\le5$ and  $\alpha\le0.1~(0.01)$ for the
uncorrelated (totally correlated/anti-correlated) case.
These are also translated into constraints on 
some limiting scenarios which can be realized in models
based on particle physics such as SUSY axion model.
%Assuming there exist some extra radiation components,
%we also investigated if there is any evidence for 
%socurvature perturbations in them.
%For the fixed $N_\mathrm{eff}=4$, we found that 
%the initial perturbations in the extra radiations are 
%consistent with adiabatic ones.

Our model will be tested more precisely by
cosmological observations which are ongoing or projected
in the near future. In particular, CMB power spectrum from the Planck survey 
is expected to improve constraints on both $N_\mathrm{eff}$ and $\alpha$
by an order of magnitude. Furthermore, 
non-Gaussianities and other observational signatures  would be also 
informative. 
%Improvements both in theoretical predictions 
%and observations are important to
%extract irreplaceable implications for cosmology
%and particle physics. 

%%%%%%%%%%%%%%%%%%%%%%%%%%%%%%%%%%%%
\section*{Acknowledgment}
%%%%%%%%%%%%%%%%%%%%%%%%%%%%%%%%%%%%

This work is also supported by Grant-in-Aid for Scientific research from
the Ministry of Education, Science, Sports, and Culture (MEXT), Japan,
No.\ 14102004 (M.K.), No.\ 21111006 (M.K. and K.N.), No. 23.10290 (K.M.), No.\ 22244030 (K.N.) and also 
by World Premier International Research Center
Initiative (WPI Initiative), MEXT, Japan. 
K.M. and T.S. would like to thank the Japan
Society for the Promotion of Science for financial support.

%%%%%%%%%%%%%%%%%%%%%%%%%%%%%%%%%%%%%%%%%%%%
%\end{acknowledgements}
%%%%%%%%%%%%%%%%%%%%%%%%%%%%%%%%%%%%%%%%%%%%

%%%%%%%%%%%%%%%%%%%%%%%%%%%%%%%%%%%
\appendix
%%%%%%%%%%%%%%%%%%%%%%%%%%%%%%%%%%%

%%%%%%%%%%%%%%%%%%%%%%%%%%%%%%%%%%%
\section{Extra radiation production after BBN}   
\label{sec:app}   
%%%%%%%%%%%%%%%%%%%%%%%%%%%%%%%%%%%

So far we have assumed that the $\sigma$ decays before the BBN and hence
the neutrino freezeout epoch.
This may not necessarily be the case. If the $\sigma$ mainly decays 
into the $X$ particle
without producing a significant amount of visible particles,
the decay of $\sigma$ does not upset BBN.
Thus it may be possible that the $\sigma$ decays after the neutrino 
freezeout but 
well before the recombination.
In this case the produced $X$ particles contribute to 
$\Delta N_{\rm eff}$ measured from  CMB
but does not affect $\Delta N_{\rm eff}$ measured from BBN.
Examples of concrete models were proposed in 
Ref.~\cite{Ichikawa:2007jv,Fischler:2010xz}. 
In this appendix we derive a formula in such a case.

The scenario we are considering here is summarized 
in Table~\ref{component:2}.
The curvaton $\sigma$ decays after BBN begins.
if the decay is too late, the effects of extra radiation on the CMB 
and LSS are much more different
and we do not consider such a case.
Thus we restrict the lifetime of the curvaton to be less than 
$\sim 10^8~{\rm sec}$.
By taking the uniform density slice at the curvaton decay, we have following relations
\begin{gather}
	\rho_X^{(\phi)}(\vec x) + (1-r_\sigma^{(\gamma)}-r_\sigma^{(\nu)}) \rho_\sigma (\vec x) = \rho_X(\vec x),\\
	\rho_\gamma^{(\phi)}(\vec x)+ r_\sigma^{(\gamma)} \rho_\sigma (\vec x) = \rho_\gamma(\vec x), \\
	\rho_\nu^{(\phi)}(\vec x)+ r_\sigma^{(\nu)} \rho_\sigma (\vec x) = \rho_\nu(\vec x), \\
	\rho_\gamma^{(\phi)}(\vec x)+\rho_\nu^{(\phi)}(\vec x)+\rho_X^{(\phi)}(\vec x)+ \rho_\sigma (\vec x) = 
	\rho_{\rm tot}(\vec x) (=\bar \rho_{\rm tot}).
\end{gather}
Here $r_\sigma^{(\gamma)}$ and $r_\sigma^{(\nu)}$ are the branching 
ratios of $\sigma$ into the photon and neutrino.\footnote{%
The decay products of $\sigma$, except for the neutrino and $X$, 
are thermalized.
They are collectively called ``photon'' here.
}
The curvature perturbation of each component is related to 
its background value as
\begin{gather}
	\rho_X^{(\phi)}(\vec x) =\bar \rho_X^{(\phi)} e^{4(\zeta_\phi-\hat\zeta)},\\
	\rho_\gamma^{(\phi)}(\vec x) =\bar \rho_\gamma^{(\phi)} e^{4(\zeta_\phi-\hat\zeta)},\\
	\rho_\nu^{(\phi)}(\vec x) =\bar \rho_\nu^{(\phi)} e^{4(\zeta_\phi-\hat\zeta)},\\
	\rho_\sigma(\vec x) =\bar \rho_\sigma e^{3(\zeta_\sigma-\hat\zeta)}.
\end{gather}

The requirement that the curvaton decay should not spoil BBN 
sets strong constraint on the branching ratio into 
the photon~\cite{Kawasaki:2004yh} and neutrino~\cite{Kanzaki:2007pd}.
Therefore we can safely set $r_\sigma^{(\gamma)}=r_\sigma^{(\nu)}=0$
since we are considering the case where the curvaton energy density is not negligible 
at the curvaton decay.
Then we soon find 
\begin{equation}
	\zeta_\gamma = \zeta_\phi,~~~\zeta_\nu=\zeta_\phi.
\end{equation}
The total curvature perturbation and the curvature perturbation in $X$ are found to be
\begin{equation}
\begin{split}
	\hat\zeta &= \zeta_\phi + R (\zeta_\sigma -\zeta_\phi) + \frac{1}{2}R(1-R)(3+R)(\zeta_\phi-\zeta_\sigma)^2, \\
	\zeta_X&=\frac{1}{4R_X}\left[ (4R_X^{(\phi)}+(1-R)R_X^{(\sigma)})\zeta_\phi 
	+(3+R)R_X^{(\sigma)}\zeta_\sigma  \right] \\
		&~~ +\frac{(3+R)R_X^{(\sigma)}}{8R_X^2}
		\left[ (1+R)(3-R)R_X^{(\phi)} + (1-R)R R_X^{(\sigma)} \right](\zeta_\phi-\zeta_\sigma)^2,
		\label{zeta_from_zetaphi_app}
\end{split}
\end{equation}
where $R$ is defined by Eq.~(\ref{R-Rsigma}) with 
$R_\sigma \equiv \bar \rho_{\sigma}/\bar \rho_{\rm tot}|_{H=\Gamma_\sigma}$.
Notice also that $R_X^{(\sigma)}=R_\sigma$ and $R_r^{(\sigma)}=0$ in the present case.
Other quantities are also defined as Eq.~(\ref{RX-Rr}).
Since there are no changes in the relativistic degrees of freedom after the curvaton decay,
the expression (\ref{zeta_from_zetaphi_app}) gives the final curvature perturbation.
The DR energy density is defined by
\begin{equation}
	\rho_{\nu}^{(\phi)}(\vec x)+\rho_{X}^{(\phi)}(\vec x)+\rho_\sigma(\vec x) = \rho_{\rm DR} (\vec x),
\end{equation}
on the uniform density slice,
where $\rho_{\rm DR}(\vec x) = \bar \rho_{\rm DR} e^{4(\zeta_{\rm DR}-\hat\zeta)}$. 
The DR isocurvature perturbation is calculated as
\begin{equation}
\begin{split}
	\hat S_{\rm DR} &\equiv 3(\zeta_{\rm DR}-\hat\zeta)  \\
	& = -3\frac{R_r R}{R_{\rm DR}}
	(1-\hat c_\nu) 
	\left [ -(\zeta_\sigma-\zeta_\phi)
	   +\left\{ \frac{(R-3)(R+1)}{2}+\frac{2R}{R_{\rm DR}}
		(1+ R_{\rm DR})
	\right\} (\zeta_\sigma-\zeta_\phi)^2
	\right],
	\label{SDR_app}
\end{split}
\end{equation}
where $\hat c_\nu$ is given in Eq.~(\ref{hatcnu}),
$R_{r}=(\bar\rho_{\gamma}+\bar\rho_\nu)/\bar\rho_{\rm tot}$
and $R_{\rm DR}=(\bar\rho_{X}+\bar\rho_\nu)/\bar\rho_{\rm tot}$ evaluated after the curvaton decay.

%%%%%%%%%%%%%%%% table %%%%%%%%%%%%%%%%%%%%%%
\begin{table}[t]
  \begin{center}
    \begin{tabular}{ | c | c | c |}
      \hline 
         epoch  & component &  energy transfer  \\ \hline \hline
         $\Gamma_\phi < H  $  
            & $\phi$, $\sigma$  
            & $\phi \to X^{(\phi)}+r^{(\phi)} $ \\ \hline 
         $\Gamma_\nu < H < \Gamma_\phi$  
            & $X^{(\phi)}$, $r^{(\phi)}$, $\sigma$ 
            &  $r^{(\phi)} \to \nu + r_e$ \\ \hline 
         $\Gamma_{e^\pm} < H <\Gamma_\nu$   
            & $X^{(\phi)}$, $\nu$, $r_e$, $\sigma$
            &  $e^\pm \to \gamma$   \\ \hline
         $\Gamma_\sigma < H<\Gamma_{e^\pm}$  
            & $X^{(\phi)}$, $\nu$, $\gamma$, $\sigma$
            & $\sigma \to X^{(\sigma)} $ \\ \hline
         $H<\Gamma_{\sigma}$         
            & $X$, $\nu$, $\gamma$  (${\rm DR}=X+\nu$) 
            &   \\ \hline
    \end{tabular}
    \caption{ 
    	Same as Table~\ref{component} but for the case of $\Gamma_\sigma < \Gamma_{e^\pm}$.
	The $\sigma$ is assumed to dominantly decay into $X$. 
    }
    \label{component:2}
  \end{center}
\end{table}
%%%%%%%%%%%%%%%%%%%%%%%%%%%%%%%%%%%%%%%%%%%%%% 

The extra effective number of neutrino species, $\Delta N_{\rm eff}$, is given by
\begin{equation}
	\Delta N_{\rm eff} =3\frac{\rho_X}{\rho_\nu} = \frac{3 R_\sigma}{\hat c_\nu R_r}.
\end{equation}
This is explicitly evaluated as
\begin{equation}
	\Delta N_{\rm eff} \simeq 1.1 \left( \frac{1{\rm keV}}{T_\sigma} \right)
	\left( \frac{\rho_\sigma/s}{10^{-7}{\rm GeV}} \right).
\end{equation}
%%

%%%%%%%%%%%%%%%%%%%%%%%%%%%%%%%%%%%%
{}
%%%%%%%%%%%%%%%%%%%%%%%%%%%%%%%%%%%%


\begin{thebibliography}{99}
%%%%%%%%%%%%%%%%%%%%%%%%%%%%%%%%%%%%

  
%\cite{Komatsu:2010fb}
\bibitem{Komatsu:2010fb}
  E.~Komatsu {\it et al.} [ WMAP Collaboration ],
  %``Seven-Year Wilkinson Microwave Anisotropy Probe (WMAP) Observations: Cosmological Interpretation,''
  Astrophys.\ J.\ Suppl.\  {\bf 192}, 18 (2011).
  [arXiv:1001.4538 [astro-ph.CO]].

%\cite{Dunkley:2010ge}
\bibitem{Dunkley:2010ge}
  J.~Dunkley, R.~Hlozek, J.~Sievers {\it et al.},
  %``The Atacama Cosmology Telescope: Cosmological Parameters from the 2008 Power Spectra,''
  [arXiv:1009.0866 [astro-ph.CO]].
  
  %\cite{Keisler:2011aw}
\bibitem{Keisler:2011aw}
  R.~Keisler, C.~L.~Reichardt, K.~A.~Aird, B.~A.~Benson, L.~E.~Bleem, J.~E.~Carlstrom, C.~L.~Chang, H.~M.~Cho {\it et al.},
  %``A Measurement of the Damping Tail of the Cosmic Microwave Background Power Spectrum with the South Pole Telescope,''
  [arXiv:1105.3182 [astro-ph.CO]].

%\cite{GonzalezMorales:2011ty}
\bibitem{GonzalezMorales:2011ty}
  A.~X.~Gonzalez-Morales, R.~Poltis, B.~D.~Sherwin and L.~Verde,
  %``Are priors responsible for cosmology favoring additional neutrino
  %species?,''
  arXiv:1106.5052 [astro-ph.CO].
  %%CITATION = ARXIV:1106.5052;%%
  
%\cite{Hamann:2011hu}
\bibitem{Hamann:2011hu}
  J.~Hamann,
  %``Evidence for extra radiation? Profile likelihood versus Bayesian
  %posterior,''
  arXiv:1110.4271 [astro-ph.CO].
  %%CITATION = ARXIV:1110.4271;%%  
  
 %\cite{Mangano:2006ur}
\bibitem{Mangano:2006ur}
  G.~Mangano, A.~Melchiorri, O.~Mena, G.~Miele, A.~Slosar,
  %``Present bounds on the relativistic energy density in the Universe from cosmological observables,''
  JCAP {\bf 0703}, 006 (2007).
  [astro-ph/0612150].
  
  %\cite{Cirelli:2006kt}
\bibitem{Cirelli:2006kt}
  M.~Cirelli, A.~Strumia,
  %``Cosmology of neutrinos and extra light particles after WMAP3,''
  JCAP {\bf 0612}, 013 (2006).
  [astro-ph/0607086].
    
  %\cite{Simha:2008zj}
\bibitem{Simha:2008zj}
  V.~Simha, G.~Steigman,
  %``Constraining The Early-Universe Baryon Density And Expansion Rate,''
  JCAP {\bf 0806}, 016 (2008).
  [arXiv:0803.3465 [astro-ph]].
  
 %\cite{Ichikawa:2008pz}
\bibitem{Ichikawa:2008pz}
  K.~Ichikawa, T.~Sekiguchi, T.~Takahashi,
  %``Probing the Effective Number of Neutrino Species with Cosmic Microwave Background,''
  Phys.\ Rev.\  {\bf D78}, 083526 (2008).
  [arXiv:0803.0889 [astro-ph]].
  
  %\cite{Izotov:2010ca}
\bibitem{Izotov:2010ca}
  Y.~I.~Izotov, T.~X.~Thuan,
  %``The primordial abundance of 4He: evidence for non-standard big bang nucleosynthesis,''
  Astrophys.\ J.\  {\bf 710}, L67-L71 (2010).
  [arXiv:1001.4440 [astro-ph.CO]].
  
  %\cite{Aver:2010wq}
\bibitem{Aver:2010wq}
  E.~Aver, K.~A.~Olive, E.~D.~Skillman,
  %``A New Approach to Systematic Uncertainties and Self-Consistency in Helium Abundance Determinations,''
  JCAP {\bf 1005}, 003 (2010).
  [arXiv:1001.5218 [astro-ph.CO]];
    %\cite{Aver:2010wd}
%\bibitem{Aver:2010wd}
  E.~Aver, K.~A.~Olive, E.~D.~Skillman,
  %``Mapping systematic errors in helium abundance determinations using Markov Chain Monte Carlo,''
  [arXiv:1012.2385 [astro-ph.CO]].
  
 %\cite{Hamann:2010bk}
\bibitem{Hamann:2010bk}
  J.~Hamann, S.~Hannestad, G.~G.~Raffelt, I.~Tamborra, Y.~Y.~Y.~Wong,
  %``Cosmology seeking friendship with sterile neutrinos,''
  Phys.\ Rev.\ Lett.\  {\bf 105}, 181301 (2010).
  [arXiv:1006.5276 [hep-ph]].
  
%\cite{Hansen:2001hi}
\bibitem{Hansen:2001hi}
  S.~H.~Hansen, G.~Mangano, A.~Melchiorri, G.~Miele, O.~Pisanti,
  %``Constraining neutrino physics with BBN and CMBR,''
  Phys.\ Rev.\  {\bf D65}, 023511 (2002).
  [astro-ph/0105385].
  
  %\cite{Kawasaki:2002hq}
\bibitem{Kawasaki:2002hq}
  M.~Kawasaki, F.~Takahashi, M.~Yamaguchi,
  %``Large lepton asymmetry from Q balls,''
  Phys.\ Rev.\  {\bf D66}, 043516 (2002).
  [hep-ph/0205101].
  
  %\cite{Popa:2008tb}
\bibitem{Popa:2008tb}
  L.~A.~Popa, A.~Vasile,
  %``WMAP 5-year constraints on lepton asymmetry and radiation energy density: Implications for Planck,''
  JCAP {\bf 0806}, 028 (2008).
  [arXiv:0804.2971 [astro-ph]].
  
 %\cite{Mangano:2010ei}
\bibitem{Mangano:2010ei}
  G.~Mangano, G.~Miele, S.~Pastor, O.~Pisanti, S.~Sarikas,
  %``Constraining the cosmic radiation density due to lepton number with Big Bang Nucleosynthesis,''
  JCAP {\bf 1103}, 035 (2011).
  [arXiv:1011.0916 [astro-ph.CO]].
    
  %\cite{Ichikawa:2007jv}
\bibitem{Ichikawa:2007jv}
  K.~Ichikawa, M.~Kawasaki, K.~Nakayama, M.~Senami, F.~Takahashi,
  %``Increasing effective number of neutrinos by decaying particles,''
  JCAP {\bf 0705}, 008 (2007).
  [hep-ph/0703034 [HEP-PH]].
  
  %\cite{Krauss:2010xg}
\bibitem{Krauss:2010xg}
  L.~M.~Krauss, C.~Lunardini, C.~Smith,
  %``Neutrinos, WMAP, and BBN,''
  %Submitted to: Phys.Rev.D.
  [arXiv:1009.4666 [hep-ph]].
  
 %\cite{Nakayama:2010vs}
\bibitem{Nakayama:2010vs}
  K.~Nakayama, F.~Takahashi, T.~T.~Yanagida,
  %``A theory of extra radiation in the Universe,''
  Phys.\ Lett.\  {\bf B697}, 275-279 (2011).
  [arXiv:1010.5693 [hep-ph]].
  
%\cite{Fischler:2010xz}
\bibitem{Fischler:2010xz}
  W.~Fischler, J.~Meyers,
  %``Dark Radiation Emerging After Big Bang Nucleosynthesis?,''
  Phys.\ Rev.\  {\bf D83}, 063520 (2011).
  [arXiv:1011.3501 [astro-ph.CO]].
  
  %\cite{deHolanda:2010am}
\bibitem{deHolanda:2010am}
  P.~C.~de Holanda, A.~Y.~.Smirnov,
  %``Solar neutrino spectrum, sterile neutrinos and additional radiation in the Universe,''
  [arXiv:1012.5627 [hep-ph]].
  
  %\cite{Lyth:2001nq}
\bibitem{Lyth:2001nq}
  D.~H.~Lyth, D.~Wands,
  %``Generating the curvature perturbation without an inflaton,''
  Phys.\ Lett.\  {\bf B524}, 5-14 (2002).
  [hep-ph/0110002];
  %\cite{Moroi:2001ct}
%\bibitem{Moroi:2001ct}
  T.~Moroi, T.~Takahashi,
  %``Effects of cosmological moduli fields on cosmic microwave background,''
  Phys.\ Lett.\  {\bf B522}, 215-221 (2001).
  [hep-ph/0110096];
  %\cite{Enqvist:2001zp}
%\bibitem{Enqvist:2001zp}
  K.~Enqvist, M.~S.~Sloth,
  %``Adiabatic CMB perturbations in pre - big bang string cosmology,''
  Nucl.\ Phys.\  {\bf B626}, 395-409 (2002).
  [hep-ph/0109214].
  
%\cite{Kawasaki:2008sn}
\bibitem{Kawasaki:2008sn}
  M.~Kawasaki, K.~Nakayama, T.~Sekiguchi, T.~Suyama, F.~Takahashi,
  %``Non-Gaussianity from isocurvature perturbations,''
  JCAP {\bf 0811}, 019 (2008).
  [arXiv:0808.0009 [astro-ph]].

%\cite{Kawasaki:2008jy}
\bibitem{Kawasaki:2008jy}
  M.~Kawasaki, K.~Nakayama, F.~Takahashi,
  %``Non-Gaussianity from Baryon Asymmetry,''
  JCAP {\bf 0901}, 002 (2009).
  [arXiv:0809.2242 [hep-ph]].
  
  %\cite{Langlois:2008vk}
\bibitem{Langlois:2008vk}
  D.~Langlois, F.~Vernizzi, D.~Wands,
  %``Non-linear isocurvature perturbations and non-Gaussianities,''
  JCAP {\bf 0812}, 004 (2008).
  [arXiv:0809.4646 [astro-ph]].
  
  %\cite{Kawasaki:2008pa}
\bibitem{Kawasaki:2008pa}
  M.~Kawasaki, K.~Nakayama, T.~Sekiguchi, T.~Suyama, F.~Takahashi,
  %``A General Analysis of Non-Gaussianity from Isocurvature Perturbations,''
  JCAP {\bf 0901}, 042 (2009).
  [arXiv:0810.0208 [astro-ph]].
  
  %\cite{Hikage:2008sk}
\bibitem{Hikage:2008sk}
  C.~Hikage, K.~Koyama, T.~Matsubara, T.~Takahashi, M.~Yamaguchi,
  %``Limits on Isocurvature Perturbations from Non-Gaussianity in WMAP Temperature Anisotropy,''
  Mon.\ Not.\ Roy.\ Astron.\ Soc.\  {\bf 398}, 2188-2198 (2009).
  [arXiv:0812.3500 [astro-ph]].
  
  %\cite{Kawakami:2009iu}
\bibitem{Kawakami:2009iu}
  E.~Kawakami, M.~Kawasaki, K.~Nakayama, F.~Takahashi,
  %``Non-Gaussianity from Isocurvature Perturbations : Analysis of Trispectrum,''
  JCAP {\bf 0909}, 002 (2009).
  [arXiv:0905.1552 [astro-ph.CO]].
  
  %\cite{Hikage:2009rt}
\bibitem{Hikage:2009rt}
  C.~Hikage, D.~Munshi, A.~Heavens, P.~Coles,
  %``Adiabatic versus Isocurvature Non--Gaussianity,''
  Mon.\ Not.\ Roy.\ Astron.\ Soc.\  {\bf 404}, 1505-1511 (2010).
  [arXiv:0907.0261 [astro-ph.CO]].
  
  %\cite{Nakayama:2009cr}
\bibitem{Nakayama:2009cr}
  K.~Nakayama, F.~Takahashi,
  %``The R-axion and non-Gaussianity,''
  Phys.\ Lett.\  {\bf B679}, 436-439 (2009).
  [arXiv:0907.0834 [hep-ph]].
  
  %\cite{Langlois:2010dz}
\bibitem{Langlois:2010dz}
  D.~Langlois, A.~Lepidi,
  %``General treatment of isocurvature perturbations and non-Gaussianities,''
  JCAP {\bf 1101}, 008 (2011).
  [arXiv:1007.5498 [astro-ph.CO]].
  
 %\cite{Langlois:2010fe}
\bibitem{Langlois:2010fe}
  D.~Langlois, T.~Takahashi,
  %``Primordial Trispectrum from Isocurvature Fluctuations,''
  JCAP {\bf 1102}, 020 (2011).
  [arXiv:1012.4885 [astro-ph.CO]].
  
  %\cite{Langlois:2011hn}
\bibitem{Langlois:2011hn}
  D.~Langlois, B.~van Tent,
  %``Hunting for Isocurvature Modes in the CMB non-Gaussianities,''
  [arXiv:1104.2567 [astro-ph.CO]].

%\cite{Hu:1998gy}
\bibitem{Hu:1998gy}
  W.~Hu,
  %``An Isocurvature mechanism for structure formation,''
  Phys.\ Rev.\  D {\bf 59}, 021301 (1999)
  [arXiv:astro-ph/9809142].
  %%CITATION = PHRVA,D59,021301;%%
    
    %\cite{Sasaki:1995aw}
\bibitem{Sasaki:1995aw}
  M.~Sasaki, E.~D.~Stewart,
  %``A General analytic formula for the spectral index of the density perturbations produced during inflation,''
  Prog.\ Theor.\ Phys.\  {\bf 95}, 71-78 (1996).
  [astro-ph/9507001].
  
  %\cite{Lyth:2004gb}
\bibitem{Lyth:2004gb}
  D.~H.~Lyth, K.~A.~Malik, M.~Sasaki,
  %``A General proof of the conservation of the curvature perturbation,''
  JCAP {\bf 0505}, 004 (2005).
  [astro-ph/0411220].
  
  %\cite{Wands:2000dp}
\bibitem{Wands:2000dp}
  D.~Wands, K.~A.~Malik, D.~H.~Lyth, A.~R.~Liddle,
  %``A New approach to the evolution of cosmological perturbations on large scales,''
  Phys.\ Rev.\  {\bf D62}, 043527 (2000).
  [astro-ph/0003278].
  
  %\cite{Sasaki:2006kq}
\bibitem{Sasaki:2006kq}
  M.~Sasaki, J.~Valiviita, D.~Wands,
  %``Non-Gaussianity of the primordial perturbation in the curvaton model,''
  Phys.\ Rev.\  {\bf D74}, 103003 (2006).
  [astro-ph/0607627].

%\cite{Enqvist:2005pg}
\bibitem{Enqvist:2005pg}
  K.~Enqvist, S.~Nurmi,
  %``Non-gaussianity in curvaton models with nearly quadratic potential,''
  JCAP {\bf 0510}, 013 (2005).
  [astro-ph/0508573];
  %\cite{Enqvist:2008gk}
%\bibitem{Enqvist:2008gk}
  K.~Enqvist, T.~Takahashi,
  %``Signatures of Non-Gaussianity in the Curvaton Model,''
  JCAP {\bf 0809}, 012 (2008).
  [arXiv:0807.3069 [astro-ph]];
  %\cite{Huang:2008bg}
%\bibitem{Huang:2008bg}
  Q.~-G.~Huang, Y.~Wang,
  %``Curvaton Dynamics and the Non-Linearity Parameters in Curvaton Model,''
  JCAP {\bf 0809}, 025 (2008).
  [arXiv:0808.1168 [hep-th]].
  
  %\cite{Kawasaki:2008mc}
\bibitem{Kawasaki:2008mc}
  M.~Kawasaki, K.~Nakayama, F.~Takahashi,
  %``Hilltop Non-Gaussianity,''
  JCAP {\bf 0901}, 026 (2009).
  [arXiv:0810.1585 [hep-ph]];
  %\cite{Chingangbam:2009xi}
%\bibitem{Chingangbam:2009xi}
  P.~Chingangbam, Q.~-G.~Huang,
  %``The Curvature Perturbation in the Axion-type Curvaton Model,''
  JCAP {\bf 0904}, 031 (2009).
  [arXiv:0902.2619 [astro-ph.CO]];
  %\cite{Huang:2010cy}
%\bibitem{Huang:2010cy}
  Q.~-G.~Huang,
  %``Negative spectral index of $f_{NL}$ in the axion-type curvaton model,''
  JCAP {\bf 1011}, 026 (2010).
  [arXiv:1008.2641 [astro-ph.CO]].
  
  %\cite{Ichikawa:2008iq}
\bibitem{Ichikawa:2008iq}
  K.~Ichikawa, T.~Suyama, T.~Takahashi, M.~Yamaguchi,
  %``Non-Gaussianity, Spectral Index and Tensor Modes in Mixed Inflaton and Curvaton Models,''
  Phys.\ Rev.\  {\bf D78}, 023513 (2008).
  [arXiv:0802.4138 [astro-ph]].
  
%\cite{Bucher:1999re}
\bibitem{Bucher:1999re}
  M.~Bucher, K.~Moodley, N.~Turok,
  %``The General primordial cosmic perturbation,''
  Phys.\ Rev.\  {\bf D62}, 083508 (2000).
  [astro-ph/9904231].
  
%\cite{Trotta:2004qj}
\bibitem{Trotta:2004qj}
  R.~Trotta,
  %``Cosmic microwave background anisotropies: Beyond standard parameters,''
  [astro-ph/0410115].

%\cite{Lewis:1999bs}
\bibitem{Lewis:1999bs}
  A.~Lewis, A.~Challinor, A.~Lasenby,
  %``Efficient computation of CMB anisotropies in closed FRW models,''
  Astrophys.\ J.\  {\bf 538}, 473-476 (2000).
  [astro-ph/9911177].
  
%\cite{Gordon:2002gv}
\bibitem{Gordon:2002gv}
  C.~Gordon, A.~Lewis,
  %``Observational constraints on the curvaton model of inflation,''
  Phys.\ Rev.\  {\bf D67}, 123513 (2003).
  [astro-ph/0212248].
  
%\cite{Kawasaki:2011ze}
\bibitem{Kawasaki:2011ze}
  M.~Kawasaki, T.~Sekiguchi, T.~Takahashi,
  %``Differentiating CDM and Baryon Isocurvature Models with 21 cm Fluctuations,''
  [arXiv:1104.5591 [astro-ph.CO]].

%\cite{Hu:1994uz}
\bibitem{Hu:1994uz}
  W.~Hu, N.~Sugiyama,
  %``Anisotropies in the cosmic microwave background: An Analytic approach,''
  Astrophys.\ J.\  {\bf 444}, 489-506 (1995).
  [astro-ph/9407093].

%\cite{Zunckel:2010mm}
\bibitem{Zunckel:2010mm}
  C.~Zunckel, P.~Okouma, S.~M.~Kasanda, K.~Moodley and B.~A.~Bassett,
  %``Fundamental Uncertainty in the BAO Scale from Isocurvature Modes,''
  Phys.\ Lett.\  B {\bf 696}, 433 (2011)
  [arXiv:1006.4687 [astro-ph.CO]].
  %%CITATION = PHLTA,B696,433;%%
  
%\cite{Bashinsky:2003tk}
\bibitem{Bashinsky:2003tk}
  S.~Bashinsky, U.~Seljak,
  %``Neutrino perturbations in CMB anisotropy and matter clustering,''
  Phys.\ Rev.\  {\bf D69}, 083002 (2004).
  [astro-ph/0310198].

%\cite{Gold:2010fm}
\bibitem{Gold:2010fm}
  B.~Gold {\it et al.},
  %``Seven-Year Wilkinson Microwave Anisotropy Probe (WMAP) Observations:
  %Galactic Foreground Emission,''
  Astrophys.\ J.\ Suppl.\  {\bf 192}, 15 (2011)
  [arXiv:1001.4555 [astro-ph.GA]].
  %%CITATION = APJSA,192,15;%%
  
%\cite{Larson:2010gs}
\bibitem{Larson:2010gs}
  D.~Larson {\it et al.},
  %``Seven-Year Wilkinson Microwave Anisotropy Probe (WMAP) Observations: Power
  %Spectra and WMAP-Derived Parameters,''
  Astrophys.\ J.\ Suppl.\  {\bf 192}, 16 (2011)
  [arXiv:1001.4635 [astro-ph.CO]].
  %%CITATION = APJSA,192,16;%%
    
%\cite{Jarosik:2010iu}
\bibitem{Jarosik:2010iu}
  N.~Jarosik {\it et al.},
  %``Seven-Year Wilkinson Microwave Anisotropy Probe (WMAP) Observations: Sky
  %Maps, Systematic Errors, and Basic Results,''
  Astrophys.\ J.\ Suppl.\  {\bf 192}, 14 (2011)
  [arXiv:1001.4744 [astro-ph.CO]].
  %%CITATION = APJSA,192,14;%%

%\cite{Hajian:2010fj}
\bibitem{Hajian:2010fj}
  A.~Hajian {\it et al.},
  %``The Atacama Cosmology Telescope: Calibration with WMAP Using
  %Cross-Correlations,''
  arXiv:1009.0777 [astro-ph.CO].
  %%CITATION = ARXIV:1009.0777;%%
      
%\cite{Das:2010ga}
\bibitem{Das:2010ga}
  S.~Das {\it et al.},
  %``The Atacama Cosmology Telescope: A Measurement of the Cosmic Microwave
  %Background Power Spectrum at 148 and 218 GHz from the 2008 Southern Survey,''
  Astrophys.\ J.\  {\bf 729}, 62 (2011)
  [arXiv:1009.0847 [astro-ph.CO]].
  %%CITATION = ASJOA,729,62;%%

%\cite{Percival:2009xn}
\bibitem{Percival:2009xn}
  B.~A.~Reid {\it et al.}  [SDSS Collaboration],
  %``Baryon Acoustic Oscillations in the Sloan Digital Sky Survey Data Release 7
  %Galaxy Sample,''
  Mon.\ Not.\ Roy.\ Astron.\ Soc.\  {\bf 401}, 2148 (2010)
  [arXiv:0907.1660 [astro-ph.CO]].
  %%CITATION = MNRAA,401,2148;%%
  
%\cite{Riess:2009pu}
\bibitem{Riess:2009pu}
  A.~G.~Riess {\it et al.},
  %``A Redetermination of the Hubble Constant with the Hubble Space Telescope
  %from a Differential Distance Ladder,''
  Astrophys.\ J.\  {\bf 699}, 539 (2009)
  [arXiv:0905.0695 [astro-ph.CO]].
  %%CITATION = ASJOA,699,539;%%

%\cite{Sehgal:2009xv}
\bibitem{Sehgal:2009xv}
  N.~Sehgal {\it et al.},
  %``Simulations of the Microwave Sky,''
  Astrophys.\ J.\  {\bf 709}, 920 (2010)
  [arXiv:0908.0540 [astro-ph.CO]].
  %%CITATION = ASJOA,709,920;%%
  
  %%% NID %%%

%\cite{Trotta:2002iz}
\bibitem{Trotta:2002iz}
  R.~Trotta, A.~Riazuelo and R.~Durrer,
  %``The cosmological constant and general isocurvature initial conditions,''
  Phys.\ Rev.\  D {\bf 67}, 063520 (2003)
  [arXiv:astro-ph/0211600].
  %%CITATION = PHRVA,D67,063520;%%

%\cite{Moodley:2004nz}
\bibitem{Moodley:2004nz}
  K.~Moodley, M.~Bucher, J.~Dunkley, P.~G.~Ferreira and C.~Skordis,
  %``Constraints on isocurvature models from the WMAP first-year data,''
  Phys.\ Rev.\  D {\bf 70}, 103520 (2004)
  [arXiv:astro-ph/0407304].
  %%CITATION = PHRVA,D70,103520;%%

%\cite{Beltran:2004uv}
\bibitem{Beltran:2004uv}
  M.~Beltran, J.~Garcia-Bellido, J.~Lesgourgues and A.~Riazuelo,
  %``Bounds on CDM and neutrino isocurvature perturbations from CMB and LSS
  %data,''
  Phys.\ Rev.\  D {\bf 70}, 103530 (2004)
  [arXiv:astro-ph/0409326].
  %%CITATION = PHRVA,D70,103530;%%

%\cite{Bean:2006qz}
\bibitem{Bean:2006qz}
  R.~Bean, J.~Dunkley and E.~Pierpaoli,
  %``Constraining Isocurvature Initial Conditions with WMAP 3-year data,''
  Phys.\ Rev.\  D {\bf 74}, 063503 (2006)
  [arXiv:astro-ph/0606685].
  %%CITATION = PHRVA,D74,063503;%%
  
%\cite{Trotta:2006ww}
\bibitem{Trotta:2006ww}
  R.~Trotta,
  %``The isocurvature fraction after WMAP 3-year data,''
  Mon.\ Not.\ Roy.\ Astron.\ Soc.\  {\bf 375}, L26 (2007)
  [arXiv:astro-ph/0608116].
  %%CITATION = MNRAA,375,L26;%%
  
%\cite{Kawasaki:2007mb}
\bibitem{Kawasaki:2007mb}
  M.~Kawasaki and T.~Sekiguchi,
  %``Cosmological Constraints on Isocurvature and Tensor Perturbations,''
  Prog.\ Theor.\ Phys.\  {\bf 120}, 995 (2008)
  [arXiv:0705.2853 [astro-ph]].
  %%CITATION = PTPKA,120,995;%%  

  %%% NID %%%
  
%\cite{Lewis:2002ah}
\bibitem{Lewis:2002ah}
  A.~Lewis, S.~Bridle,
  %``Cosmological parameters from CMB and other data: A Monte Carlo approach,''
  Phys.\ Rev.\  {\bf D66}, 103511 (2002).
  [astro-ph/0205436].

%\cite{Kim:1986ax}
\bibitem{Kim:1986ax}
  For reviews, see J.~E.~Kim,
  %``Light Pseudoscalars, Particle Physics and Cosmology,''
  Phys.\ Rept.\  {\bf 150}, 1-177 (1987);
  %\cite{Kim:2008hd}
%\bibitem{Kim:2008hd}
  J.~E.~Kim, G.~Carosi,
  %``Axions and the Strong CP Problem,''
  Rev.\ Mod.\ Phys.\  {\bf 82}, 557-602 (2010).
  [arXiv:0807.3125 [hep-ph]].

%\cite{Rajagopal:1990yx}
\bibitem{Rajagopal:1990yx}
  K.~Rajagopal, M.~S.~Turner, F.~Wilczek,
  %``Cosmological implications of axinos,''
  Nucl.\ Phys.\  {\bf B358}, 447-470 (1991);
  %\cite{Goto:1991gq}
%\bibitem{Goto:1991gq}
  T.~Goto, M.~Yamaguchi,
  %``Is axino dark matter possible in supergravity?,''
  Phys.\ Lett.\  {\bf B276}, 103-107 (1992).
  
  %\cite{Chun:1995hc}
\bibitem{Chun:1995hc}
  E.~J.~Chun, A.~Lukas,
  %``Axino mass in supergravity models,''
  Phys.\ Lett.\  {\bf B357}, 43-50 (1995).
  [hep-ph/9503233].

%\cite{Asaka:1998ns}
\bibitem{Asaka:1998ns}
  T.~Asaka, M.~Yamaguchi,
  %``Hadronic axion model in gauge mediated supersymmetry breaking,''
  Phys.\ Lett.\  {\bf B437}, 51-61 (1998).
  [hep-ph/9805449];
  %\cite{Asaka:1998xa}
%\bibitem{Asaka:1998xa}
  %T.~Asaka, M.~Yamaguchi,
  %``Hadronic axion model in gauge mediated supersymmetry breaking and cosmology of saxion,''
  Phys.\ Rev.\  {\bf D59}, 125003 (1999).
  [hep-ph/9811451].
  
  %\cite{Abe:2001cg}
\bibitem{Abe:2001cg}
  N.~Abe, T.~Moroi, M.~Yamaguchi,
  %``Anomaly mediated supersymmetry breaking with axion,''
  JHEP {\bf 0201}, 010 (2002).
  [hep-ph/0111155].
  
%\cite{Kawasaki:2007mk}
\bibitem{Kawasaki:2007mk}
  M.~Kawasaki, K.~Nakayama, M.~Senami,
  %``Cosmological implications of supersymmetric axion models,''
  JCAP {\bf 0803}, 009 (2008).
  [arXiv:0711.3083 [hep-ph]];
  %\cite{Kawasaki:2008jc}
%\bibitem{Kawasaki:2008jc}
  M.~Kawasaki, K.~Nakayama,
  %``Solving Cosmological Problems of Supersymmetric Axion Models in Inflationary Universe,''
  Phys.\ Rev.\  {\bf D77}, 123524 (2008).
  [arXiv:0802.2487 [hep-ph]].
  
  %\cite{Nakamura:2008ey}
\bibitem{Nakamura:2008ey}
  S.~Nakamura, K.~-i.~Okumura, M.~Yamaguchi,
  %``Axionic Mirage Mediation,''
  Phys.\ Rev.\  {\bf D77}, 115027 (2008).
  [arXiv:0803.3725 [hep-ph]].
  
  %\cite{Kim:2008yu}
\bibitem{Kim:2008yu}
  S.~Kim, W.~-I.~Park, E.~D.~Stewart,
  %``Thermal inflation, baryogenesis and axions,''
  JHEP {\bf 0901}, 015 (2009).
  [arXiv:0807.3607 [hep-ph]].
    
  %\cite{Kim:1979if}
\bibitem{Kim:1979if}
  J.~E.~Kim,
  %``Weak Interaction Singlet and Strong CP Invariance,''
  Phys.\ Rev.\ Lett.\  {\bf 43}, 103 (1979);
  %\cite{Shifman:1979if}
%\bibitem{Shifman:1979if}
  M.~A.~Shifman, A.~I.~Vainshtein, V.~I.~Zakharov,
  %``Can Confinement Ensure Natural CP Invariance of Strong Interactions?,''
  Nucl.\ Phys.\  {\bf B166}, 493 (1980).
  
%\cite{Dine:1981rt}
\bibitem{Dine:1981rt}
  M.~Dine, W.~Fischler, M.~Srednicki,
  %``A Simple Solution to the Strong CP Problem with a Harmless Axion,''
  Phys.\ Lett.\  {\bf B104}, 199 (1981);
%\cite{Zhitnitsky:1980tq}
%\bibitem{Zhitnitsky:1980tq}
  A.~R.~Zhitnitsky,
  %``On Possible Suppression Of The Axion Hadron Interactions. (In Russian),''
  Sov.\ J.\ Nucl.\ Phys.\  {\bf 31}, 260 (1980)
  [Yad.\ Fiz.\  {\bf 31}, 497 (1980)].
  %%CITATION = YAFIA,31,497;%%

%\cite{Graf:2010tv}
\bibitem{Graf:2010tv}
  P.~Graf, F.~D.~Steffen,
  %``Thermal axion production in the primordial quark-gluon plasma,''
  Phys.\ Rev.\  {\bf D83}, 075011 (2011).
  [arXiv:1008.4528 [hep-ph]].

%\cite{Kawasaki:2004yh}
\bibitem{Kawasaki:2004yh}
  M.~Kawasaki, K.~Kohri, T.~Moroi,
  %``Hadronic decay of late - decaying particles and Big-Bang Nucleosynthesis,''
  Phys.\ Lett.\  {\bf B625}, 7-12 (2005).
  [astro-ph/0402490];
  %\cite{Kawasaki:2004qu}
%\bibitem{Kawasaki:2004qu}
  %M.~Kawasaki, K.~Kohri, T.~Moroi,
  %``Big-Bang nucleosynthesis and hadronic decay of long-lived massive particles,''
  Phys.\ Rev.\  {\bf D71}, 083502 (2005).
  [astro-ph/0408426].

%\cite{Kanzaki:2007pd}
\bibitem{Kanzaki:2007pd}
  T.~Kanzaki, M.~Kawasaki, K.~Kohri, T.~Moroi,
  %``Cosmological Constraints on Neutrino Injection,''
  Phys.\ Rev.\  {\bf D76}, 105017 (2007).
  [arXiv:0705.1200 [hep-ph]].


%%%%%%%%%%%%%%%%%%%%%%%%%%%%%%%%%%%%
\end{thebibliography}
\end{document}